\documentclass[oneside,numbers,spanish]{ezthesis}
\usepackage{amsmath}
\usepackage{amsfonts}
\usepackage{cancel}
\usepackage{xfrac}
\pdfoutput=1
\usepackage{epsfig,graphicx,bm,amsmath}
\usepackage{amssymb}
\usepackage{float}
\usepackage{subfigure}

\renewcommand{\(}{\left(}
\renewcommand{\)}{\right)}
\renewcommand{\[}{\left[}
\renewcommand{\]}{\right]}

\newcommand{\pa}{\partial}

\newcommand{\be}{\begin{equation}}
\newcommand{\ee}{\end{equation}}
\newcommand{\bba}{\begin{align}}
\newcommand{\eea}{\end{align}}

\author{Ra\'ul Fabi\'an Rojas Mej\'ias}
\title{Termodin\'amica de Agujeros Negros \\ y campos escalares}
\degree{Doctor en Ciencias F\'isicas}
\supervisor{Dumitru Astefanesei}
\institution{Pontificia Universidad Cat\'olica de Valpara\'iso}
\faculty{Facultad de Ciencias}
\department{INSTITUTO DE F\'ISICA}

\hyperlinking
\begin{document}



\begin{titlepage}
\TitleBlock{\includegraphics[height=9cm]{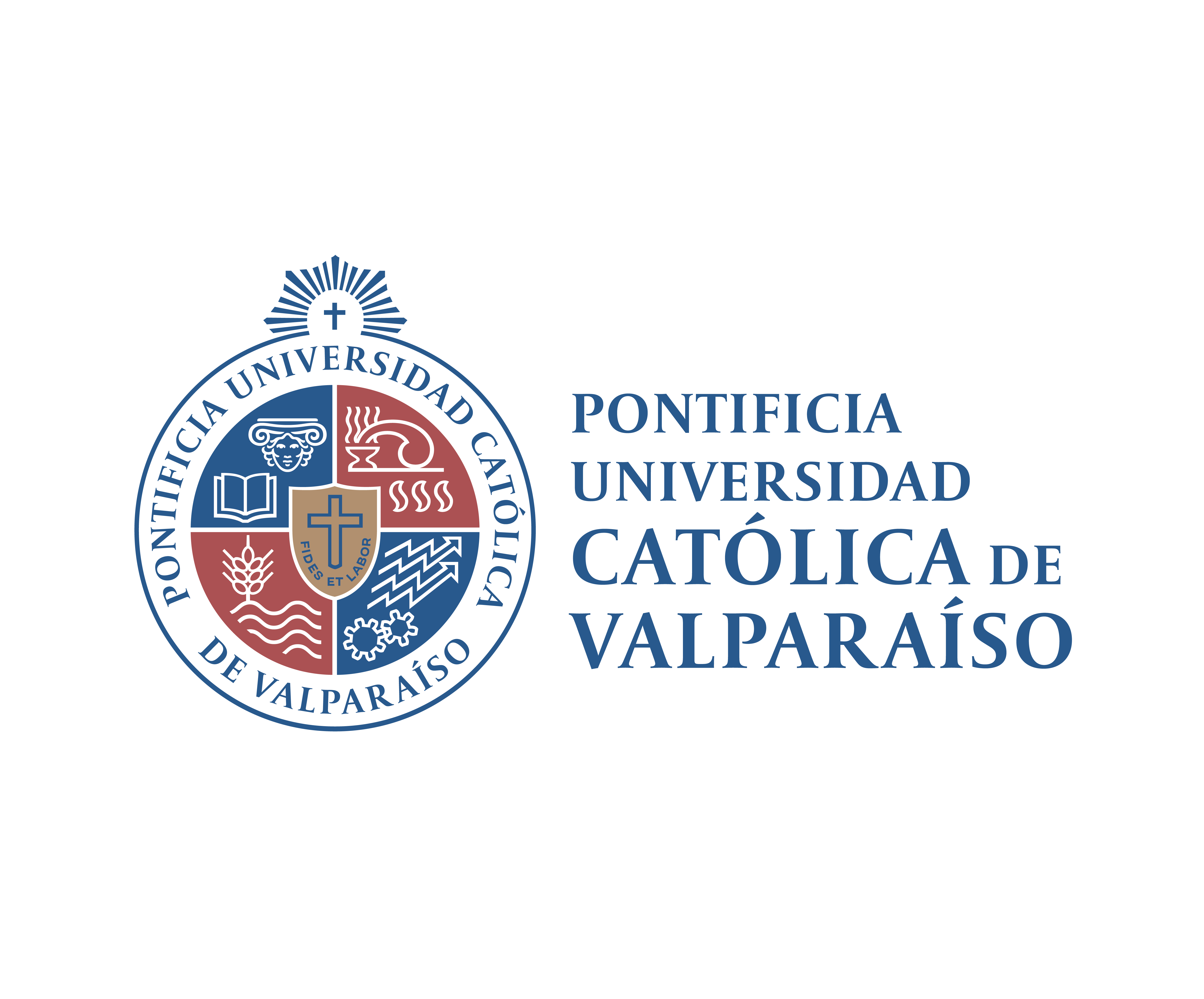}}
  \TitleBlock{\scshape\insertinstitution}
  \TitleBlock[\bigskip]{\scshape\insertfaculty}
  \TitleBlock{\LARGE\scshape\inserttitle}
  \TitleBlock{\scshape
    Tesis presentada por \insertauthor \\
    para obtener el grado de \insertdegree}
  \TitleBlock{\insertsubmitdate}
  \TitleBlock[\bigskip]{\scshape\insertdepartment}
\end{titlepage}




\prefacesection{}


\prefacesection{\scshape{Miembros del Comit'e Examinador}}


\begin{align*}
\text{\scshape{Tutor:\qquad\qquad\qquad\qquad\;\;\;\;\;\;\;\;\;\,}}
&\text{\scshape{Dr. Dumitru Astefanesei (PUCV)}}\\
\text{\scshape{Co-Tutor:\qquad\qquad\qquad\qquad\;\;\;\,}} 
&\text{\scshape{Dr. Andr'es Anabal'on (UAI)}}\\
\text{\scshape{Evaluador Interno:\qquad\qquad\;\;}}
&\text{\scshape{Dr. Joel Saavedra (PUCV)}}\\
\text{\scshape{Evaluador Externo:\qquad\qquad\;}}
&\text{\scshape{Dr. Marco Astorino (UAI)}}
\end{align*}

\prefacesection{Resumen}

Desde el descubrimiento realizado por Stephen Hawking en la d'ecada de los 70 sobre la radiaci'on t'ermica de los agujeros negros, la termodin'amica de agujeros negros se ha vuelto un activo campo de investigaci'on en la f'isica te'orica, por ser la primera predicci'on (en un contexto semi-cl'asico) entre la relatividad general y la teor'ia cu'antica de campos. 

En esta tesis, se estudia la influencia de los campos escalares en la termodin'amica de agujeros negros cargados en $D=4$ dimensiones.
Por un lado, dilucidamos el rol que juegan las cargas escalares en la primera ley de la termodin'amica, usando el formalismo cuasilocal de Brown y York, basado en un correcto principio variacional, y proveemos una serie de ejemplos concretos en los que aplican nuestros resultados.

Por otra parte, estudiamos la estabilidad termodin'amica de soluciones exactas de agujeros negros el'ectricamente cargados y acoplados a un campo escalar con una auto-interacci'on no trivial, las cuales pueden ser puestas en teor'ias de supergravedad. Mostramos expl'icitamente que, cuando el espaciotiempo en que est'an inmersos es asint'oticamente plano, estos pueden estar en un equilibrio termodin'amico estable.

\prefacesection{Contribuciones}

Los cap'itulos \ref{cap3}, \ref{cap4} y \ref{estabilidad} de esta tesis est'an basados en los trabajos publicados \cite{Astefanesei:2018vga} y \cite{Astefanesei:2019mds}, y un tercero que, estando en etapas finales, ser\'a publicado pronto.


\prefacesection{Agradecimientos}

Quisiera agradecer al programa de becas internas PUCV, por brindarme apoyo financiero durante parte de mis estudios, y al programa de becas de CONICYT (de doctorado nacional, 21140024), por financiar la mayor parte de mi permanencia en el programa, permiti'endome concretar satisfactoriamente dos pasant'ias de investigaci'on.

Durante mis a\~nos como estudiante en la Pontificia Universidad Cat'olica de Valpara'iso, he tenido el gran placer de poder comprender, un poco mejor, c'omo funciona la naturaleza, de personas muy apasionadas y profesionales quienes alimentaron mi curiosidad y amor al conocimiento con cada asignatura.
En correspondencia, vaya mi m'as sincera gratitud hacia todos los/as profesores/as del Instituto de F'isica, PUCV.
En especial, quisiera agradecer a la memoria de Dr. Sergio del Campo, con quien di mis primeros pasos en relatividad general y cosmolog'ia, durante mis a\~nos finales en la Licenciatura en F'isica.
\bigskip

Esta tesis est'a realizada bajo la tutela de Dumitru Astefanesei, quien, adem'as de ser un excelente cient'ifico y una gran persona, constituy'o una gu'ia y un apoyo fundamentales para mi formaci'on. Sus conocimientos en f'isica y su extraordinaria intuici'on me han permitido iluminar mi camino para seguir en la investigaci'on.
Asimismo, quisiera expresar una gran gratitud hacia Andr'es Anabal'on, por permitirme iniciar junto a 'el y ense\~narme, con paciencia, elementos indispensables para la investigaci'on, tanto en lo te'orico como en lo pr'actico, durante mis primeros a\~nos en el programa de postgrado.

A Joel Saavedra, quien fue una fuente de inspiraci'on durante mis a\~nos en licenciatura y quien, con la amabilidad que le caracteriza, me ofreci'o ingresar a un proyecto de investigaci'on sobre agujeros negros, con Andr'es, al comienzo de mis estudios de postgrado. Ello marc'o mis inicios en el campo de la \textit{gravitaci'on}.
\bigskip

Quisiera agradecer tambi'en a mis amigos/as en el equipo de trabajo que se ha ido formando con los a\~nos, y con quienes la investigaci'on ha resultado m'as divertida e interactiva. A David Choque, quien ha sido un importante apoyo, especialmente durante mis comienzos; a Romina Ballesteros y Francisco G'omez, con quienes he tenido interesantes y fruct'iferas discusiones, y tambi'en a Jorge Maggiolo, Fabrizzio Merello y Paulo Rojas.
\bigskip

Quiero destacar y agradecer, muy cari\~nosamete, a mi familia por su apoyo incondicional y constante durante estos a\~nos, sin el cual no hubiera logrado completar esta primera etapa de lo que en un momento fue un sue\~no. A mis padres, Juan y Marta, y a mi hermana, Susana, siempre presentes. Tambi'en agradezco el cari\~no omnipresente de mis abuelos: mi tata Ra\'ul y abuelita Mar\'ia, y mi tata Nano, quien nos acompa\~na en nuestros recuerdos, y mi abuelita Chela.

Quisiera agradecer tambi'en la compa\~n'ia dulce de Noemy durante estos meses. Tu amor ha significado mucho para m'i.

\bigskip

\prefacesection{Convenciones}

En f'isica te'orica, es \'util trabajar en el sistema de unidades \textit{naturales}, en el cual los valores num'ericos de las constantes fundamentales son iguales a $1$.
En los cap'itulos \ref{cap1} y \ref{cap2}, por claridad, se ha decidido mantener las constantes f'isicas intactas. En cambio, en los cap'itulos posteriores, la constante de gravitaci'on universal, la velocidad m'axima de propagaci'on para la informaci'on (velocidad de la luz), la constante de Planck y la constante de Boltzmann, ser'an fijadas $G_N=c=\hbar=k_B=1$, respectivamente. 
La permitividad el'ectrica del vac'io se tomar'a como $\epsilon_0=\frac{1}{4\pi}$, tal que la constante de Coulomb $k_e=(4\pi\epsilon_0)^{-1}=1$. En tal caso, la ley de Gauss para el campo el'ectrico $\vec F$ sobre una superficie esf'erica en infinito es
\begin{equation*}
\frac{1}{4\pi}\oint_{s^2_\infty}{\vec F\cdot d\vec A}=Q
\label{gauss1}
\end{equation*}
donde $Q$ es la carga f'isica. En ocasiones, se consider'a tambi'en cargas magn'eticas $P$, y usaremos formas diferenciales, tal que las cargas, el'ectrica y magn'etica, son\footnote{Puesto que las ecuaciones de Maxwell pueden escribirse como $dF=0$ y $d(\star F)=0$, las cargas f'isicas resultan de la integraci'on de ellas.}
\begin{equation*}
Q=\frac{1}{4\pi}\oint_{s_\infty^2}{\star F}\;, \qquad
P=\frac{1}{4\pi}\oint_{s_\infty^2}{F}
\end{equation*}
donde $F=\frac{1}{2}F_{\mu\nu}\,dx^\mu\wedge dx^\nu$, $\star F\equiv\frac{1}{4}\sqrt{-g}F^{\mu\nu}\epsilon_{\mu\nu\alpha\beta}\,dx^\alpha\wedge dx^\beta$, en teor'ias de Einstein-Maxwell definidas en el espaciotiempo $\mathcal{M}$ con la m'etrica $g_{\mu\nu}$, dadas por la acci'on
\begin{equation*}
I=\frac{1}{2\kappa}\int_{\mathcal{M}}
{d^4x\sqrt{-g}\(R-F_{\mu\nu}F^{\mu\nu}\)}
\end{equation*}
donde $\kappa=8\pi G_N/c^4=8\pi$.

\tableofcontents



\chapter{Introducci'on}
\label{cap1}


Un agujero negro podr'ia verse como una part'icula elemental en el sentido de que, despu'es del colapso gravitacional de estrellas masivas, s'olo un peque\~no conjunto de par'ametros lo caracteriza, t'ipicamente su masa $M$, momento angular $J$ y carga el'ectrica $Q$, cantidades que tambi'en son cargas conservadas que est'an protegidas por una ley de Gauss en la regi'on asint'otica. La existencia de estos par'ametros es de gran importancia puesto que, desde el descubrimiento de que los agujeros negros emiten radiaci'on t'ermica, realizado por S. Hawking en 1974 \cite{Hawking:1974sw}, es ampliamente aceptado que los agujeros negros son sistemas termodin'amicos y, en consecuencia, las leyes de la mec'anica de agujeros negros \cite{Bardeen:1973gs}, que tienen un origen geom'etrico, son en su derecho propio leyes termodin'amicas. 

Las cantidades conservadas aparecen en la primera ley de la termodin'amica de agujeros negros estacionarios
\begin{equation}
dE=TdS+\Phi dQ+\Omega dJ,
\end{equation}
donde $E=Mc^2$, $T$ es la temperatura de Hawking, $S$ es la entrop'ia de Bekenstein-Hawking \cite{Bekenstein:1973ur}, proporcional con el 'area $A$ de la superficie del horizonte de eventos del agujero negro
\begin{equation}
S=\(\frac{k_Bc^3}{\hbar G_N}\)\frac{A}{4},
\end{equation}
mientras que $\Phi$ es la diferencia de potencial electrost'atico entre el horizonte de eventos y el infinito, y $\Omega$ la velocidad angular del agujero negro. Todas estas cantidades juegan un papel clave en el comportamiento termodin'amico de los agujeros negros. En este sentido, un agujero negro est'a muy lejos de ser considerado una part'icula elemental. Concretamente, la existencia de la entrop'ia de Bekenstein-Hawking es un indicador fuerte de que existen estados microsc'opicos que dan lugar a la entrop'ia de los agujeros negros. Estos microestados 
est'an intr'insecamente conectados con la naturaleza cu'antica del campo gravitacional, y s'olo una teor'ia cu'antica de la gravedad puede explicar con exactitud d'onde est'an, y c'omo son, los grados de libertad que dan lugar a esta entrop'ia.
Al respecto, se han realizado ciertos progresos en el contexto de teor'ias de cuerdas 
 \cite{Strominger:1996sh,Strominger:1997eq,Maldacena:1997de}.

Otros par'ametros asociados a la estrella antes del colapso gravitacional, como su composici'on qu'imica, su espectro electromagn'etico o los campos magn'eticos cerca de su superficie, entre otros, quedan ocultos detr'as del horizonte de eventos. Esta es la esencia del teorema de no-pelo propuesto originalmente \cite{Bekenstein:1972ny}. Un agujero negro 'unicamente quedar'ia descrito por su masa, carga el'ectrica y su momento angular.
La construcci\'on de soluciones caracterizadas por otros par'ametros, por ejemplo, soluciones no-abelianas de las ecuaciones de Einstein-Yang-Mills \cite{Volkov:1989fi,Bizon:1990sr}, sin embargo, cuestionaron la validez de la conjetura inicial. Un nueva versi'on del teorema de no-pelo se propuso \cite{Bekenstein:1995un}, considerando otras hip'otesis y asunciones (para una agradable revisi'on sobre este t'opico, vea, por ejemplo, \cite{Herdeiro:2015waa}).

Las soluciones de agujeros negros acopladas a un campo escalar existen siempre que una o m'as suposiciones del teorema de no-pelo sea evadida. En este contexto, a estos agujeros negros les llama \textit{agujeros negros con pelo} o \textit{hairy black holes}, en ingl'es.

En esta tesis, estudiamos principalmente soluciones exactas de agujeros negros en $D=4$ dimensiones, el'ectricamente (y magn'eticamente) cargados, acoplados con campos escalares reales, sin y con auto-interacci'on.

\newpage

Los campos escalares se estudian en la relatividad general por un n'umero de razones. En primer lugar, constituyen una de las formas m'as simples de materia y su estudio constituye un primer acercamiento al estudio de sitemas m'as complejos. Remarcablemente, la existencia de campos escalares, en particular el campo de Higgs, cuenta ahora con evidencia experimental \cite{Aad:2012tfa,Chatrchyan:2012xdj}, y otros campos escalares podr'ian tambi'en existir en la naturaleza. En este sentido, los campos escalares juegan un rol importante en f'isica de part'iculas y de altas energ'ias. Los campos escalares son usados en algunos modelos para entender, por ejemplo, la naturaleza de la energ'ia oscura y la materia oscura \cite{Padmanabhan:2002sh}. Adem'as, desde hace unas d'ecadas son considerados en varios modelos de inflaci'on \cite{Guth:1980zm,Linde:1981mu}. Por otra parte, los campos escalares pueden usarse para construir teor'ias efectivas.

Un campo escalar que aparece naturalmente en el contexto de la teor'ia de cuerdas en el l'imite de bajas energ'ias es el dilat'on\cite{Garfinkle:1990qj}, que figura acoplado no trivialmente a otros campos. Ya que la teor'ia de cuerdas es una teor'ia fundamental, es importante entender la influencia de estos campos escalares en sistemas gravitacionales, en particular, en la f'isica de agujeros negros y en su comportamiento termodin'amico.\footnote{Los campos escalares pueden tambien condensarse para formar objetos compactos suaves y sin horizonte, las llamadas `estrellas de bosones', por ejemplo, en espaciotiempos asint'oticamente planos \cite{Schunck:2003kk}, y en espaciotiempo asin'oticamente AdS \cite{Astefanesei:2003qy}, \cite{Astefanesei:2003rw}, \cite{Buchel:2013uba}}

Uno podr'ia preguntar si la existencia campos escalares en teor'ias gravitacionales, como el dilat'on, introduce una nueva constante de integraci'on en la soluci'on. Resulta que los campos escalares que estudiaremos en esta tesis, est'aticos y sim'etricamente esf'ericos, constituyen un pelo secundario, es decir, no aportan con una constante de integraci'on independiente a la soluci'on y, por ello, no tienen una carga conservada asociada. En teor'ias de cuerdas, el acoplamiento de las cuerdas es dado por un par'ametro adimensional $g_s$ que es controlado por el valor de expectaci'on del dilat'on, mediante $g_s=e^{\left<\phi\right>}$. Variar el valor del campo al infinito, $\phi_{\infty}$ equivale a cambiar las constantes de acoplamiento de la teor'ia y entonces podemos considerar diferentes teor'ias para una misma configuraci'on de agujero negro. Esta consideraci'on modifica la primera ley con un t'ermino extra, asociadas al campo escalar, que, sin embargo, con contiene cargas conservadas. En el cap'itulo \ref{cap3}, exploramos de nuevo este desaf'io y contribu'imos dilucidando el papel que juegan las cantidades asociadas al campo escalar en la primera ley de la termodin'amica de agujeros negros. En el cap'itulo \ref{cap4} ofrecemos una serie de ejemplos concretos donde verificamos nuestros resultados previos.

Desde una perspectiva f'isica, puede parecer sorprendente la existencia de campos de materia que, en el r'egimen est'atico (despu'es de un tiempo infinito tras el colapso), coexisten en equilibrio alrededor del agujero negro. Una pregunta natural es si aquel equilibrio es estable frente a perturbaciones mec'anicas y termodin'amicas. En el cap'itulo \ref{estabilidad}, estudiamos detalladamente estas 'ultimas para un conjunto de agujeros negros cuyo campo escalar experimenta una interacci'on no trivial consigo mismo.


Nos enfocaremos mayormente en teor'ias de Einstein-Maxwell-dilat'on cuya acci'on gravitacional tiene la forma
\begin{equation}
I=\frac{1}{2\kappa}
\int_{\mathcal{M}}{d^4x\sqrt{-g}
\[R-e^{\gamma\phi}F_{\mu\nu}F^{\mu\nu}-\frac{1}{2}\pa_\mu\phi\pa^\mu\phi-V(\phi)\]}
\label{a0}
\end{equation}
donde $F_{\mu\nu}=\pa_{\mu}A_\nu-\pa_{\nu}A_\mu$ es el campo de Maxwell, $A_\mu$ el potencial de gauge, y $\phi$ es el dilat'on con su potencial no trivial $V(\phi)$. 
\bigskip

La organizaci'on de esta tesis es entonces la siguiente. En el cap'itulo \ref{cap2}, revisaremos el principio de acci'on basado en en la formulaci'on lagrangiana de la relatividad general, de donde derivan las ecuaciones de campo. Tambi'en discutiremos algunos elementos importantes en la termodin'amica de agujeros negros, y mostraremos una manera computacionalmente simple de obtener la temperatura de Hawking.

En el cap'itulo \ref{cap3}, presentaremos la formulaci'on variacional de teor'ias de Einstein-Maxwell-dilat'on en espaciotiempos asint'oticamente planos, cuando el valor asint'otico del campo escalar no est'a fijo. Obtenemos t'erminos de borde compatibles con el principio de aci'on, y calculamos la acti'on gravitacional y el correspondiente tensor de Brown-York. Mostramos que la energ'ia tiene una nueva contribuci'on que depende del valor asint'otico del campo escalar y discutimos el rol de las cargas escalares en la primera ley de la termodin'amica.
Tambi'en extendemos nuestro an'alisis a agujeros negros con campo escalar en espaciotiempos asint'oticamente Anti-de Sitter.
En el cap'itulo \ref{cap4}, verificamos los resultados previos para distintas teor'ias.

En el cap'itulo \ref{estabilidad}, presentamos un an'alisis detallado de la termodin'amica de soluciones exactas asint'oticamente planas de agujeros negros con campo escalar en una teor'ia de Einstein-Maxwell-dilat'on. 
Calcularemos la acci'on regularizada, el tensor de estr'es cuasilocal y las cargas conservadas usando el m'etodo de contrat'erminos. En presencia de un potencial dilat'onico no trivial que se anula en el borde, probamos que, para un cierto rango de par'ametros, existen agujeros negros termodin'amicamente estables en los ensambles can'onico y gran can'onico. Concluimos con una interpretaci'on f'isica de los resultados.

\chapter{Los agujeros negros: principio de acci'on y termodin'amica}
\label{cap2}

\section{El principio de acci'on en gravedad}
\label{principioaccion}

Consideremos que el espaciotiempo est'a descrito por un n'umero de campos, por simplicidad, el campo gravitacional, dado por las componentes de un tensor de rango 2 veces covariante $g_{\mu\nu}$, y una colecci'on de campos de materia $\phi_i$. 

La acci'on $I$ es un funcional de los campos y es una cantidad construida con invariantes, es decir, independientes de sistemas de coordenadas,
\be
I\[g_{\mu\nu},\phi_i\]
=\frac{1}{2\kappa}\int_{\mathcal{M}}{d^4x\sqrt{-g}
\(R-\mathcal{L}_m\)}
\label{action}
\ee
donde $R\equiv g^{\alpha\beta}R^{\mu}{}_{\alpha\mu\beta}$ es el escalar de curvatura (de Ricci), contru\'ido mediante contracciones de las componentes del tensor de curvatura (de Riemann) $R^{\mu}{}_{\nu\alpha\beta}$. Entonces, $R$ contiene primera y segundas derivadas de la m'etrica. $\mathcal{L}_{m}$ es la densidad lagrangiana para los campos de materia y energ\'ia, es decir, $\mathcal{L}_m=\mathcal{L}_m\(\phi_i,\pa_\mu\phi_i\)$. La constante $\kappa=8\pi$ ya que estamos considerando un sistema de unidades donde $G_N=1$, $c=1$, tal que trabajamos con una \'unica dimensi\'on fundamental, por ejemplo, masa $[M]$.

El principio de acci\'on establece que asumir'an aquellos campos $(g_{\mu\nu},\phi_i)$ alrededor de los cuales la acci\'on sea un extremo, es decir,
\be
\label{principle}
\delta I
=\(\frac{\delta I}{\delta g^{\mu\nu}}\)
\delta g^{\mu\nu}
+\(\frac{\delta I}{\delta\phi_i}\)\delta\phi_i=0
\ee
donde $\delta g_{\mu\nu}$ y $\delta\phi_i$ son funciones peque\~nas y arbitrarias de las coordenadas $x^\mu$. Realizar la variaci'on de la acci'on (\ref{action}) equivale a realizar la variaci'on sobre las cantidades $\sqrt{-g}R$ y $\sqrt{-g}\mathcal{L}_m(\phi_i,\pa_\mu\phi_i)$.

Para el c\'alculo de la variaci'on del determinante de la m'etrica, la f'ormula de Jacobi, $$\delta\sqrt{-g}=-\frac{1}{2}\sqrt{-g}g_{\mu\nu}\delta g^{\mu\nu}$$ es particularmente 'util y puede seguirse, a grandes rasgos, considerando la identidad $\delta(\det M)=(\det M)\,\text{Tr}\(M^{-1}\delta M\)$, donde $M$ representa la matriz con elementos $g_{\mu\nu}$ tal que $\det M=g$, y $\text{Tr}\(M^{-1}\delta M\)=g^{\mu\nu}\delta g_{\mu\nu}$. Por una parte, tenemos entonces que
\be
\delta\left(\sqrt{-g}R\right)
=\sqrt{-g}\left(R_{\mu\nu}-\frac{1}{2}g_{\mu\nu}R\right)
\delta g^{\mu\nu}+\sqrt{-g}g^{\mu\nu}\delta R_{\mu\nu}
\label{var1} \ee
y, por otra parte, tenemos que
\begin{align}
\delta\left(\sqrt{-g}\mathcal{L}_{mat}\right)
= -&\,\frac{1}{2}\sqrt{-g}\,T_{\mu\nu}\,\delta g^{\mu\nu}
+\sqrt{-g}
\[\frac{\pa\mathcal{L}_m}{\pa\phi_i}
-\frac{1}{\sqrt{-g}}\pa_\mu
\(\frac{\pa\mathcal{L}_m}{\pa(\pa_\mu\phi_i)}\)\]
\delta\phi_i \notag\\
+&\,\pa_\mu
\[\sqrt{-g}\frac{\pa\mathcal{L}_m}{\pa(\pa_\mu\phi_i)}\delta\phi_i\]
\label{var2} \end{align}
N'otese que el tensor de energ'ia-momento es definido como
\begin{equation}
T_{\mu\nu}:=g_{\mu\nu}\mathcal{L}_{m}-
2\(\frac{\pa\mathcal{L}_{m}}{\delta g^{\mu\nu}}\)
\end{equation}
Poniendo todo junto, observamos que
\begin{align}
\delta I&=\frac{1}{2\kappa}\int_{\mathcal{M}}
{d^4x\sqrt{-g}E_{\mu\nu}\delta g^{\mu\nu}}
+\int_{\mathcal{M}}{d^4x\sqrt{-g}L_i}
\delta\phi_i \label{motion1} \\
&+\frac{1}{2\kappa}\int_{\mathcal{M}}{d^4x\sqrt{-g}g^{\mu\nu}\delta R_{\mu\nu}}
+\int_{\pa\mathcal{M}}{d^3x\sqrt{|h|}\frac{\pa\mathcal{L}_m}{\pa(\pa_\mu\phi_i)}\delta\phi_i} \notag
\end{align}
donde $E_{\mu\nu}:=R_{\mu\nu}-\frac{1}{2}g_{\mu\nu}R-\kappa T_{\mu\nu}$ y $L_i:=\frac{\pa\mathcal{L}_m}{\pa\phi_i}
-\frac{1}{\sqrt{-g}}\pa_\mu
\[\frac{\pa\mathcal{L}_m}{\pa(\pa_\mu\phi_i)}\]$.
Puesto que $\delta g_{\mu\nu}\neq 0$ y $\delta\phi_i\neq 0$, por ser arbitrarias, el principio de acci'on $\delta I=0$ est'a, en parte, garantizado siempre que $E_{\mu\nu}=0$ y $L_i=0$. Estas son las ecuaciones de Einstein
\begin{equation}
R_{\mu\nu}-\frac{1}{2}g_{\mu\nu}R=\kappa T_{\mu\nu}
\end{equation}
y las ecuaciones de Euler-Lagrange para los otros campos de materia,
\begin{equation}
\frac{\pa\mathcal{L}_m}{\pa\phi_i}
-\frac{1}{\sqrt{-g}}\pa_\mu
\[\frac{\pa\mathcal{L}_m}{\pa(\pa_\mu\phi_i)}\]=0
\end{equation}
  
Los t'erminos en la segunda l'inea de (\ref{motion1}) son de naturaleza diferente. Por ejemplo, $\delta R_{\mu\nu}$ no puede ser puesto en t'erminos de $\delta g_{\mu\nu}$, mientras que el 'ultimo t'ermino es una integral de superficie, o un t'ermino de borde\footnote{Si uno demanda que $\phi_i$ est'an fijas al borde, entonces esta integral se anula autom'aticamente.}. El principio de acci'on est'a bien definido siempre que la acci'on (\ref{action}) est'e suplementada con t'ermino de borde. Para anular el t'ermino $\delta R_{\mu\nu}$, debe agregarse un t'ermino de borde conocido como el t'ermino de Gibbons-Hawking (en el ap'endice \ref{apeA} se provee una revisi'on sobre este t'ermino), que est'a 'intimamente relacionada con condiciones de borde para la m'etrica.

Por ejemplo, como veremos en el cap'itulo \ref{cap3}, para un correcto principio variacional, un t'ermino de borde debe agregarse cuando el campo escalar posee un valor asint'otico din'amico y diferente de cero al infinito, bajo determinadas condiciones.

\newpage
\section{La termodin'amica de agujeros negros}
\label{fisicaestadistica}


Perm'itanos realizar un repaso elemental de f'isica estad'istica. Esta introducci'on es 'util para observar c'omo se conectan los grados de libertad de un sistema f'isico con las propiedades termodin'amicas del mismo\footnote{Esto, si bien la f'isica estad'istica de los agujeros negros no es materia de la presente tesis.}. Finalmente, comentaremos sobre la funci'on de partici'on para sistemas gravitacionales y c'omo se conecta con la acci'on que estudiamos en la secci'on previa.
\bigskip

Imagine un sistema compuesto por un n'umero $N$ de part\'iculas, cada una de las cuales puede estar en un nivel de energ\'ia dado, $E_i$ (donde $i=0,1,2,\,\dots,M$). Digamos que $E_0$ es la m'inima energ\'ia y que las dem'as est'an ordenadas como
\begin{equation}
E_0<E_1<\dots<E_i<\dots<E_M
\label{orden}
\end{equation}

Si $N_i$ el n\'umero de part\'iculas en el nivel de energ\'ia $E_i$, entonces $\sum_{i=0}^M{N_i}=N$.
La probabilidad $p_i$ de que las $N_i$ part\'iculas est\'en en el nivel $E_i$ es simplemente
\begin{equation}
p_i
=\frac{N_i}{N}
\label{prob1}
\end{equation}
y, naturalmente, suman la unidad, $\sum_{i=0}^M{p_i}=1$.

Desde un punto de vista f\'isico, la probabilidad de que las $N_i$ part\'iculas est\'en en el nivel $E_i$ debe ser una funci\'on del valor de la energ'ia de aquel nivel y tambi\'en de una propiedad del sistema que, convenientemente, llamamos temperatura $T$,\footnote{M'as cantidades intensivas pueden caracterizar al sistema. Para esta subsecci'on, y por simplicidad, 'unicamente consideramos la temperatura.}
\begin{equation}
p_i\equiv f(E_i,T)
\label{ans}
\end{equation}

Por 'ultimo, asuma que la probabilidad de que $N_i$ part'iculas est'en en el nivel $E_i$ es independiente de la probabilidad de que $N_j$ part'iculas est'en en $E_j$ (con $i\neq j$). Bajo esta hip'otesis, la probabilidad $p$ de que $N_0$ est'en con $E_0$, $N_1$ con $E_1$, y as'i sucesivamente, es el producto de todas las respectivas probabilidades por separado.

Usando la identificaci'on $(\ref{ans})$, obtenemos
\begin{equation}
f\(\sum_{i=1}^{M}{E_i},T\)=\prod_{i=1}^{M}f(E_i,T)
\end{equation}
donde $\sum_{i=1}^{M}{E_i}$ no es la energ'ia total del sistema, sino la suma de las energ'ias de cada nivel.
La expresi'on m\'as simple que reproduce este resultado es
\begin{equation}
f(E,T)=\frac{\exp\(-\beta E\)}{Z(\beta)}
\end{equation}
donde $\beta$, definida positiva\footnote{La condici'on $\beta(T)>0$ viene como consecuencia de la normalizaci'on de la probabilidad. En otras palabras, $\exp(-\beta E_i)$ no puede ser arbitrariamente grande para alg'un $E_i$, en virtud de las desigualdades (\ref{orden}) que no imponen un m'aximo para $E_i$.}, es una funci\'on 'unicamente de la temperatura del sistema y $Z(\beta)$ una funci'on de la temperatura por determinar. Introduciendo la constante de Boltzmann, $k_B\approx1.38\times10^{-23}$ [J$\cdot$K$^{-1}]$, por an'alisis dimensional se sigue que la funci'on m'as sencilla para $\beta$ es 
\begin{equation}
\beta(T)=\frac{1}{k_BT}
\end{equation}
La condici'on de normalizaci'on para la probabilidad, $\sum_{i=0}^M{p_i}=1$, implica que 
\begin{equation}
Z(\beta)=\sum_{i=0}^M\exp\(-\beta{E_i}\)
\end{equation}

Una expresi'on final para $Z(\beta)$, que es conocida como la \textit{funci'on de partici'on} del sistema, es posible solo conociendo los detalles sobre los niveles de energ'ia para un sistema f'isico concreto. De esta manera, la funci'on de partici'on codifica toda la informaci'on relevante, por ejemplo, el valor de expectaci'on para la energ'ia de una part'icula o la entrop'ia del sistema
\begin{equation}
\left<E\right>
=-\frac{\pa \ln Z}{\pa\beta}
, \quad
S\equiv -k_B\sum_{i=0}^M{p_i\ln p_i}
=k_B\(\ln Z+\beta\left<E\right>\)
\end{equation}

En la secci'on previa (\ref{principioaccion}), hemos visto el principio de acci'on en gravedad. Aqu'i, estamos interesados en agujeros negros como sistemas termodin'amicos. Por lo tanto, estamos interesados en determinar la funci'on de partici'on apropiada para sistemas gravitacionales. El objetivo es entonces mostrar brevemente c'omo se conecta la acci'on gravitacional con la funci'on de partici'on.
\bigskip

Estudiando c'omo se comportan los campos de materia en las cercan'ias de los agujeros negros, Hawking mostr'o que, usando la teor'ia cu'antica de campos, los agujeros negros deben emitir part'iculas a un ritmo constante \cite{Hawking:1974sw}, con un espectro de emisi'on t'ermico. 
Para un sistema cu'antico descrito por un hamiltoniano $H$, funci'on de partici'on es
\begin{equation}
Z=\text{Tr}\(e^{-\beta H}\),
\end{equation}
donde $\beta=(k_BT)^{-1}$. N'otese que, por ejemplo, si $\left.|E_i\right>$ son autoestados con energ'ia $E_i$, es decir, $H\left.|E_i\right>=E_i\left.|E_i\right>$, entonces puede seguirse f'acilmente que
\begin{equation}
\text{Tr}\(e^{-\beta H}\)=\sum_i{
\left<E_i|\right.e^{-\beta H}\left.|E_i\right>}
=\sum_i e^{-\beta E_i},
\end{equation}
como ten'iamos antes.

Tomemos ahora un campo escalar. En la aproximaci'on de la integral de caminos, podemos considerar la amplitud para ir de una configuraci'on $\phi_1$ al tiempo $t_1=0$ a otra configuraci'on $\phi_2$ al tiempo $t_2=t$, dada por
\begin{equation}
\left<\phi_2,t_2|\right.\left.\phi_1,t_1\right>
=\int{\mathcal{D}\phi \;e^{iI[g_{\mu\nu},\phi]}},
\end{equation}
donde $\mathcal{D}\phi$ es una medida sobre el espacio del campo $\phi$. Esta amplitud puede tambi'en ser expresada en t'erminos del operador evoluci'on $U=e^{-iHt/\hbar}$, cuya traza es
\begin{equation}
\text{Tr}\(U\)=\left<\phi_2,t_2|\right.\left.\phi_1,t_1\right>
=\int{d\phi \left<\phi|\right.e^{-iHt/\hbar}\left.|\phi\right>}
\label{path1}
\end{equation}
La funci'on de partici'on puede ser asociada con la traza del operador evoluci'on, realizando una rotaci'on de Wick
\begin{equation}
t\rightarrow -i\tau
\end{equation}
y evaluando $\tau=\hbar\beta$. Mientras que $t\in\mathbb{R}$, el \textit{tiempo imaginario} $\tau$ es peri'odico, $0<\tau<\hbar\beta$, como mostraremos en la pr'oxima subsecci'on. Tenemos entonces la siguiente identificaci'on
\begin{equation}
\text{Tr}(U)
\underbrace{=}_{{\left.t\rightarrow -i\tau\right|_{\tau=\hbar\beta}}}
\text{Tr}(e^{-\beta H})=Z,
\label{UZ}
\end{equation}
con lo cual la funci'on de partici'on para el sistema gravitacional es identificada como
$Z=\int{\mathcal{D}\phi\;e^{-I^E}}$,
donde $I^E$ es la acci'on en la secci'on Euclidiana\footnote{La rotaci'on de Wick cambia la signatura de la m'etrica desde $(-,+,+,+)$ a $(+,+,+,+)$.}. La relaci'on entre la acci'on $I$ y su versi'on euclidea $I^E$ debe ser
\begin{equation}
I^E=-\left.iI\right|_{t\rightarrow -i\tau_E}.
\label{actions}
\end{equation}

Ahora, debemos considerar la contribuci'on dominante en la funci'on de partici'on gravitacional. Esta es la aproximaci'on semicl'asica (o de punto silla), en la cual tomamos 'unicamente 
\begin{equation}
Z\approx e^{-I^E},
\end{equation}
donde la acci'on es evaluada en los campos que resuelven las ecuaciones de movimiento (\textit{evaluaci'on ``on shell"}). Esto equivale a tomar los campos cu'anticos en un background cl'asico (campos cu'anticos en espaciotiempos curvos descritos por la relatividad general).

Es importante comentar que la periodicidad en el tiempo imaginario $\tau$ viene como un requerimiento de regularidad en la m'etrica.
Al calcular la periodicidad en el tiempo imaginario, estamos obteniendo una temperatura.
 
A continuaci'on, entonces, veremos que tras la rotaci'on de Wick la m'etrica de una soluci'on adquiere una \textit{singularidad c'onica} que puede ser removida proveyendo al tiempo $\tau$ de una periodicidad que es inversa proporcional de la temperatura asociada a la radiaci'on que emite un agujero negro.

\newpage
\section{La temperatura de Hawking}
\label{sub:hawkingtemp}

Con el descubrimiento de Hawking de la radiaci'on de los agujeros negros se inicia la termodin'amica de agujeros negros. Como sistema gravitacional, los agujeros negros pueden alcanzar estados de equilibrio t'ermico con el espaciotiempo. 

Obtendremos la temperatura por medio del c'alculo de la periodicidad de $\tau$, de acuerdo a los discutido previamente. 

Desde ahora, llamaremos $\tau^E$ al tiempo imaginario.
\bigskip 

Considere un elemento de l\'inea conectando dos eventos infinitesimalmente pr'oximos en un espaciotiempo est'atico descrito por una m'etrica est'atica y con simetr'ia esf'erica,
\be
ds^2=g_{tt}dt^2+g_{rr}dr^2+g_{\theta\theta}
\(d\theta^2+\sin^2\theta d \varphi^2\)
\label{Hawtemp1}
\ee
donde $g_{tt}\leq 0$.
Ahora, llevemos la m'etrica a la secci'on Euclidiana, es decir, realizando el cambio $t=-i\tau^E$.
En seguida, definamos la funci\'on $N(r)^2\equiv -g_{tt}$,
y la coordenada $R\equiv \sqrt{N(r)^2}$.
Escribiendo la m'etrica en la coordenada $R$, empleando $dr^2=\frac{dR^2}{\(N'\)^2}$, donde $N'=\frac{dN}{dr}$,
se obtiene
\begin{equation}
ds^2
=\frac{g_{rr}}{\left(N'\right)^2}\left[ dR^2+R^2d\left(\frac{N'}{\sqrt{g_{rr}}}
\,\tau^E\right)^2\right]+g_{\theta\theta}
\(d\theta^2+\sin^2\theta d \varphi^2\)
\end{equation}
Escrita en esta manera conveniente, es sencillo identificar una singularidad de coordenada en $R=0$, muy similar a aquella que aparece en coordenadas polares $dr^2+r^2d\theta^2$, donde $0\leq \theta <2\pi$, puesto que, si menor que $2\pi$, se forma una singularidad c'onica (una variedad f'isica debe ser diferenciable en cada punto).

Por lo tanto, mientras que $t\in\mathbb{R}$, el tiempo euclidiano $\tau^E$ debe ser peri'odico
\begin{equation}
0\leqslant\tau^E<
\left.\frac{2\pi \sqrt{g_{rr}}}{N'}\right|_{R=0}=\hbar\beta
\end{equation}
Esto, seg'un la discusi'on en la secci'on previa, provee de una temperatura asociada al horizonte de evento ($R=0\leftrightarrow g_{tt}=0$)
\begin{equation}
T=\left.
\frac{\hbar}{2\pi k_B}\frac{N'}{\sqrt{g_{rr}}}
\right|_{r=r_+}
\end{equation}
donde $r=r_+$ es la localizaci'on del horizonte de eventos.

Note que la definici'on de $N$ es tal que uno puede considerar la expresi'on positiva o negativa, a fin de que $T>0$.
\bigskip 

La aplicaci'on m'as trivial es el c'alculo de la temperatura del agujero negro de Schwarzschild,  donde $N(r)=c\(1-\frac{r_+}{r}\)^{1/2}$, siendo $r_+$ la coordenada del horizonte de eventos. El resultado es
\begin{equation}
T_{Schw}=\(\frac{\hbar c^3}{k_BG_N}\)
\frac{1}{{8\pi M}}
\end{equation}

Para una m'etrica est'atica del tipo
$ds^2=-f(r)dt^2+\frac{dr^2}{f(r)}+b(r)^2d\sigma^2$,
la temperatura de Hawking es
\begin{equation}
T=\frac{\hbar}{4\pi k_B}\left.\frac{df(r)}{dr}\right|_{r=r_+}
\end{equation}
donde $f(r_+)=0$ es la ecuaci'on del horizonte.

\chapter{Las cargas escalares y la primera ley de la termodin'amica}
\label{cap3}

\newpage

Los campos escalares juegan un papel central en la f'isica de part'iculas y  en la cosmolog'ia, y aparecen naturalmente en teor'ias de unificaci'on en f'isica de altas energ'ias. Es entonces importante entender sus propiedades generales en teor'ias gravitacionales acopladas a escalares (y otros campos de materia), particularmente el rol de estos campos en la f'isica de agujeros negros.

En particular, el dilat'on es un campo escalar que aparece en el l'imite de bajas energ'ias en teor'ias de cuerdas. Algunos de los conocimientos aceptados en la relatividad general podr'ian ser reconsiderados en este contexto. Una de las diferencias importantes es que, contrario a fijar las condiciones de borde como en la relatividad general, las condiciones de borde en teor'ias de cuerdas son determinadas por valores de expectaci'on din'amicos de los campos escalares. Una consecuencia importante e inusual es que, para agujeros negros no extremos ($T\neq 0$) en teor'ias de cuerdas, tanto la masa como el 'area del horizonte de eventos dependen de una forma no trivial del valor asint'otico de los campo escalares, $\phi^a_\infty$ (donde $a$ etiqueta diferentes escalares), lo que conduce a una dr'astica modificaci'on a la primera ley de la termodin'amica de agujeros negros\cite{Gibbons:1996af}:
%
\begin{equation}
\label{1stlaw}
dM = TdS + \Psi dQ + \Upsilon dP + \left(\frac{\partial M}{\partial \phi_{\infty}^a} \right) d\phi_{\infty}^a
\end{equation}
donde $\Psi$ y $\Upsilon$ son los potenciales conjugados el'ectrico y magn'etico, y los coeficientes de $\phi_{\infty}^a$ son calculados a cargas y entrop'ia fijas,
\begin{equation}
\left(\frac{\partial M}{\partial \phi_{\infty}^a} \right)_{S,Q,P} = -G_{ab}(\phi_\infty)\Sigma^b
\end{equation}
Usando las notaciones de \cite{Gibbons:1996af}, $G_{ab}(\phi_\infty)$ es la m'etrica del espacio de los escalares y $\Sigma^a$ son las cargas escalares, que pueden ser obtenidas mediante expansi'on asint'otica (en el infinito espacial) de los campos escalares:
\begin{equation}
\phi^a = \phi_{\infty}^a + \frac{\Sigma^a}{r}+\mathcal{O}\(\frac{1}{r^2}\)
\end{equation}  
Una propuesta similar aparece en el contexto de la dualidad AdS/CFT donde, para una soluci'on exacta de agujero negro con campo escalar que es asint'oticamente AdS, se encontr'o que la primera ley deber'ia er modificada por un par $(X,Y)$ conjugado adicional de variables termodin'amicas \cite{Lu:2013ura}:
\begin{equation}
dM = TdS + \Psi dQ + \Upsilon dP + XdY
\end{equation}
Estas cantidades, $(X,Y)$, son expresables como funciones de las cargas conservadas $(M,P,Q)$ y fueron interpretadas en su propio derecho como una carga escalar y su potential conjugado \cite{Lu:2013ura}.

Un problema con la primera ley de la termodin'amica (\ref{1stlaw}) para agujeros negros en teor'ias de cuerdas es que las cargas escalares no son cargas conservadas. Ellas corresponden a grados de libertad que viven fuera del horizonte (el `pelo') y no est'an asociadas a una nueva e independiente constante de integraci'on (por lo que se les llaman `pelo secundario'). En teor'ias de cuerdas, los campos escalares (o `moduli') se interpretan como constantes de acoplamiento locales y una variaci'on en sus valores en el borde es equivalente a cambiar los acoplamientos de la teor'ia. Una resoluci'on fue propuesta en \cite{Astefanesei:2006sy} (o, tambi'en, \cite{Astefanesei:2009wi}): uno puede en principio redefinir las cargas tal que la masa y las cargas escalares no dependan de $\phi_\infty$, pero el precio que se paga es que las nuevas cargas el'ectricas y magn'eticas definidas (o cargas `vestidas') ya no son cargas f'isicas. Si el valor asint'otico del campo escalar es diferente de cero, pero fijadas directamente en la acci'on,  $\phi_\infty= const.$, ello corresponde a una diferente teor'ia con diferente acoplamiento (el factor $e^{\phi_\infty}$ es absorbido in la constante de acoplamiento, no en los valores de las cargas) para el campo gauge y, dentro de la teor'ia, el t'ermino $\Sigma d\phi_{\infty}$ se anula. Esta propuesta se hizo concreta en \cite{Hajian:2016iyp} donde, usando un m'etodo de espacio de fase, se mostr'o que esto es una condici'on de integrabilidad v'alida y que no hay necesidad de una contribuci'on extra del campo escalar en la primera ley.


Sin embrago, quisi'eramos enfatizar que la propuesta de Gibbons, Kallosh, y Kol \cite{Gibbons:1996af} es sobre la variaci'on de las condiciones de borde para los campos escalares y, de esta forma, a pesar de los argumentos en \cite{Astefanesei:2006sy, Hajian:2016iyp}, permanece robusto e intrigante. La cuesti'on principal que todav'ia permanece, es entonces, ?`por qu'e las cargas escalares que act'uan como fuente para los campos escalares, pero que no son cargas conservadas, aparecen en la primera ley de la termodin'amica de agujeros negros cuando se consideran variaciones de $\phi_\infty$?


En este cap'itulo de la presente tesis,  investigamos el rol de las condiciones de borde no triviales de los campos escalares en teor'ias Einstein-Maxwell-dilat'on. Estamos interesados en soluciones exactas, asint'oticamente planas, de agujeros negros para las cuales el valor asint'otico del campo escalar pueda variar, y en soluciones asint'oticamente AdS para agujeros negros cargados tanto el'ectricamente como magn'eticamente, para las cuales los escalares rompen la simetr'ia conforme en el borde. En espacios asint'oticamente planos, obtenemos un principio variacional bien posicionado, agregando un nuevo t'ermino de borde a la acci'on, lo cual permite calcular la energ'ia total correcta, clarificando con ello el rol de las cargas escalares (no conservadas) en la primera ley \cite{Gibbons:1996af}. Una vez con la intuici'on desarrollada en espacios planos, mostraremos que una vez que la energ'ia es tambi'en correctamente obtenida en espaciotiempos AdS \cite{Anabalon:2014fla}, cuando las condiciones de borde del campo escalar no preservan las isometr'ias de AdS en el borde \cite{Henneaux:2006hk}, la primera ley es satisfecha y no hay necesidad de considerar una contribuci'on extra del campo escalar. Estas consideraciones son de especial inter'es cuando se consideran incrustaciones (\textit{embedding}) en teor'ias de cuerdas y el campo escalar (dilat'on) se vuelve din'amico y, para aplicaciones hologr'aficas (AdS), los agujeros negros con campos escalares pueden ser usados para describir rompimiento de simetr'ias o transiciones de fase en la teor'ia cu'antica de campos dual.


\newpage
\section{Agujeros negros con pelo escalar en espaciotiempos asint'oticamente planos}

En esta secci'on, proponemos un principio variacional para agujeros negros en espacios asint'oticamente planos\footnote{En espaciotiempo asint'oticamente plano, existe una clase distinta de agujeros negros con campos escalar (agujeros negros escalarizados), vea, por ejemplo, \cite{Astefanesei:2019pfq}, y las referencias dentro. En esta tesis, trabajamos con campos escalares dilat'onicos, no con agujeros negros escalarizados.} cuando el valor del borde del campo escalar puede variar y mostramos que la energ'ia total tiene una nueva contribuci'on que es relevante para la termodin'amica. El objetivo es discutir este desaf'io en un marco no trivial y lo m'as simple posible, esto es, usaremos el \textit{formalismo cuasilocal} de Brown-York \cite{Brown:1992br} para una teor'ia con solo un campo escalar que est'a acoplado a un campo gauge.
\subsection{La primera ley de la termodin'amica}
Comenzamos con una breve revisi'on del trabajo \cite{Gibbons:1996af} y, para mayor claridad, expl'icitamente obtenemos los t'erminos de carga escalar en la primera ley.
Aparte del gravit'on, cada teor'ia de cuerdas contiene otro estado universal, un campo sin masa llamado dilat'on $\phi$. Consiramos la acci'on de Einstein-Maxwell-dilat'on
\begin{align}
I\left[g_{\mu\nu},A_\mu,\phi\right]
&= I_{bulk} + I_{GH} \\
&=\frac{1}{2\kappa}\int_{\mathcal{M}}
{d^4x\sqrt{-g}\left(R-e^{\alpha\phi}F_{\mu\nu}F^{\mu\nu}
-2\pa_\mu\phi\pa^\mu\phi\right)}
+\frac{1}{\kappa}\int_{\pa\mathcal{M}}
{d^3x\sqrt{-h}K}
\label{action0}
\end{align}
donde $\kappa=8\pi$, de acuerdo a nuetras convenciones $c=G_N=1$. El segundo t'ermino es el t'ermino de borde de Gibbons-Hawking y $K$ es la traza de la curvatura extr'inseca $K_{ab}$ definida sobre el borde $\pa\mathcal{M}$ con la m'etrica inducida $h_{ab}$. 

El acoplamiento entre el campo escalar y el campo gauge en la acci'on (\ref{action0}) aparece en acciones de bajas energ'ia de teor'ias de cuerdas para valores particulares de $\alpha$, aunque en nuestro an'alisis podemos mantener $\alpha$ arbitrario. Las ecuaciones de movimiento para la m'etrica, campo escalar y campo gauge son
\begin{equation}
E_{\mu\nu}:=R_{\mu\nu}
-2\pa_\mu\phi\pa_\nu\phi
-2\text{e}^{\alpha\phi}
\left(F_{\mu\alpha}F_{\nu}{}^{\alpha}
-\frac{1}{4}g_{\mu\nu}
F_{\alpha\beta}F^{\alpha\beta}\right)=0 
\end{equation}
\begin{equation}\frac{1}{\sqrt{-g}}\pa_\mu\left(\sqrt{-g}
g^{\mu\nu}\pa_\nu\phi\right)
-\frac{1}{4}\alpha \text{e}^{\alpha\phi}F_{\mu\nu}F^{\mu\nu}=0
\end{equation}
\begin{equation}
\pa_\mu
\left(\sqrt{-g}\text{e}^{\alpha\phi}
F^{\mu\nu}\right)=0 \label{maxw}
\end{equation}
El ansatz general para la m'etrica de una soluci'on est'atica de agujero negro cargado es
\begin{equation}
ds^{2}=-a^2dt^2+a^{-2}dr^2+b^2(d\theta^{2}+\sin^{2}{\theta}\, d\varphi^{2})
\end{equation}
donde $a=a(r)$ y $b=b(r)$. El campo de gauge compatible con este ansatz y con las ecuaciones de movimiento es
\begin{equation}
\label{gauge}
F = -\frac{qe^{-\alpha\phi}}{b^2} dt\wedge dr
-p\sin\theta\, d\theta\wedge d\varphi 
\end{equation}
Ahora, la combinaci'on $E_{t}^{t}+E_{\theta}^{\theta}$ conduce a una ecuaci'on integrable $(a^2b^2)''=2$ (donde la prima $(~)'$ significa derivada con respecto a la coordenada radial) con la soluci'on general
\begin{align}
a^{2}=\frac{(r-r_{+})(r-r_{-})}{b^2}
\end{align}
donde las constantes $r_{\pm}$, con una de ellas indicando la localizaci'on del horizonte de eventos, deben ser determinadas.

Ya que estamos interesados en soluciones asint'oticamente planas, consideramos la expansi'on $a^2=1+\mathcal{O}(r^{-1})$ que determina de la expansi'on de la otra funci'on m'etrica
\begin{equation}
b^2=r^2+\beta r+\gamma+\mathcal{O}(r^{-1})
\end{equation}
donde $\beta$ y $\gamma$ son constantes. Usando esta expresi'on en la combinaci'on $E_{t}^{t}-E_{r}^{r}=0$, que conduce a la ecuaci'on $b''+b\phi'^2=0$, obtenemos la siguiente forma asint'otica del campo escalar
\begin{equation}
\phi=\phi_{\infty}+\frac{\Sigma}{r}+O(r^{-2})
\label{scalar}
\end{equation}
donde $\phi_{\infty}$ es la condici'on de borde para la teor'ia que hemos considerado (para el campo escalar) y $\Sigma$ es la carga escalar. Uno puede obtener de manera simple que $4\Sigma^2=\beta^2-4\gamma$ y que, para  $\beta=0$, tenemos que $b^2=r^2-\Sigma^2$, que corresponde, de hecho, al caso para la soluci'on exacta en la teor'ia con $\alpha=-2$ que ser'a presentada a continuaci'on.

Para verificar concretamente los pasos en la obtenci'on de (\ref{1stlaw}), vamos a usar la soluci'on exacta \cite{Gibbons:1987ps}, donde la carga magn'etica es nula y el campo escalar est'a acoplado con el par'ametro exponencial $\alpha=-2$. Las ecuaciones de movimiento pueden ser resueltas anal'iticamente (\cite{Garfinkle:1990qj}) y la soluci'on exacta es
\begin{equation}
a^2=\frac{(r-r_+)(r-r_-)}{r^2-\Sigma^2} \,,\qquad
b^2=r^2-\Sigma^2 
\,,\qquad
\phi=\phi_\infty
+\frac{1}{2}\ln\left(\frac{r+\Sigma}{r-\Sigma}\right)
\end{equation}
donde 
\begin{equation}
r_-=-\Sigma,\qquad r_+=\Sigma
-\frac{(qe^{\phi_{\infty}})^2}{\Sigma}
\end{equation}

La masa ADM \cite{Arnowitt:1959ah,Arnowitt:1960es} es obetnida expandiendo la componente $g_{tt}$ de la m'etrica,
\begin{equation}
-g_{tt}=a^2
=\frac{\Sigma r+(qe^{\phi_\infty})^2-\Sigma^2}
{\Sigma(r-\Sigma)}
=1+\frac{\left(qe^{\phi_\infty}\right)^2}{\Sigma{r}}
+O\left(r^{-2}\right) 
\end{equation}
que conduce a la identificaci'on
\begin{equation}
\label{mass}
M=-\frac{\left(qe^{\phi_\infty}\right)^2}{2\Sigma}
\end{equation}
con la carga escalar negativa, $\Sigma < 0$. 
El mismo resultado se obtiene

Note que $\Sigma$ no es una constante de integraci'on independiente y que la soluci'on es regular\footnote{Cuando ambas cargas (el'ectrica y magn'etica) son diferente de cero, existen dos horizontes. Sin embargo, en este caso especial con s'olo un campo el'ectrico no cero, existe s'olo un horizonte $r_+$, puesto que $r=r_-$ corresponde a una singularidad real. La conduci'on de regularidad $r_{+} >r_{-}$ es, desde un punto de vista f'isico, equivalente con el hecho de que hay una carga m'axima que puede ser llevada por el agujero negro.} siempre que $2M^2-Q^2e^{2\phi_\infty}> 0$.

La carga el'ectrica $Q$ es calculada, como es usual, integrando la ecuaci'on del campo el'ectrico. Con nuestras convenciones, 
\begin{equation}
Q = \frac{1}{4\pi}\oint{e^{-2\phi}\star{F}}
=\frac{1}{4\pi}\oint{e^{-2\phi}	\left(\frac{1}{4}\sqrt{-g}\epsilon_{\alpha\beta\mu\nu}
	F^{\alpha\beta}dx^\mu\wedge dx^\nu\right)} =q
\end{equation}
Con la masa (\ref{mass}), uno puede expl'icitamente verificar la primera ley
\begin{equation}
dM=TdS+\Phi dQ-\Sigma d\phi_{\infty}
\end{equation}
la cual contiene el t'ermino extra propuesto en \cite{Gibbons:1996af}, y la f'ormula de Smarr,
\begin{equation}
M=2TS+Q\Psi
\label{smarr0}
\end{equation}
que no contiene expl'Icitamente una contribuci'on $\Sigma\phi_\infty$.

\subsection{La energ'ia total y el formalismo de Brown y York}


Para aplicar el formalismo de Brown y York\cite{Brown:1992br}, la acci'on gravitacional debe satisfacer el principio de acci'on, lo que implica que la acci'on a considerar debe estar regularizada y debe estar suplementada con t'erminos de bordes consistentes con las condiciones de borde para los campos. Para una aplicaci'on directa al agujero negro de Reissner-Nordstr\"om, se ha provisto de detalles en el ap'endice \ref{aB}.

La regularizaci'on de la acci'on implica la incorporaci'on de un contrat'ermino gravitacional, en el borde, para espaciotiempos asint'oticamente planos\cite{Lau:1999dp,Mann:1999pc,Kraus:1999di,Mann:2005yr},
\begin{equation}
I=I_{bulk}+I_{GH}+I_{ct} \,\,,\qquad I_{ct}=-\frac{1}{\kappa}\int_{\pa\mathcal{M}}
{d^3x\sqrt{-h}\sqrt{2\mathcal{R}^{(3)}}}
\end{equation}
donde $\mathcal{R}^{(3)}$ es el escalar de Ricci de la m'etrica en el borde ($3$-dimensional), $h_{ab}$. Este contrat'ermino cancela las {divergencias infrarojas} de la teor'ia. 


Para emplear el formalismo de Brown y York (vea en ap'endice \ref{aB} para una aplicaci'on al agujero negro de Reissner-Nordstr\"om) para teor'ias Einstein-Maxwell-dilat'on, debido a la variaci'on de $\phi_\infty$, para obtener un principio variacional bien definido cuando la carga escalar $\Sigma$ es mantenida fija, se tiene que agregar un nuevo t'ermino de borde
\begin{equation}
I_\phi=-\frac{2}{\kappa}\int_{\pa\mathcal{M}}
{d^3x\sqrt{-h}\left[
\frac{\phi_\infty}{\Sigma}(\phi-\phi_\infty)^2\right]}
\label{Iphi}
\end{equation}

Para obtener la energ'ia libre $\mathcal F$ de los agujeros negros con campo escalar, debemos calcular la acci'on on-shell en la secci'on Euclidiana:
\begin{equation}
I^E =\beta F =\beta\left(M-TS-Q\Psi-P\Upsilon+\Sigma\phi_\infty\right)
\end{equation}
donde la periodicidad del tiempo Euclidiano est'a relacionada con la temperatura mediante
$\beta = 1/T$. 

Observe que hay un t'ermino extra $\Sigma\phi_\infty$ que, de hecho, viene del contrat'ermino para el campo escalar $I_{\phi}$, aunque, como hemos visto, un t'ermino similar no aparece en la f'ormula de Smarr (\ref{smarr0}). Esto es una importante se\~nal de que un c'alculo de la energ'ia total podr'ia er diferente de la masa $ADM$ cuando se considera el t'ermino de borde $I_{\phi}$. Con todos los t'erminos requeridos para un correcto principio variacional, obtenemos el tensor cuasilocal regularizado de \cite{Astefanesei:2005ad}, pero esta vez suplementado con la contribuci'on del campo escalar
\begin{equation}
\label{stress1}
\tau_{ab}=\frac{1}{\kappa}\left[
K_{ab}-h_{ab}K-\Phi(\mathcal{R}^{(3)}_{ab}-h_{ab}\mathcal{R}^{(3)})-h_{ab}\Box\Phi+\Phi_{;ab}\right]
+\frac{2h_{ab}}{\kappa}\left[\frac{\phi_{\infty}}{\Sigma}(\phi-\phi_{\infty})^2\right]
\end{equation}
donde
\begin{equation}
\Phi=\sqrt\frac{2}{\mathcal{R}^{(3)}}
\end{equation}
Si consideramos una colecci'on de observadores en el borde de un espaciotiempo est'atico, conteniendo un agujero negro, puesto que $\xi^\mu=\delta^\mu_t$ es un vector de Killing, ellos van a medir la misma energ'ia (total), que es la carga conservada asociada con este vector de Killing espec'ifico, que es definida como\cite{Brown:1992br}:
\begin{equation}
\label{BYcharge}
E=\oint_{s^2_\infty}{d^2\sigma\sqrt{\sigma}n^a\xi^b\tau_{ab}}
\end{equation}
Aqu'i, $\Xi$ es una superficie cerrada 2-dimensional con la normal unitaria $n^a$ y la m'etrica inducida
\begin{equation}
\sigma_{ij}dx^idx^j=b^2(d\theta^2+\sin^2\theta d\varphi^2)
\end{equation}
Puesto que estamos interesados en configuraciones est'aticas, no hay ondas gravitacionales y entonces no hay necesidad de considerar el infinito nulo en nuetro an'alisis. Evaluando esta cantidad conservada, (\ref{BYcharge}), al infinito espacial, obtenemos la siguiente expresi'on para la energ'ia total:
\begin{equation}
E_{\text{total}}=M+\phi_{\infty}\Sigma
\end{equation}
Este resultado fue obtenido para una carga escalar fija, $\Sigma=constant$, y esto conduce a la siguiente primera ley de la termodin'amica
\begin{equation}
dE_{\text{total}} = T dS + \Psi dQ + \Upsilon dP
\end{equation}
con  $\Sigma d\phi_\infty$ reabsorbido en la energ'ia total del espaciotiempo, el cual es diferente de la masa $ADM$.

Debemos ahora considerar una condici'on de borde m'as general de la forma
\begin{equation}
\Sigma\equiv \frac{dW(\phi_{\infty})}{d\phi_{\infty}}
\end{equation}
la cual es muy similar con la propuesta en \cite{Hertog:2004ns} para agujeros negros con campo escalar en AdS. El t'ermino de borde general es (para una derivaci'on de este termino de borde, vea el ap'endice \ref{apendaux})
\begin{equation}
\tilde I^\phi_{ct}=-\frac{2}{\kappa}\int_{\pa\mathcal{M}}
{d^3x\sqrt{-h}
\[\frac{\(\phi-\phi_\infty\)^2}{\Sigma^2}W(\phi_{\infty})\]}
\end{equation}
y se reduce a (\ref{Iphi}) cuando $\Sigma$ es constante. Como es esperado, un c'alculo similar de la energ'ia total conduce al siguiente resultado
\begin{equation}
E_{total}=M+W(\phi_{\infty})
\end{equation}
donde $M$ es la masa $ADM$ obtenida de la expansi'on de $g_{tt}$ en el infinito espacial.

\newpage
\section{Agujeros negros con campo escalar, asint'oticamente AdS}

En esta secci'on, calculamos la energ'ia del agujero negro di'onico (con carga el'ectrica y magn'etica), propuesto en \cite{Lu:2013ura} y verificamos la primera ley de la termodin'amica.\footnote{En los 'ultimos a\~nos, se han construido soluciones regulares de agujeros negros\cite{Acena:2013jya,Anabalon:2013qua,Anabalon:2013sra,Acena:2012mr,Anabalon:2013eaa, Anabalon:2016izw} para un potencial espec'ifico, el que finalmente se ha mostrado corresponder a modelos extendidos de supergravedad\cite{Anabalon:2017yhv,DallAgata:2012mfj} y nuetro an'alisis puede tambi'en ser aplicado a estos casos.} Seguimos de cerca\cite{Anabalon:2014fla} porque es t'ecnicamente m'as sencillo desde un punto de vista pr'actico\footnote{El m'etodo de contrat'erminos en AdS fue desarrollado en\cite{Balasubramanian:1999re,Skenderis:2000in,Henningson:1998gx} y en presencia de campos escalares con condiciones de borde mixtas en\cite{Papadimitriou:2007sj, Anabalon:2015xvl,Marolf:2006nd}, y se pueden aplicar a soluciones que son localmente asint'oticamente AdS, por ejemplo, \cite{Astefanesei:2005yj,Astefanesei:2005eq,Astefanesei:2004kn,Astefanesei:2004ji,Balasubramanian:2002am,Balasubramanian:2005bg}}.
Vamos a mostrar que
la primera ley es, de nuevo, satifecha sin introducir t'erminos extras dependientes de las cargas escalares.

\subsection{Agujeros negros di'onicos con campo escalar}
Considere la teor'ia descrita por la acci'on\cite{Lu:2013ura}
\begin{equation}
I=\frac{1}{2\kappa}\int_{\mathcal{M}}
{d^4x\sqrt{-g}
	\left[
	R-\frac{1}{2}(\pa\phi)^2
	-\frac{1}{4}e^{-\sqrt3\phi}F^2
	+\frac{6}{l^2}
	\cosh\left(\frac{1}{\sqrt3}\phi\right)
	\right]}
\end{equation}
y la siguiente soluci'on regular,
\begin{align}
ds^2 &= 
-(H_1 H_2)^{-\frac{1}{2}}f dt^2
+(H_1 H_2)^{\frac{1}{2}}
\left[\frac{dr^2}{f}
+r^2\left(d\theta^2+\sin^2\theta\, d\varphi^2\right)\right] 
\label{lupopemetric}\\
\phi &=\frac{\sqrt{3}}{2} \ln\left(\frac{H_2}{H_1}\right) \\
A_{\mu} dx^\mu &=\sqrt2 \left(\frac{1-\beta_1 f_0}
{\sqrt{\beta_1\gamma_2}\, H_1}\, dt + 2\mu\,\gamma_2^{-1}\sqrt{\beta_2\gamma_1}\, \cos\theta\, d\varphi\right)
\label{lupopegauge}
\end{align}
donde las funciones relevantes son
\begin{align}
f&=f_0 + \frac{r^2}{l^2} H_1 H_2\,,\qquad f_0=1 - \frac{2\mu}{r} \\
H_1&=\gamma_1^{-1} (1-2\beta_1 f_0 + \beta_1\beta_2 f_0^2)\,,\qquad
H_2=\gamma_2^{-1}(1 - 2\beta_2 f_0 + \beta_1\beta_2 f_0^2) \\
\gamma_1&= 1- 2\beta_1 + \beta_1\beta_2\,,\qquad \gamma_2 = 1-2\beta_2 + \beta_1\beta_2\,.
\end{align}
Aqu'i, los par'ametros $\beta_1$ y $\beta_2$ est'an relacionados con las cargas el'ectrica y magn'etica, de acuerdo con la ley de Gauss
\begin{align}
Q
&=\frac{1}{4\pi}
\oint_{s_\infty^2}{e^{-\sqrt{3}\phi}\star \(\frac{1}{4}F\)}
=\frac{\mu\sqrt{\beta_{1}\gamma_{2}}}{\gamma_{1}\sqrt{2}} \\ 
P&=\frac{1}{4\pi}\oint_{s_\infty^2}{\(\frac{1}{4}F\)}
=\frac{\mu\sqrt{\beta_{2}\gamma_{1}}}{\gamma_{2}\sqrt{2}}
\end{align}
y los potenciales conjugados son
\begin{align}
\Psi&=A_t(\infty)-A_t(r_+)=
\sqrt{\frac{2}{\beta_1\gamma_2}}
\left[
1-\beta_1-\frac{1-\beta_1f_0(r_+)}{H_1(r_+)}
\right] \\
\Upsilon&=A^p_t(\infty)-A^p_t(r_+)=
\sqrt{\frac{2}{\beta_2\gamma_1}}
\left[
1-\beta_2-\frac{1-\beta_2f_0(r_+)}{H_2(r_+)}
\right]
\end{align} 
donde $F^p\equiv e^{\sqrt{3}\phi}\star F=dA^p$.

Para calcular la masa como en\cite{Lu:2013ura}, debemos usar coordenadas can'onicas para las cuales el factor en frente de la parte angular de la m'etrica se vuelve $b^2 = \rho^2 + O(\rho^{-1})$ en el l'imite asint'otico. El cambio de coordenadas en aquel l'imite es
\begin{equation}
r=\rho+c_1+\frac{c_2}{\rho}+O(\rho^{-2})
\end{equation}
donde 
\begin{equation}
c_1=\frac{\mu(2\beta_1^2\beta_2^2-3\beta_1^2\beta_2-3\beta_1\beta_2^2+6\beta_1\beta_2-\beta_1-\beta_2)}
{\gamma_1\gamma_2} \,, \quad c_2=\frac
{3\mu^2(1-\beta_1\beta_2)^2(\beta_1-\beta_2)^2}
{2\gamma_1^2\gamma_2^2}
\end{equation}
Con este cambio de coordenadas, obtenemos el siguiente comportamiento para la componente $g_{tt}$,
\begin{equation}
-g_{tt}= 1+ \frac{\rho^2}{l^2}-\frac{2(1-\beta_1)(1-\beta_2)(1-\beta_1\beta_2)\mu}{\gamma_1\gamma_2\,\rho}+O(\rho^{-2})
\end{equation}
de donde se lee la masa $ADM$,
\begin{equation}
M=\frac{(1-\beta_1)(1-\beta_2)(1-\beta_1\beta_2)\mu}{\gamma_1\gamma_2}
\end{equation}

Es sencillo verificar que la primera ley se satisface, pero con la adici'on de un t'ermino extra, $XdY$\cite{Lu:2013ura}
\begin{equation}
dM=TdS+\Psi dQ+\Upsilon dP+XdY
\label{first1}
\end{equation}
donde
\begin{equation}
X=\frac{4\mu^3(\beta_1-\beta_2)
	\sqrt{\beta_1\beta_2^3}}
{l^2(1-\beta_1\beta_2)\gamma_2^2},
\qquad
Y=\frac{\sqrt{\beta_1}\gamma_2}
{\sqrt{\beta_2}\gamma_1}
\end{equation}
%

\subsection{Masa hamiltoniana y energ'ia conservada}

Como en el caso asint'oticamente plano, quisi'eramos entender si el t'ermino $XdY$ puede ser reabsorbido en una definici'on correcta de energ'ia total. Ahora seguimos de cerca\cite{Anabalon:2014fla, Anabalon:2015xvl},
El campo escalar se comporta en el borde como
\begin{equation}
\phi(\rho)=\frac{A}{\rho}+\frac{B}{\rho^2}
+O(\rho^{-3})
\end{equation}
donde 
\begin{align}
A&=\frac
{2\sqrt{3}\mu(\beta_2-\beta_1)(1-\beta_1\beta_2)}
{\gamma_1\gamma_2} \\
B
&=\frac{2\sqrt{3}\mu^2(\beta_1-\beta_2)
	(\beta_1^3\beta_2^2+\beta_1^2\beta_2^3
	-8\beta_1^2\beta_2^2+6\beta_1^2\beta_2
	+6\beta_1\beta_2^2-8\beta_1\beta_2+\beta_1+\beta_2)}
{\gamma_1^2\gamma_2^2}
\end{align}
Observe que $A=A(\mu,\beta_1,\beta_2)$ y $B=B(\mu,\beta_1,\beta_2)$.
La masa hamiltoniana, que puede ser le'ida directamente de $g_{\rho\rho}$ (y no de $g_{tt}$), puede tener una nueva contribuci'on debido a la \textit{back-reaction} del campo escalar en el borde (vea las ecuaciones (12) y (18) de \cite{Anabalon:2014fla}):
\begin{equation}
g_{\rho\rho}=\frac{l^2}{\rho^2}
+\frac{Cl^4}{\rho^4}
+\frac{Dl^5}{\rho^5}
+O(\rho^{-6})
\end{equation}
donde los coeficientes $C=C(\mu,\beta_1,\beta_2)$ y $D=D(\mu,\beta_1,\beta_2)$ est'an dados por:
\begin{align}
C&=-1-\frac
{3\mu^2(\beta_2-\beta_1)^2
	(1-\beta_1\beta_2)^2}
{l^2\gamma_1^2\gamma_2^2} \label{metric-a} \\
D&=\frac
{2\mu(1-\beta_1)(1-\beta_2)(1-\beta_1\beta_2)}
{l\gamma_1\gamma_2} \notag \\
&+\frac{8\mu^3(\beta_2-\beta_1)^2
	(1-\beta_1\beta_2)
	(\beta_1^3\beta_2^2+\beta_1^2\beta_2^3
	-8\beta_1^2\beta_2^2+6\beta_1^2\beta_2
	+6\beta_1\beta_2^2-8\beta_1\beta_2+\beta_1+\beta_2)}
{l^3\gamma_1^3\gamma_2^3}
\label{metric-b}
\end{align}
Como una verificaci'on extra, es importante enfatizar que las relaciones
\begin{align}
C&=-1-\frac{A^2}{4l^2} \\
D&=\frac{2M}{l}-\frac{2AB}{3l^3}
\end{align}
son satisfechas y entonces podemos usar el resultado general para la masa presentada en \cite{Anabalon:2014fla, Anabalon:2015xvl}:
\begin{align}
\label{EtAdS}
E_{\text{total}} &= \frac{(1-\beta_1)(1-\beta_2)(1-\beta_1\beta_2)\mu}
{\gamma_1\gamma_2} + \frac{1}{4l^2}
\left(W-\frac{1}{3}AB\right) \\ &= M+\frac{1}{4l^2}
\left(W-\frac{1}{3}AB\right)
\end{align}
donde $W$ se introduce por medio de las condiciones de borde del campo escalar $B\equiv \frac{dW}{dA}$. Note que las convenciones para la acci'on en \cite{Anabalon:2015xvl} son ligeramente diferentes de las nuestras y, para hacer concidir los resultados, uno deber'ia rescalar el campo escalar apropiadamente.

Ahora es sencillo mostrar que el t'ermino $XdY$ es precisamente la variaci'on del t'ermino extra en (\ref{EtAdS}). Una vez m'as, usando la energ'ia total correcta, la primera ley de la termodin'amica puede ser rescrita como
\begin{equation}
dM=TdS+\Psi dQ+\Upsilon dP+XdY \quad
\Leftrightarrow \quad
dE_{\text{total}}=TdS+\Psi dQ+\Upsilon dP
\end{equation}
y no hay necesidad de agregar una contribuci'on extra dependiente de las cargas escalares.

\section{Conclusiones}

En este cap'itulo, usando ideas del formalismo cuaislocal de la energ'ia y el m'etodo de contrat'erminos, hemos revisitado la primera ley de la termodin'amica de agujeros negros con campo escalar y hemos mostrado que las cargas escalares (no conservadas) no pueden aparecer como t'erminos independientes, ni en espaciotiempos asint'oticamente planos ni AdS.
El trabajo de \cite{Anabalon:2017eri,Anabalon:2016yfg} en el cual fue probado que no hay constantes de integraci'on independientes asociadas con el campo escalar para agujeros negros en AdS soporta nuetra conclusi'on para los agujeros negros di'onicos\footnote{Para otros intentos de explicar el t'ermino extra $XdY$ en la primera ley de la termodin'amica para soluciones de agujeros negros di'onicos con campo escalar presentados en \cite{Lu:2013ura}, vea \cite{Anabalon:2016ece,Cardenas:2016uzx}. }

En teor'ias de cuerdas, existe un par'ametro adimensional $g_s$ (el acoplamiento de las cuerdas) que es controlado por el valor de expectaci'on del dilat'on, $g_s= e^{<\phi>}$, el cual no est'a fijo por la ecuaciones de movimiento. De hecho, la teor'ia de cuerdas no tiene par'ametros libres porque todas las constantes de acoplamientos est'an fijas por valores de expectaci'on. Por lo tanto, los valores de los campos al infinito, $\phi_\infty$, pueden ser interpretados como que etiquetan una familia continua de vac'ios de la teor'ia. Cambiar los valores asint'oticos es similar con cambiar las constantes de acoplamiento de la teor'ia y, entonces, una misma configuraci'on de agujero negro puede ser interpretada en teor'ias diferentes.
Esto es inusual en relatividad general, puesto que las condiciones de borde son fijas, pero es bastante com'un en teor'ia de cuerdas. Por ejemplo, para calcular la entrop'ia de agujeros negros extremos supersim'etricos en teor'ia de cuerdas, uno hace un c'alculo de $D$-brana  en el r'egimen de acoplamiento d'ebil y, ya que este resultado est'a protegido por supersimetr'ia, permanece igual en el r'egimen de acoplamiento fuerte en el cual los agujeros negros exiten. Desde el punto de vista de la relatividad general, hemos mostrado que una variaci'on de $\phi_{\infty}$, tanto si se mantiene o no la carga escalar fija, produce una nueva contribuci'on a la energ'ia total del sistema y la primera ley de la termodin'amica es satisfecha sin la necesidad de incluir la contribuci'on del campo escalar.

Con esta nueva expresi'on para la energ'ia, las cargas escalares no contribuyen y la primera ley usual de la termodin'amica que contiene s'olo cargas conservadas es, de nuevo, satisfecha para agujeros negros con cargas escalares.

	\newpage
\chapter{Aplicaciones concretas}
\label{cap4}

En este cap'itulo, proveemos aplicaciones directas del resultado previo, para lo cual consideraremos teor'ias dadas por la acci'on del tipo
\begin{equation}
I=\frac{1}{2\kappa}\int_{\mathcal{M}}		{d^4x\sqrt{-g}\[R-Z(\phi)F^2-\frac{1}{2}(\pa\phi)^2\]}
\end{equation}
para diferentes funciones de acoplamiento $Z(\phi)$. Re-obtenemos las soluciones para un acoplamiento general, $Z=e^{a\phi}$, para cualquier valor del par'ametro $a$, y luego obtenemos soluciones di'onicas para $a=1$. Utilizamos t'ecnicas desarrolladas en \cite{Astefanesei:2006sy}. En cada caso, estudiamos la primera ley, verificando que esta se cumple sin la contribuci'on expl'icita de las cargas escalares, las cuales son reabsorbidas en la energ'ia conservada del sistema, como mostramos en el cap'itulo anterior.

\newpage
\section{Acoplamiento general $Z=e^{a\phi}$}
\label{General2}

Considere las teor'ias
\begin{equation}
I\[\,g_{\mu\nu},A_\mu,\phi\]=
\frac{1}{2\kappa}\int_{\mathcal{M}}
{d^4x\sqrt{-g}\[R-Z(\phi)F^2-\frac{1}{2}\(\pa\phi\)^2\]}
\label{action00}
\end{equation}
con las correspondientes ecuaciones de movimiento
\begin{align}
R_{\mu\nu}-\frac{1}{2}g_{\mu\nu} R
&=\kappa\(T_{\mu\nu}^{EM}+T_{\mu\nu}^{\phi}\),
\label{eins0}\\
\frac{1}{\sqrt{-g}}\pa_\mu\(\sqrt{-g}g^{\mu\nu}\pa_\nu\phi\)
&=\frac{dZ(\phi)}{d\phi}F^2,
\label{klein0}\\
\pa_\mu
\left(\sqrt{-g}Z(\phi)F^{\mu\nu}\right)&=0,
\label{max0}
\end{align}
donde los tensores de energ'ia-momentum son
\begin{equation}
T_{\mu\nu}^{EM}=
\frac{2}{\kappa}Z(\phi)
\(F_{\mu\alpha}F_{\nu}{}^{\alpha}-\frac{1}{4}g_{\mu\nu}F^2\), \quad T_{\mu\nu}^{\phi}
=\frac{1}{2\kappa}\[\pa_\mu\phi\,\pa_\nu\phi-g_{\mu\nu}\(\pa\phi\)^2\]
\end{equation}

El campo escalar presenta el siguiente comportamiento asint'otico\footnote{Esta forma del campo escalar  al borde para la teor'ia considerada se sigue de la ecuaci'on de Klein-Gordon (\ref{klein0}).}
\begin{equation}
\label{sca}
\phi(r)=\phi_{\infty}+\frac{4\Sigma}{r}+\mathcal{O}\(r^{-2}\)
\end{equation}
donde $r$ es la coordenada radial est'andar, $\phi_\infty$ es el valor asint'otico del campo escalar, tratado aqu'i como una cantidad din'amica, de acuerdo a la discusi'on en el cap'itulo previ, y $\Sigma$ es una constante no independiente. El factor 4 en la expansi'on en (\ref{sca}) es introducido por conveniencia, puesto que hemos cambiado la notaci'on usada previamente. Como antes, definimos la condici'on de borde del campo escalar mediante
\begin{equation}
\Sigma\(\phi_{\infty}\)\equiv\frac{dW(\phi_{\infty})}{d\phi_{\infty}}
\end{equation}

Primero, vamos a reobtener la soluci'on a estas teor'ias, primero presentadas en \cite{Garfinkle:1990qj}, usando m'etodos desarrollados en \cite{Astefanesei:2006sy}, que involucran la propuesta de anzats adecuados para desacoplar las ecuaciones de movimiento. 
%
Vamos a considerar acoplamientos de la forma
\begin{equation}
Z(\phi)=e^{a\phi}
\end{equation}
donde $a$ es una contante parametrizando las teor'ias. Usemos ahora el siguiente ansatz para la m'etrica y el campo de gauge,
\begin{align}
ds^2&=
\frac{1}{\eta^2(u-1)^2}\[-{h(u)\Omega(u)}dt^2
+\frac{\eta^2du^2}{h(u)\Omega(u)}+\Omega(u)\(
d\theta^2+\sin^2\theta \,d\varphi^2\)\] \label{a1}\\
F&=-\frac{qe^{-a\phi(u)}}{\Omega(u)}\, dt \wedge du \label{a2}
\end{align}
donde $q$, un par'ametro de carga, y $\eta$ son las dos constantes de integraci'on independientes. No hay p'erdida de generalidad en tomar $\eta\geq 0$. Note que los ansatz (\ref{a1}) y (\ref{a2}) autom'aticamente satisfacen las ecuaciones de Maxwell (\ref{max0}).

Obseve, adem'as, que el borde del espaciotiempo est'a en el l'imite $u=1$, donde el factor conforme en la m'etrica diverges. Esta observaci'on permite dividir la soluci'on en dos espaciotiempos desconectados, uno donde $u$ toma valores entre $0<u<1$ (le llamamos rama negativa) y otro donde $u>1$ (la rama positiva).

La carga f'isica $Q$, salvo un signo global, puede ser obtenida mediante la ley de Gauss, es decir, integrando las ecuaciones de Maxwell sobre una 2-esfera en infinito
\begin{equation}
\label{charg0}
Q
=\frac{1}{4\pi}\oint_{s^2_\infty}{Z(\phi)\star F}
=\frac{1}{4\pi}
\oint{\sqrt{-g}e^{a\phi}F^{tu}d\theta\wedge d\varphi}=\frac{q}{\eta}
\end{equation}

Para resolver las ecuaciones diferenciales, consideremos la combinaci'on $E_ t^t-E_u^u$ , donde $E_{\mu\nu}:={R}_{\mu\nu}-\frac{1}{2}g_{\mu\nu}{R}-\kappa(T^{EM}_{\mu\nu}+T^{\phi}_{\mu\nu})=0$, lo cual da
\begin{equation}
\label{ttuu}
\phi'^2=\(\frac{\Omega'}{\Omega}\)^2
-\frac{2\Omega''}{\Omega},
\end{equation}
donde el s'imbolo de prima indica derivada respecto de $u$. La funci'on $\Omega(u)$ puede ser elegida de dos maneras diferentes, dando lugar a dos familias de soluciones. La familia 1 se obtiene al considerar
\begin{equation}
\label{omeg1}
\Omega(u)=\exp\[-a\(\phi-\phi_{\infty}\)\],
\end{equation}
mientras que la familia 2, al considerar
\begin{equation}
\label{omeg2}
\Omega(u)=\exp\[\frac{1}{a}\(\phi-\phi_{\infty}\)\].
\end{equation}
\textbf{Familia 1:}
Integrando la ecuaci'on (\ref{ttuu}), usando (\ref{omeg1}), obtenemos el siguiente campo escalar
\begin{equation}
\label{scalar1}
\phi(u)=\phi_\infty-\frac{2a}{1+a^2}\ln(u)
\end{equation}
Ahora, usando (\ref{omeg1}) y (\ref{scalar1}), las ecuaciones de Einstein restantes pueden ser integradas para obtener la funci'on m'etrica $h(u)$\footnote{N'otese que las constantes de integraci'on ya fueron elegidas a ser $\eta$ y $q$ (desde el ansatz), por lo tanto, cualquier constante que aparezca al integrar las ecuaciones de movimiento, debe ser una apropiada combinaci'on de $\eta$ y $q$.},
\begin{equation}
\label{metrf1}
h(u)=\(u-1\)^2\eta^2u^{-\frac{3a^2-1}{a^2+1}}
\[\(u-1\)\(1+a^2\)\(qe^{-\frac{1}{2}a\phi_\infty}\)^2+1\]
\end{equation}
Observe que $$\lim_{u=1}{g_{tt}}=1, \quad
\lim_{u=1}{\phi(u)=\phi_{\infty}}$$ como se espera para el borde localizado en $u=1$. Por otro lado, como es sabido, a pesar de la carga el'ectrica, solamente existe un horizonte para estas soluciones
\begin{equation}
u_+=1-\frac{e^{a\phi_\infty}}{(1+a^2)q^2}
\end{equation} 
mientras que $u=0$ corresponde a la singularidad del agujero negro.
\\
\textbf{Familia 2:}
Integrando la ecuaci'on (\ref{ttuu}), usando (\ref{omeg1}), obtenemos
\begin{equation}
\label{scalar21}
\phi(u)=\phi_\infty+\frac{2a}{1+a^2}\ln(u)
\end{equation}
y, usando (\ref{omeg2}) junto con (\ref{scalar21}), podemos integrar la restante ecuaci'on de Einstein para obtener
\begin{equation}
\label{metrf2}
h(u)=\(u-1\)^2\eta^2u^{-\frac{4}{a^2+1}}
\[-\(u-1\)\(1+a^2\)\(qe^{-\frac{1}{2}a\phi_\infty}\)^2+u\]
\end{equation}
donde el horizonte est'a localizado en
\begin{equation}
\label{solfam2}
u_+
=\frac{q^2e^{-a\phi_\infty}(1+a^2)}
{q^2e^{-a\phi_\infty}(1+a^2)-1}
\end{equation}

Comentaremos brevemente sobre estas dos familias.
Como vimos antes, para la familia 1, s'olo la rama negativa contiene agujeros negros. Por otro lado, la familia 2 solamente contiene agujeros negros en la rama positiva. Es una cuesti'on de convenci'on que la rama negativa recibe su nombre debido a que $(\phi-\phi_\infty)<0$, entonces, siguiendo esta convenci'on, debemos asociar la familia 1 con valores negativos de $a$, y a la familia 2 con valores positivos de $a$.

Permitanos verificar la primera ley de la termodin'amica para la familia 1 (con $a<0$). Puesto que los agujeros negros s'olo existen para la rama negativa $0<u<1$, consideremos el siguiente cambio de coordenadas
\begin{equation}
\label{coord}
u=1-\frac{1}{\eta r}
\end{equation}
y usemos este cambio para expandir asint'oticamente el campo escalar. Esto nos permite encontrar una relaci'on entre $\eta$, $a$ y $\Sigma$\footnote{N'otese que para $a<0$, $\Sigma<0$, lo cual es consistente para esta rama, donde $\phi<\phi_\infty$.},
\begin{equation}
\label{sigm}
\Sigma=\frac{a}{2\(a^2+1\)\eta}
\end{equation}
AHora, usamos el mismo cambio para leer la masa $ADM$, mediante la expansi'on de la m'etrica,
\begin{equation}
M=\frac{1}{2\eta}\(qe^{-\frac{1}{2}a\phi_\infty}\)^2
-\frac{a^2-1}{2\eta\(a^2+1\)}
\end{equation}

La temperatura es obtenida en la manera usual
\begin{align}
T&=\frac{\Omega(u_+)}{4\pi\eta}\left.\frac{dh(u)}{du}\right|_{u_+} \\
&=\frac{\eta\(u_{+}-1\)^2}{4\pi}
u_{+}^{-\frac{4a^2}{a^2+1}}
\[\(3a^2+4u_{+}-1\)\(qe^{-\frac{1}{2}a\phi_\infty}\)^2
-\frac{\(a^2-3\)u_{+}-3a^2+1}{\(u_{+}-1\)\(a^2+1\)}\]
{u_{+}}^{{\frac{2a^2}{a^2+1}}}
\end{align}
donde $h(u_+)=0$. La entrop'ia del agujero negro y el potencial conjugado son
\begin{equation}
S=\frac{\pi\Omega(u_+)}{\eta^2\(u_{+}-1\)^2}, \qquad
\Phi\equiv A_t(u_+)-A_t(u=1)={\eta}\(u_{+}-1\)Qe^{-a\phi_\infty}
\end{equation}

Note que, en la rama negativa, valores positivos para $\Phi$ y $Q$ se corresponden con $q<0$ y, por lo tanto, la carga el'ectrica que debe ser considerada es $Q=-\frac{q}{\eta}$.
Usando la ecuaci'on del horizonte $h(u_+)=0$, es sencillo mostrar que la primera ley se cumple, justo como mostramos en el cap'itulo anterior,
\begin{equation}
dE=TdS+\Phi dQ
\end{equation}
donde $E$ no es la masa $ADM$, sino $M+W$, con $dW=\Sigma d\phi_\infty$. $E$ es la energ'ia conservada del sistema, obtenida usando el formalismo cuiasilocal de Brown y York\cite{Brown:1992br}.

El mismo resultado se verifica para la familia 2.

Por 'ultimo, por completitud, quisi'eramos reescribir esta soluci'on en la forma original\cite{Garfinkle:1990qj}. Para ello, consideramos el cambio de coordenadas (\ref{coord}) una vez m'as, para la rama negativa.

Puesto que $\eta\geq 0$ y $a<0$, $\Sigma$, dado por (\ref{sigm}), es negativa tambi'en. La soluci'on puede ser escrita como
\begin{equation}
-g_{tt}=g^{rr}
=\frac{(r-r_+)(r-r_0)^\frac{1-a^2}{1+a^2}}
{r^\frac{2}{1+a^2}} \; , \qquad
g_{\theta\theta}
=r^2\(1-\frac{r_0}{r}\)^{\frac{2a^2}{a^2+1}}
=r^2+\mathcal{O}(r)
\end{equation}
donde $g^{rr}=g^{uu}\(\frac{du}{dr}\)^{-2}$ y
\begin{equation}
\label{hor_sing}
r_+\equiv \frac{aQ^2}{\Sigma e^{a\phi_\infty}}, \qquad
r_0\equiv\frac{2\Sigma\(a^2+1\)}{a}.
\end{equation}

Como es claro, $r_+$ es la coordenada del horizonte de eventos del agujero negro. Por otra parte, a pesar de que $r_{0}>0$, $r_0$ es la ubicaci'on de la singularidad central, puesto que el escalar de Ricci $R$ 
\begin{equation}
R=\frac{2a^2(r-r_+)r_0^2r^{-\frac{2(a^2+2)}{a^2+1}}}
{\(a^2+1\)^2(r-r_0)^{\frac{1+3a^2}{1+a^2}}}
\end{equation}
y el propio campo escalar divergen en el l'imite $r=r_0$.

Escribiendo la m'etrica en la coordenada radial, obtenemos la soluci'on como en \cite{Garfinkle:1990qj},
\begin{equation}
ds^2=-a(r)^2dt^2+\frac{dr^2}{a(r)^2}
+b(r)^2\(d\theta^2+\sin^2\theta d\varphi^2\)
\end{equation}
donde
\begin{equation}
\label{sol1}
a(r)^2=\frac{(r-r_+)(r-r_0)^\frac{1-a^2}{1+a^2}}
{r^\frac{2}{1+a^2}} \;\; ,\qquad
b(r)^2=\(1-\frac{r_0}{r}\)^{\frac{2a^2}{a^2+1}}r^2
\end{equation}

\section{Agujeros negros di'onicos, $Z(\phi)=e^{\phi}$}

Aqu'i, consideramos el caso $Z(\phi)=e^{\phi}$, con ambas cargas, el'ectrica y magn'etica.

Nos proponemos obtener la soluci'on exacta, presentada en\cite{Astefanesei:2006sy}, aunque mediante otro m'etodo, usando el mismo m'etodo usado en la secci'on previa. Nos centraremos en la rama positiva esta vez, aunque el an'alisis es similar como en la rama negativa.

Consideramos el ansatz
\begin{align}
ds^2&=\Omega(u)\[-h(u)dt^2+\frac{\eta^2 du^2}{u^2h(u)}
+d\theta^2+\sin^2\theta d\varphi^2\]  \\
F&=-\frac{q}{ue^{\phi}} \,dt \wedge du
+p\sin\theta \,d\theta\wedge d\varphi
\end{align}
donde $\eta$, $q$ y $p$ son las tres constantes de integraci'on de la soluci'on. Las cargas f'isicas son
\begin{align}
Q=\frac{1}{4\pi}\oint_{s^2_\infty}{e^{\phi}\star F}
=\frac{q}{\eta}, \qquad
P=\frac{1}{4\pi}\oint_{s^2_\infty}{F}=p\,,
\end{align}
respectivamente.
La combinaci'on de las ecuaciones de Einstein $E_t^t-E_u^u$ da
\begin{equation}
\label{comb1}
\phi'{}^2
=3\(\frac{\Omega'}{\Omega}\)^2
-\frac{2}{u}
\(\frac{\Omega''}{\Omega}+\frac{\Omega'}{\Omega}\)
\end{equation}
Escogiendo
\begin{equation}
\Omega(u)=\frac{u}{\eta^2(u-1)^2}\,,
\end{equation}
la ecuaci'on (\ref{comb1}) puede ser integrada para obtener
\begin{equation}
\phi(u)=\ln(u)+\phi_\infty
\end{equation}
Las ecuaciones de movimiento restantes son resueltas por
\begin{equation}
h(u)=\frac{\eta^2(u-1)^2}{u^2}
\[u-(u-1)\(qe^{-\frac{1}{2}\phi_\infty}\) 
+2\eta^2u(u-1)\(pe^{\frac{1}{2}\phi_\infty}\)^2\]
\end{equation}
Observe que la inclusi'on de la carga magn'etica da lugar a dos horizontes esta vez (el interno y externo)
\begin{equation}
u_\pm=\frac{1}{2}
+\frac{q^2e^{-2\phi_\infty}}{2\eta^2p^2}\pm
\frac{\sqrt{4\(\eta^2p^2e^{\phi_\infty}\)^2-8\(\eta pq\)^2+4q^4e^{-2\phi_\infty}-4\eta^2p^2e^{\phi_\infty}-4q^2
		e^{-2\phi_\infty}+1}}
{4\eta^2p^2e^{\phi_\infty}}
\end{equation}	
que, en el l'imite $p=0$, se reduce solo al horizonte exterior dado en (\ref{solfam2}) para la familia 2, ya que $a=1$, de acuerdo a la discusi'on en la secci'on previa. Debido a la existencia de dos horizontes, estas configuraciones admiten el l'imite extremo $T=0$.
%
%
Consideremos el cambio de coordenadas
%
\begin{equation}
\label{chang22}
u=\frac{2r\eta+1}{2r\eta-1}
\end{equation}
donde $\Sigma=(4\eta)^{-1}$. 
El campo escalar, bajo este cambio, toma la forma asint'otica
\begin{equation}
\phi(r)=\phi_{\infty}+\frac{4\Sigma}{r}+O(r^{-2})
\end{equation}
que es consistente con nuestras convenciones aqu'i. Ahora, expandiendo la componente $g_{tt}$ podemos leer la masa $ADM$,
\begin{equation}
-g_{tt}=1-
\frac{Q^2 {e}^{-\phi_\infty}-P^2{e}^{\phi_\infty}}
{2\Sigma r}
+\frac {Q^2{e}^{-\phi_\infty}
	+P^2{e}^{\phi_\infty}}{r^2}
+\mathcal O(r^{-3})
\end{equation}
obteniendo
\begin{equation}
M=\frac{Q^2 {e}^{-\phi_\infty}-P^2{e}^{\phi_\infty}}
{4\Sigma}
\end{equation}
La temperatura y entrop'ia son
\begin{align}
T&=-\frac{u_+}{4\pi\eta}
\left.\frac{dh(u)}{du}\right|_{u_+} \notag \\
&=\frac{\eta\(u_+-1\)^2}{2\pi u_+^2}
\[
q^2e^{-\phi_\infty}\(2+u_{+}\)
-\eta^2p^2e^{\phi_\infty}\(1+2u_{+}\)u_+-\frac{u_+(1+u_{+})}{2(u_{+}-1)}\] \\
S&=\pi\Omega(u_+)
\end{align}
mientras que los potenciales conjugados, el'ectrico y magn'etico, son
\begin{align}
\Phi&=A_t(u_+)-A(u=1)= \frac{\eta\(u_+{-}1\)}{u_+}Qe^{-\phi_{\infty}}  \\
\Psi&=A^p_t(u_+)-A^p(u=1)
=\eta\(u_{+}-1\)Pe^{\phi_\infty}
\end{align}
donde, en el lenguaje de formas diferenciales, $dA^p=F^p\equiv e^{\phi}\star F$ o, equivalentemente, $\Psi=\int_{u_+}^{u=1}{F^p_{ut}du}$.
Ahora, usando la ecuaci'on del horizonte, $h(u_+)=0$, es sencillo verificar la primera ley
\begin{equation}
dE=TdS+\Phi dQ+\Psi dP
\end{equation}
donde, de nuevo, $E=M+W$ es la energ'ia conservada del sistema.
\bigskip

Finalmente, podemos escribir la soluci'on en la coordenada $r$, usando el mismo cambio de coordenadas (\ref{chang22}),
\begin{equation}
ds^2=-a(r)^2dt^2+\frac{dr^2}{a(r)^2}+b(r)^2
\(d\theta+\sin^2\theta\,d\varphi\)
\end{equation}
donde $a(r)$ y $b(r)$ son las funciones m'etricas dadas por
\begin{equation}
a(r)^2=\frac{(r-r_+)(r-r_-)}{b(r)^2}, \qquad
b(r)^2=r^2-4\Sigma^2
\end{equation}
donde $r=\Sigma$ es la ubicaci'on de la singularidad central, mientras que $r_{\pm}$ indican el horizonte interno y externo.

\section{Comentarios finales}

En este cap'itulo, hemos re-obtenido soluciones exactas conocidas mediante unas t'ecnicas que permiten desacoplar convenientemente e integrar las ecuaciones de movimiento.
Hemos verificado que las cargas escalares no aparecen en la primera ley de la termodin'amica para una soluci'on de agujero negro cargado el'ectricamente con pelo, para cualquier coupling de la forma $e^{a\phi}$. En estos casos, la energ'ia conservada no es 'unicamente la masa $ADM$ sino que el contiene un t'ermino extra, tal y como mostramos en el cap'itulo anterior. Adicionalmente, obtuvimos la misma conclusi'on para una soluci'on exacta considerando carga el'ectrica y magn'etica tambi'en.

\newpage

\chapter{Configuraciones termodin'amicamente estables de agujeros negros con campo escalar auto-interactuante}
\label{estabilidad}

\newpage
El car'acter termodin'amico de la gravedad se vuelve aparente en el contexto de la f'isica de agujeros negros. Cuando se consideran efectos cu'anticos, el significado termodin'amica de las cantidades f'isicas, las cuales fueron originalmente consideradas puramente geom'etricas \cite{Bekenstein:1973ur}, viene naturalmente \cite{Hawking:1974sw,Hawking:1976de}.

Esta conexi'on sutil entre la termodin'amica y la f'isica de agujeros negros provee una de las m'as importantes caracter'isticas de cualquier teor'ia que intente unificar la mec'anica cu'antica con la relatividad general. Particularmente, la relaci'on entre la entrop'ia de un agujero negro y el 'area de su horizonte de eventos podr'ia ser una importante pista de que el principio hologr'afico \cite{tHooft:1993dmi, Susskind:1994vu} es fundamental para construir tal teor'ia \cite{Bigatti:1999dp}.

Un agujero negro en equilibrio t'ermico con sus alrededores es claramente relevante para entender la relaci'on entre las cantidades termodin'amicas y su contraparte geom'etrica. La sabidur'ia convencional es que los agujeros negros asint'oticamente planos no son termodin'amicamente estables \cite{Hawking:1976de}. Esto es una consecuencia directa del hecho que el agujero negro de Schwarzschild en un espaciotiempo asint'oticamente plano, sin la imposici'on de m'as condiciones, tiene una capacidad cal'orica negativa y no puede estar en equilibrio t'ermico con un reservorio indefinidamente grande de energ'ia. Esto es, una fluctuaci'on puede romper el equilibrio entre la tasa de absorci'on de radiaci'on t'ermica y la tasa de emisi'on de la radiaci'on de Hawking, lo que conducir'ia a la evaporaci'on del agujero negro o a un indenifido crecimiento (dependiendo de si la fluctuaci'on inicial hizo al agujero negro un poco m'as caliente, o un poco m'as fr'io que el ba\~no t'ermico, respectivamente. 

La capacidad cal'orica para un agujero negro neutro y asint'oticamente plano puede pasar a ser positiva en el ensamble can'onico si se le entrega carga el'ectrica (o se le suministra rotaci'on), pero esto por s'i solo no asegura una estabilidad termodin'amica completa en tanto otras funciones respuestas pueden volverse negativa en el mismo rango de par'ametros.

Entonces, en este cap'itulo, tomaremos este desaf'io e investigaremos soluciones de agujeros negros en una teor'ia m'as general con un campo escalar y su interacci'on no trivial. La ventaja de considerar teor'ias de Einstein-Maxwell-dilat'on con un potencial para el dilat'on, es que existen soluciones exactas \cite{Anabalon:2013qua} y no tenemos que recurrir a m'etodos num'ericos. La principal suposici'on para las demostraciones para teoremas de no-pelo son planitud asint'otica (asymptotic flatness) y la nauraleza del tensor de energ'ia-momento y, entonces, a primera vista, la existencia de soluciones regulares con campo escalar, asint'oticamente planas, puede ser sorprendente. Sin embargo, la existencia de estos fue conjeturada en \cite{Nucamendi:1995ex} y algunas evidencias num'ericas fueron presentadas\footnote{Una clase diferente de agujeros negros en rotaci'on con campo escalar fueron encontradas en \cite{Herdeiro:2014goa, Herdeiro:2015gia}. Para un revisi'on, vea, por ejemplo, \cite{Herdeiro:2015waa}.}. La idea detr'as de esta conjetura es que en una teor'ia con un potencial no trivial para el campo escalar, sus par'ametros pueden ser ajustados al que la constante cosmol'ogica efectiva pueda cancelarse. El potencial considerado en \cite{Anabalon:2013qua}, el cual se anula en el borde, recuerda a un potencial general que fue obtenido para una truncaci'on de un campo escalar de una supergravedad $\omega$-deformada \cite{DallAgata:2012mfj, Tarrio:2013qga, Anabalon:2013eaa} (vea, por ejemplo, \cite{Trigiante:2016mnt} para una revisi'on) y, tambi'en, de una supergravedad de gauge $N=2$, con un t'ermino electromagn'etico de Fayet-Iliopoulos \cite {Faedo:2015jqa, Anabalon:2017yhv}.

Obtendremos las cantidades termodin\'amicas usando un desarrollo relativamente reciente, el llamado m'etodo de contrat'erminos en espaciotiempos asint'oticamente planos, el cual est'a motivado por trabajos similares \cite{Henningson:1998gx, Balasubramanian:1999re, Skenderis:2000in, deHaro:2000vlm} en la dualidad AdS-CFT\cite{Maldacena:1997re}. Interesantemente, vamos a probar que la autointeracci'on del campo escalar es la clave para obtener agujeros negros termodin'amicamente estables.

Ya que los sistemas gravitacionales son intr'insecamente no lineales, las cantidades conservadas son una gran herramienta para investigar su comportamiento. Sin embargo, debido al principio de equivalencia, es claro que una definici'on local para la energ'ia gravitacional no es posible. Sin embargo, en relatividad general, para espaciotiempos asint'oticamente planos, las cantidades conservadas asociadas con simetr'ias asint'oticas han sido definidas en el infinito espacial y nulo. Los primeros trabajos sobre la energ'ia total asociada con una geometr'ia asint'otica fueron debidos a Arnowitt, Deser, y Misner ($ADM$) \cite{Arnowitt:1960es, Arnowitt:1960zzc, Arnowitt:1961zz, Arnowitt:1962hi}, que condujeron a una construcci'on bien definida de energ'ia y momento angular. 

Los sistemas de extensi'on infinita son idealizaciones de situaciones f'isicas m'as realistas, y es deseable tener disponible un an'alisis similar en el caso de sistemas f'isicos de extensi'on finita. Esta es la noci'on de la 'energ'ia cuasilocal', la que es actualmente la descripci'on de energ'ia m'as prometedora en el contexto de la relatividad general y es tambi'en relavante para nuestro trabajo. Este formalismo, iniciado por Brown y York \cite{Brown:1992br}, puede ser caractizado como sigue: la energ'ia gravitacional es asociada con una 2-superficie cerrada tipo-espacio. Una vez que esta superficie es llevada al borde, los resultados coinciden con lo obtenidos por el formalismo $ADM$.

Un problema previo aparece cuando se calcula la acci'on (y el tensor de estr'es cuasilocal) en el infinito espacial es la existencia de divergencias infrarojas asociadas con el volumen infinito del espaciotiempo. El acercamiento inicial de tratar con este desaf'io fue usar la `substracci'on de fondo'  (\textit{background subtraction}), cuya geometr'ia asint'otica calza con aquellas de la soluci'on. Sin embargo, ta lprocedimiento casa que las cantidades f'isicas resultates dependan de la elecci'on del \textit{background} de referencia. Adem'as, no es posible poner la superficie en el borde en el background de referencia incluso para las m'as simples soluciones, por ejemplo, cuando los campos de materia est'an presentes, o para agujeros negros en rotaci'on. Inesperadamente, el rescate vino por una ruta completamente diferente, por la dualidad AdS-CFT en teor'ia de cuerdas.
La observaci'on de que las divergencias infrarojas de la (super)gravedad en el espaciotiempo son equivalentes con las divergencias ultravioletas de la teor'ia de campos dual estaba en las bases del m'etodo de contrat'erminos en AdS.  El procedimiento consiste en agregar t'erminos de borde apropiados (tal que las ecuaciones de movimiento no se vean afectadas), los `contrat'erminos', para regularizar la acci'on. La dualidad impone la restricci'on de que estos contrat'erminos sean construidos solo con t'erminos de curvatura invariantes de la m'etrica en el borde, y no con cantidades extr'insecas al borde, como en el caso del t'ermino de Gibbons-Hawking.
Se observ'o que, incluso en el espaciotiempo plano, uno puede obtener un tensor de estr'es cuasilocal regularizado \cite{Astefanesei:2005ad} y el m'etodo fue exitosamente aplicado para estudiar la termodin'amica de agujeros negros en rotaci'on y anillos negros (particularmente, para obtener la primera ley correcta incluyendo la carga dipolar de  \cite{Emparan:2004wy}). Esto fue una pista importante de que, en efecto, el m'etodo de contrat'erminos es tambi'en adecuado para espaciotiempos planos y, no mucho despuès, un m'etodo general covariante fue propuesto en \cite{Mann:2005yr}. Subsecuentemente, este m'etodo fue usado para muchos ejemplos concretos, por ejemplo, \cite{Mann:2006bd,Astefanesei:2006zd,Herdeiro:2010aq,Astefanesei:2009wi,Compere:2011db,Compere:2011ve}, y, tambi'en, como se detall'o en el cap'itulo 2, para probar que las cargas escalares, contrario a lo que era considerado previamente, no pueden aparecer en la primera ley de la termodin'amica.

\newpage

Este cap'itulo est'a organizado de la siguiente manera. En la secci'on \ref{sec:sols}, presentamos la soluci'on exacta, correspondiente a los agujeros negros cargados, con campo escalar interactuante, en un espaciotiempo asint'oticamente plano de \cite{Anabalon:2013qua} y algunas de sus propiedades geom'etricas relevantes para el an'alisis de la estabilidad termodin'amica. En la secci'on \ref{contra}, aplicamos  el formalismo cuasilocal y el m'etodo de contrat'erminos para las soluciones presentada antes. Calculamos las cantidades termodin'amicas para el agujero negro de Reissner-Nordstr\"om y, luego, para la soluci'on de inter'es. 

En la secci'on \ref{sec:therm1}, brevemente repasamos las condiciones de estabilidad termodin'amica tanto en el ensamble can'onico como gran can'onico, y estudiamos el agujero negro de Reissner-Nordstr\"om, mostrando expl'icitamente que es termodin'amicamente inestable. Entonces, procedemos en la secci'on \ref{sec:therm2} a realizar el an'alisis para la soluci'on con campo escalar presentada en \ref{sec:sols}. Mostramos que, debido al car'acter autointeractuante del campo escalar, estos agujeros negros son termodin'amicamente estables en cierto rango de par'ametros. Finalizaremos con una discusi'on detallada de los resultados.


\newpage
\section{Soluciones exactas con campo escalar}\label{sec:sols}

Consideramos la acci'on Einstein-Maxwell-dilat'on\cite{Anabalon:2013qua}
\begin{equation}
I\left[g_{\mu\nu},A_\mu,\phi\right]
=\frac{1}{2\kappa}\int_{\mathcal{M}}
{d^{4}x\sqrt{-g}}\left[
R-\text{e}^{\gamma\phi}F^2
-\frac{1}{2}(\pa\phi)^2-V(\phi)\right]
\label{action1}
\end{equation}
donde $F^2=F_{\mu\nu}F^{\mu\nu}$, $(\pa\phi)^2=\pa_\mu\phi\,\pa^\mu\phi$, $V(\phi)$ es el potencial dilat'onico y, con nuestras convenciones, $c=G_N=4\pi \epsilon_0=1$ tal que $\kappa=8\pi$. 
Las ecuaciones de movimiento son
\begin{align}
R_{\mu\nu}-\frac{1}{2}g_{\mu\nu}R&=
T_{\mu\nu}^{\phi}+T_{\mu\nu}^{EM}
\label{eins} \\
\pa_{\mu}\(\sqrt{-g}e^{\gamma\phi}F^{\mu\nu}\)
&=0 \label{maxw} \\
\frac{1}{\sqrt{-g}}\partial_{\mu}
\(\sqrt{-g}g^{\mu\nu}\partial_{\nu}\phi\)
&=\frac{dV(\phi)}{d\phi}+\gamma{}\text{e}^{\gamma\phi}F^2
\label{klein}
\end{align}
donde 
$T_{\mu\nu}^{\phi}
\equiv\frac{1}{2}\partial_{\mu}\phi\partial_{\nu}\phi
-\frac{1}{2}g_{\mu\nu}
\[\tfrac{1}{2}(\partial\phi)^2+V(\phi)\]$ y
$T_{\mu\nu}^{EM}\equiv2e^{\gamma\phi}
\left(F_{\mu\alpha}F_{\nu}^{\,\,\alpha}
-\tfrac{1}{4}g_{\mu\nu}F^2\right)$ son los tensores de energ'ia-momento para el dilat'on y para el campo el'ectico (gauge). El acoplamiento no trivial entre el dilat'on y el campo de gauge da lugar a una nueva contribuci'on en el lado derecho de la ecuaci'on de movimiento para el dilat'on, y es la raz'on de porqu'e se trata de un 'pelo secundario'; no hay una constante de integraci'on independiente asociada al campo escalar. 

En una serie de trabajoas \cite{Anabalon:2012ta, Acena:2012mr, Anabalon:2013sra, Acena:2013jya,  Anabalon:2017eri},  usando ansatz's espec'ificos \cite{Anabalon:2012ta}, se desarroll'o un nuevo procedimiento para obtener soluciones exactas regulares para un potencial para el campo escalar general.\footnote{Las propiedades de estos agujeros negros con pelo fueron cuidadosamente estudiadas en trabajos relacionados \cite{Anabalon:2014fla, Anabalon:2015vda, Anabalon:2015xvl}.}
%
\newpage
\subsection{$\gamma=1$}
\label{s1}
\begin{figure}[h]
	\centering
	\includegraphics[height=4cm]{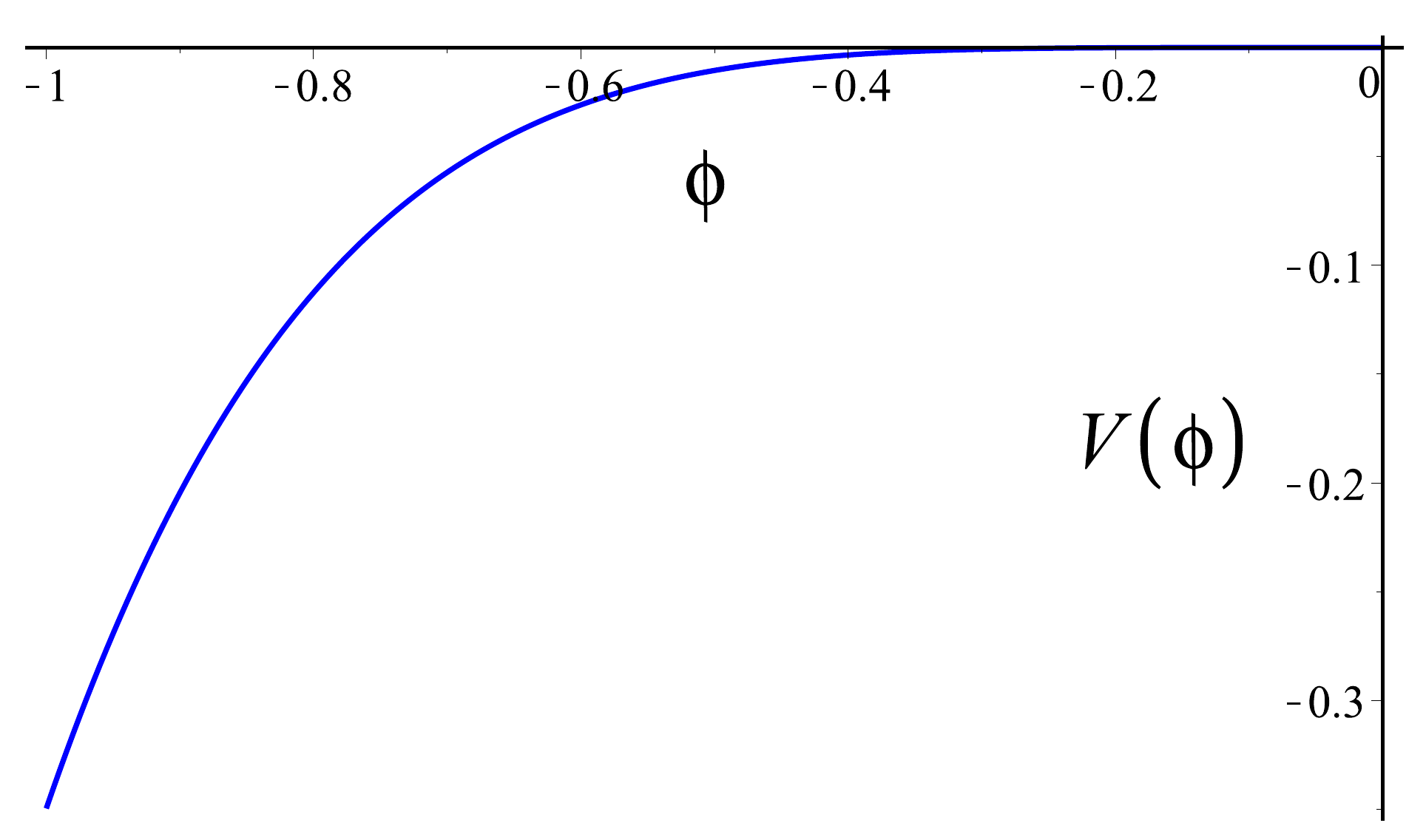}\hspace{0.5cm}
	\includegraphics[height=4cm]{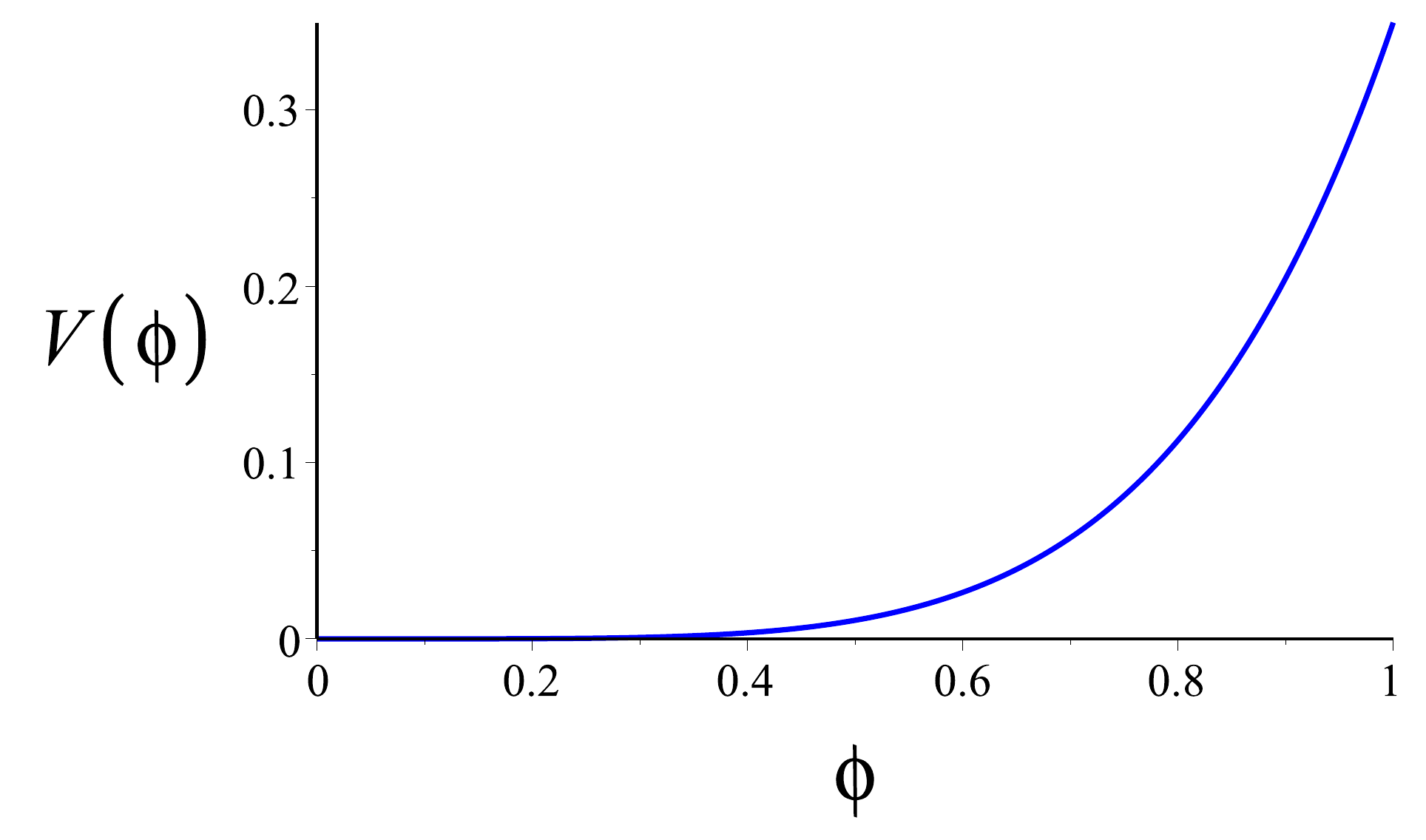}
	\begin{picture}(0,0)(0,0)
	\end{picture}
	\vspace{0.1cm}
	\caption{\small El potencial del campo escalar para la rama negativa (a la izquierda) y para la rama positiva (a la derecha), para $\alpha=10$. }
	\label{pot1}
\end{figure}

En esta secci'on, nos interesamos por la teor'ia $\gamma=1$. Las ecuaciones de movimiento son resueltas para el siguiente potencial
\begin{equation}
\label{dilaton1}
V(\phi)=2\alpha (2\phi+\phi\cosh{\phi}-3\sinh{\phi})
\end{equation}
donde $\alpha$ es un par'ametro arbitrario de la teor'ia. El comportamiento de este potencial se muestra en la Fig. \ref{pot1}, y notamos que se anula para $\phi=0$.

La soluci'on est'a dada por
\begin{align}
ds^{2}&=\Omega(x)
\[-f(x)dt^{2}+\frac{\eta^{2}dx^{2}}{x^{2}f(x)}+d\theta^{2}+\sin^2\theta{}d\varphi^{2}\]
\label{sol1} \\
A&=\left(\frac{q}{x}+C\right)dt
\label{1gammaA} \\
\phi&=\ln(x) 
\label{sol2}
\end{align}
donde $\eta$ y $q$ son los par'ametros independientes de la soluci'on y van a estar relacionados con la masa y la carga el'ectrica del agujero negro. Note que la coordenada $x$ est'a restringida al intervalo $x\in [0, \infty)$, puesto que el dilat'on es un campo escalar real. Se puede asumir, sin perder generalidad, que $\eta>0$. El factor conforme $\Omega(x)$ y la funci'on m'etrica restante $f(x)$ son
\begin{align}
\Omega(x)&=\frac{x}{\eta^{2}\left(x-1\right)^{2}} \\
f(x)&=\alpha
\[\frac{x^{2}-1}{2x}-\ln(x)\]+\frac{\eta^{2}(x-1)^2}{x}
\left[1-\frac{2q^2(x-1)}{x}\right]
\label{1gamma}
\end{align}
respectivamente.
La localizaci'on del horizonte de eventos del agujero negro $x_+$ est'a dada por $f(x_+)=0$, y la constante aditiva $C=-qx_+^{-1}$ tal que el potencial de gauge se anule al horizonte, $A=0$.

La soluci'on, como revisaremos ahora, contiene dos ramas distintas, correspondientes a dos espaciotiempos desconectados. Para ver esto, observe que el factor conforme diverge en el l'imite $x=1$, que es la regi'on asin'otica del espaciotiempo en donde tanto el dilat'on como su potencial se anulan. En este l'imite, podemos relacionar $x$ con la coordenada can'onica $r$ por medio del cambio de coordenadas
\begin{equation}
\label{expansion}
\Omega(x)=r^2+\mathcal{O}(r^{-2})
\end{equation}
Es sencillo mostrar que exiten dos posibilidades,
\begin{align}
x&=1-\frac{1}{\eta r}+\frac{1}{2\eta^2{r}^2}
-\frac{1}{8\eta^3{}r^3}+\mathcal{O}\left(r^{-5}\right)
\label{change1} \\
x&=1+\frac{1}{\eta r}+\frac{1}{2\eta^2{r}^2}
+\frac{1}{8\eta^3{}r^3}+\mathcal{O}\left(r^{-5}\right)
\label{change2}
\end{align}
Por lo tanto, el borde $x=1$ puede ser alcanzado tanto por la `izquierda'\, como por la `derecha', lo que, en efecto, divide la soluci'on en dos ramas. Mientras que el cambio de coordenadas (\ref{change1}) fija el dominio a $0<x< 1$, (\ref{change2}) lo fija a $1< x<\infty$. Puesto que el primer caso implica que el campo escalar adquiere valores negativos, se denomina la rama negativa. El segundo caso, siguiendo el mismo razonamiento, se llama la rama positiva.  Advierta que tanto en el l'imite $x=0$ como en $x=\infty$, el campo escalar diverge. Uno puede chequear que ambos corresponden a singularidades del espaciotiempo.

Estudiaremos la termodin'amica de ambas ramas de manera independiente. Si bien ambas ramas pueden contener agujeros negros, es importante estudiar bajo qu'e condiciones existen. Considere la ecuaci'on del horizonte $f(x_+)=0$ escrita como sigue
\begin{equation}
\alpha\left[1-x_+^2+2x_+\ln(x_+)\right]=
2\eta^2\left(x_{+}-1\right)^2
\left[x_{+}-2q^2(x_{+}-1)\right]
\label{condition}
\end{equation}
Para la rama negativa, $0<x_+< 1$, el lado derecho de (\ref{condition}) es claramente positivo, mientras que la combinaci'on $1-x_+^2+2x_+\ln(x_+)$ que mulplipica $\alpha$ en el lado izquierdo es una funci'on definida positiva. Por lo tanto, para la rama negativa, s'olo las teor'ias con $\alpha>0$ soportan agujeros negros (valores negativos de $\alpha$ dejar'ian singularidades desnudas). Para la rama positiva, sin embargo, no tenemos esta restricci'on sobre $\alpha$.

\subsection{$\gamma=\sqrt{3}$}

\begin{figure}[h] 
	\centering
	\includegraphics[height=4cm]{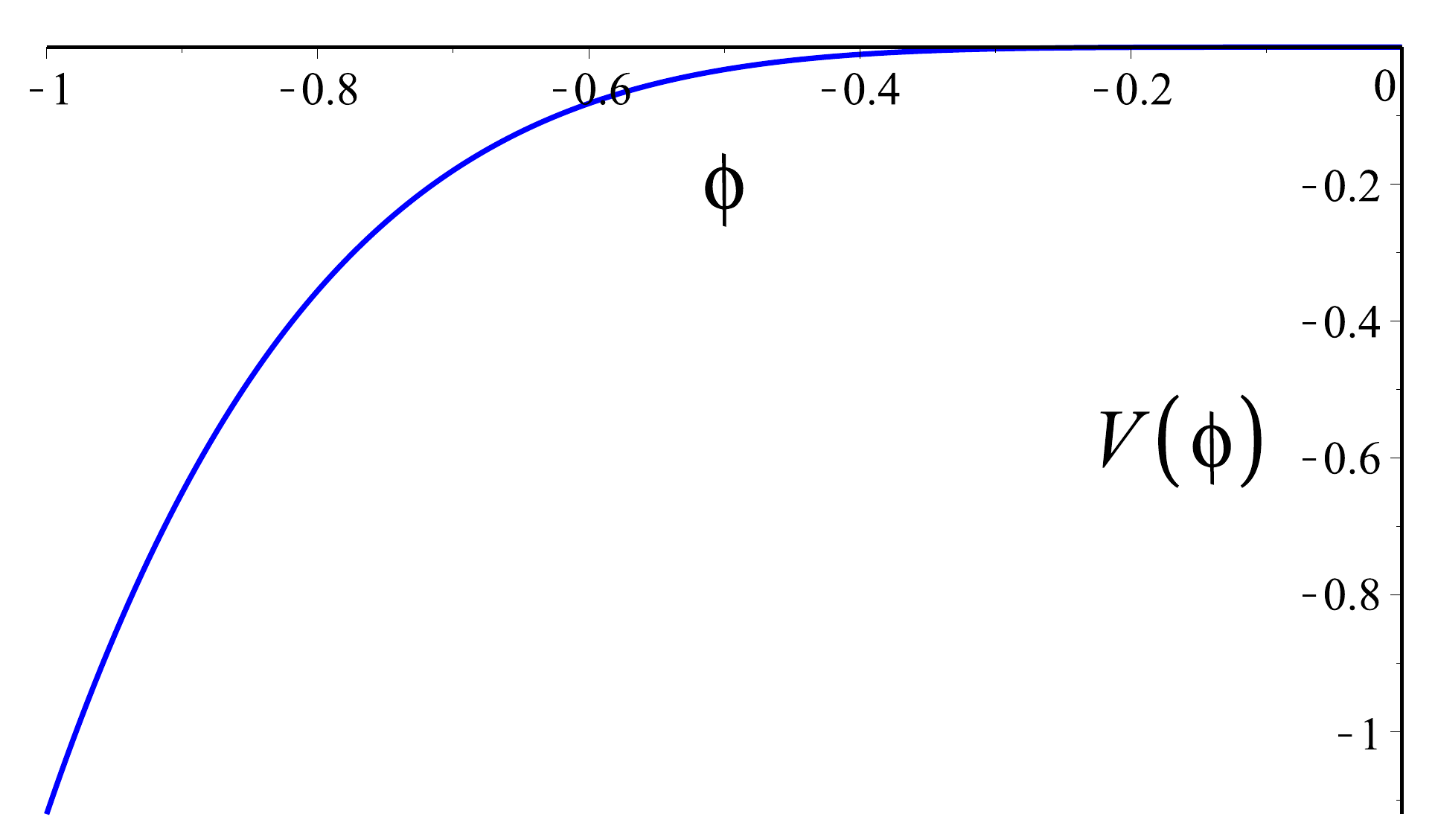}\hspace{0.5cm}
	\includegraphics[height=4cm]{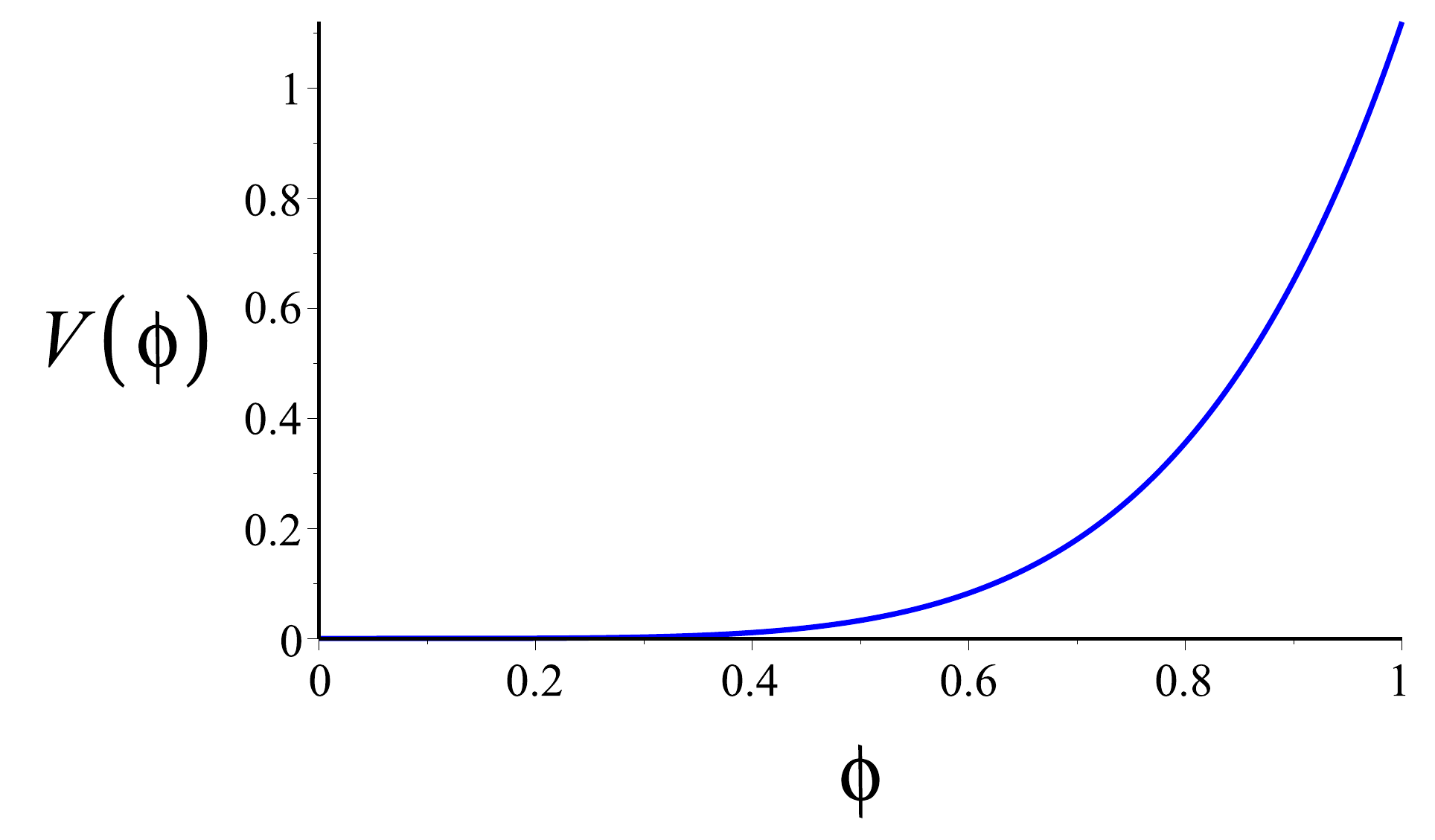}
	\begin{picture}(0,0)(0,0)
	\end{picture}
	\vspace{0.1cm}
	\caption{\small Potencial del campo escalar para la rama negativa (izquierda), y la rama positiva (derecha), para $\alpha=10$.}
	\label{pot3}
\end{figure}

Las ecuaciones de movimiento para esta constante de acoplamiento son resueltas por el potencial
\begin{equation}
\label{dilaton2}
V(\phi)=\alpha
\[\sinh\(\sqrt{3}\phi\)+9\sinh\(\frac{\phi}{\sqrt{3}}\)
-4\sqrt{3}\,\phi \cosh\(\frac{\phi}{\sqrt{3}}\)\]
\end{equation}
cuyo comportamiento se muestra en la Fig. \ref{pot3}. Se observa que tiene un comportamiento similar con el caso mostrado previamente, $\gamma=1$.

La soluci'on exacta es
\begin{equation}
ds^2=\Omega(x)\[
-f(x)dt^2+\frac{\eta^2dx^2}{f(x)}
+d\theta^{2}+\sin^2\theta{}d\varphi^{2}\]
\end{equation}
\begin{equation}
A=\left(\frac{q}{2x^2} +C\right)dt 
\end{equation}
\begin{equation}
\phi=\sqrt{3}\ln(x)
\end{equation}
donde las funciones m'etricas son
\begin{align}
f(x)&=\alpha
\left[\frac{x^4}{2}-2x^2+\frac{3}{2}+2\ln(x)\right]
+\frac{\eta^2(x^2-1)^2}{4}\[1-\frac{q^2(x^2-1)}{x^2}\]
\label{3gamma} \\
\Omega(x)&=\frac{4x}{\eta^{2}\left(x^2-1\right)^{2}}
\end{align}
La constante aditiva es $C=-q(2x_+)^{-1}$.
Al igual que antes, tambi'en existen dos ramas para esta teor'ia, donde los cambios de coordenadas son
\begin{align}
x&=1-\frac{1}{\eta r}+\frac{1}{8\eta^3 r^3}+\frac{1}{8\eta^4 r^4}
+\mathcal{O}(r^{-5}) \label{change3} \\
x&=1+\frac{1}{\eta r}-\frac{1}{8\eta^3 r^3}+\frac{1}{8\eta^4 r^4}
+\mathcal{O}(r^{-5}) \label{change4}
\end{align}
Finalmente, la ecuaci'on del horizonte $f(x_+)=0$, escrita como,
\begin{equation}
\alpha\left[-\frac{(1-x_+^2)(3-x_+^2)}{2}-2\ln(x_+)\right]
=\frac{\eta^2}{4x_+^2}
\left[x_+^2-(x_+^2-1)\,q^2\right](x_+^2-1)^2
\end{equation}
revela las mismas restricciones para $\alpha$ que en el caso $\gamma=1$, esto es, que la existencia de agujeros negros regulares dentro de la rama negativa es posible siempre que $\alpha>0$, mientras que no hay restricciones para la rama positiva.


\newpage
\subsection{M'etodo de contrat'erminos y cantidades conservadas}
\label{contra}

En esta secci'on, usamos el m'etodo de los contrat'erminos para obtener las cantidades termodin'amicas para los agujeros negros con pelo en las teor'ias $\gamma=1$ y $\gamma=\sqrt{3}$ descritas arriba. En el ap'endice \ref{aB} mostramos detalles sobre este formalismo aplicado al agujero negro de Reissner-Nordstr\"om. 

Perm'itanos comenzar con la teor'ia $\gamma=1$. En primer lugar, la carga el'ectrica es
\begin{equation}
Q=\frac{1}{4\pi}\oint_{s_\infty^2}{ e^{\phi}\star F}
=\frac{1}{4\pi}\oint_{s_\infty^2}
{e^{\phi}\(\frac{1}{4}\sqrt{-g}\epsilon_{\mu\nu\alpha\beta}
	F^{\mu\nu}dx^\alpha\wedge dx^\beta\)}=\frac{q}{\eta}
\label{charge1}
\end{equation}
La temperatura de Hawking puede ser calculada empleando los resultados de la subsecci'on (\ref{sub:hawkingtemp}). Del mismo modo, entrop'ia y potencial conjugado son obtenidos de la manera tradicional,
\begin{align}
\label{quant1}
T&=\frac{x_+}{4\pi\eta}f'(x_+) 
=\frac{(x_{+}-1)^2}{8\pi\eta x_+}
\[\alpha-\frac{4\eta^2q^2(x_{+}+2)}{x_+}
+2\eta^2\(\frac{x_{+}+1}{x_{+}-1}\)\], \notag\\
S&=\pi\,\Omega(x_{+})=\frac{\pi x_{+}}{\eta^{2}(x_{+}-1)^{2}}, \qquad
\Phi
=-\frac{Q\eta (x_{+}-1)}{x_{+}}
\end{align}
Para calcular el tensor de estr'es cuasilocal y la acci'on en la secci'on Euclidiana, debemos considerar la foliaci'on $x=constant$ con la m'etrica inducida
\begin{equation}
ds^2_{\pa\mathcal{M}}=h_{ab}\,dx^{a}dx^{b}=\Omega(x) \[-f(x)dt^{2}+d\theta^{2}+\sin^{2}{\theta}d\varphi^{2}\,\]
\label{xcte1}
\end{equation}
Las componentes del tensor de estr'es (\ref{stress1}), con $\phi_{\infty}=0$, son
\begin{align}
\tau_{tt}&
=\frac{12\eta^{2}q^{2}-\alpha}{6\eta\kappa}\,(x-1)^{2}
+\mathcal{O}\[(x-1)^{3}\] \\
\tau_{\theta\theta}&=\frac{\tau_{\varphi\varphi}}{\sin^2\theta}
=\frac{\(\alpha-12\eta^{2}q^{2}\)^{2}-36\eta^{4}\(4q^{2}-1\)}
{288\kappa\,\eta^{5}}\,(x-1)+\mathcal{O}[(x-1)^{2}]
\end{align}
%
Para calcular la energ'ia cuasilocal, usamos la ecuaci'on (\ref{BYcharge}) y el vector de Killing $\xi=\pa/\pa t$ (el procedimiento est'a detallado para RN en ap'endice \ref{aB}). Para verificar que este es el vector de Killing correcto (apropiadamente normalizado), expandimos asint'oticamente la m'etrica, alrededor de $x=1$, 
\begin{align}
ds^2&=g_{tt}dt^2+g_{xx}dx^2+g_{\theta\theta} \(d\theta^2+\sin^2\theta d\varphi^2 \)
\notag \\
&=\[-1+\mathcal{O}(x-1)\]dt^2
+\[\frac{1}{\eta^2(x-1)^4}+\mathcal{O}\[(x-1)^{-3}\]\]dx^2
+\frac{x\,\(d\theta^2+\sin^2\theta d\varphi^2 \)}{\eta^2(x-1)^2}
\end{align}
Ahora, cambiando a coordenadas can'onicas $r$, mediante la transformaci'on 
\begin{equation}
\frac {dx^2}{\eta^2(x-1)^4}=dr^2 
\quad\rightarrow\quad x=1\pm \frac{1}{\eta r}
\end{equation}
en el borde, es sencillo ver en que el l'imite $r\rightarrow\infty$, se obtiene la m'etrica de Minkowski en coordenadas esf'ericas $ds^2=-dt^2+dr^2+r^2\(d\theta^2+\sin^2\theta d\varphi^2 \)$, con el tiempo coordenado propiamente normalizada. Calculando la energ'ia total en el l'imite $x=1$, se obtiene
\begin{equation}
E = \frac{\alpha-12\eta^2q^2}{12\eta^3}
\label{mass1}
\end{equation}
Este resultado coincide con la masa $ADM$, que puede ser le'ida de la expansi'on de la componente $g_{tt}$ de la m'etrica en las coordenadas can'onicas, con $r$ dado por la ecuaci'on (\ref{expansion}),
\begin{equation}
g_{tt}
= -1+\frac{\alpha-12\eta^{2}q^{2}}{6\eta^{3}r}
+\mathcal{O}\(r^{-2}\) 
\end{equation}
Usando las cantidades termodin'amicas (\ref{charge1}), (\ref{quant1}), junto con la masa cuasilocal (\ref{mass1}) y la ecuaci'on del horizonte $f(x_+)=0$, se puede verificar que la primera ley es satisfecha,
\begin{equation}
dM=TdS+\Phi dQ
\end{equation}
El 'ultimo paso en este an'alisis es verificar la relaci'on estad'itico-cu'antica. Para ello, necesitamos calcular la acci'on en la secci'on Euclidiana,
\begin{equation}
\label{ctq3}
I^{E}_{bulk}+I_{GH}^{E}
= \beta(-TS-\Phi Q)
+\frac{8\pi\beta}{\kappa}\left[\frac{1}{\eta(x-1)}
+\frac{\alpha-12\eta^{2}q^2+3\eta^2}{6\eta^{3}}+\mathcal{O}(x-1)\right] \notag 
\end{equation}
La divergencia $\propto (x-1)^{-1}$ es cancelada por el contrat'ermino gravitacional,
\begin{eqnarray}
\label{ctgrav}
I_{ct}^{E}=\frac{8\pi\beta}{\kappa}
\lim_{x=1}\Omega \sqrt{f}
=\frac{8\pi\beta}{\kappa}
\[-\frac{1}{\eta(x-1)}
-\frac{\alpha-12\eta^{2}q^{2}+6\eta^{2}}{12\eta^{3}}
+\mathcal{O}(x-1)\]\;\;\;
\end{eqnarray} 
y tambi'en contribuye con una parte finita, al igual que para el agujero negro de Reissner-Nordstr\"on en la ecuaci'on (\ref{ctRN}). Agregando los resultados, en el l'imite $x=1$, obtenemos el siguiente resultado para la acci'on en la secci'on Euclidiana
\begin{equation}
I^E=\beta(-TS-\Phi Q)+\beta\left(\frac{\alpha-12\eta^{2}q^{2}}{12\eta^{3}}\right)
=\beta\(M-TS-\Phi Q\)
\end{equation}
Por lo tanto, la acci'on finita on-shell en la secci'on Euclidiana satisface la relaci'on estad'istico-cu'antica $I^E=\beta\mathcal{G}$ para el ensamble gran can'onico. Ahora, para obtener el potencial termodin'amico en el ensamble can'onico, debemos considerar el t'ermino de borde $I_A$ compatible con el requerimiento de que la carga sea fija, $Q=constant$, dado por
\begin{equation}\label{canterm2}I_A=\frac{2}{\kappa}\int_{\pa\mathcal{M}}{d^3x\sqrt{-h}e^{\phi}\,n_\nu F^{\mu\nu}A_\nu}\end{equation}
En la secci'on Euclidiana, tenemos,
\begin{equation}
I^E_A=\beta\, \Phi Q
\end{equation} 
y, por lo tanto, la acci'on on-shell para el ensamble can'onico es $\bar I^E=\beta\mathcal{F}$, donde $\bar I^E=I^E+I_A^E$ y $\mathcal{F}\equiv M-TS$ es el potencial termodin'amico correspondiente.
\bigskip

Por completitud, permitanos ahora realizar el mismo an'alisis para la teor'ia $\sqrt{3}$. Sin embargo, pueto que el procedimiento es exactamente como antes, presentaremos los resultados relevantes.

La carga el'ectrica es
\begin{equation}
Q=\frac{1}{4\pi}\oint_{s_\infty^2}
{\star e^{\sqrt{3}\phi}F}
=\frac{q}{\eta}
\end{equation}
y las dem'as cantidades termodin'amicas son
\begin{equation}
\label{quant2}
T=\frac{f'(x_+)}{4\pi\eta}
=\frac{(x_+^2-1)^2}{2\pi\eta\,x_+}\[
\alpha-\frac{\eta^2q^2(2x_+^2+1)}{4x_+^2}
+\frac{\eta^2x_+^2}{2(x_+^2-1)}\]
\end{equation}
\begin{equation}
S=\frac{4\pi x_{h}}{\eta^{2}(x^{2}_{h}-1)^{2}} \,,\qquad
\Phi=-\frac{Q\eta (x_{+}^{2}-1)}{2x_{+}^{2}}
\end{equation}
La energ'ia cuasilocal en este caso es
\begin{equation}
E=\frac{8\alpha+3\(1-2q^{2}\)\eta^2}{6\eta^3}
\end{equation}
y se satisface la primera ley $dM=TdS+\Phi dQ$.

La acci'on regularizada en la secci'on Euclidiana, en el l'imite $x=1$, es
\begin{equation}
I^E=I^{E}_{bulk}+I_{GH}^{E}+I_{ct}^{E}=
\beta(M-TS-\Phi Q)
\end{equation}
y se verifica $I^{E}=\beta\mathcal{G}=\beta(M-TS-\Phi Q)$ para el ensamble gran can'onico. El c'alculo de $I_A^E$ permite obtener la acci'on en el ensamble can'onico, $\bar I^{E}=\beta\mathcal{F}=\beta(M-TS)$


\newpage
\section{Estabilidad termodin'amica} 
\label{sec:therm1}

En la seci'on previa, hemos dado una descripci'on termodin'amica de agujeros negros con pelo. La idea clave de nuestro an'alisis fue considerar condiciones de borde para construir los ensambles apropiados. 
Sin embargo, la termodin'amica de agujeros negros s'olo tiene sentido si los agujeros negros pueden estar en equilibrio localmente estable en el ensamble correspondiente.

La estabilidad t'ermica en un ensamble con un agujero negro debe aplicar para el sitema completo, puesto que tal sistema claramente no puede ser subdividido en partes espacialmente separadas como se hace usualmente cuando se trata la estabilidad termodin'amica. Las funciones respuesta relevantes para la estabilidad t'ermica de un tipo de ensamble dado, por lo tanto, son aquellas que pueden ser obtenidas mediante la variaci'on de par'ametros termodin'amicos que no est'an fijos por las condiciones de borde que definen el ensamble en cuesti'on. En la descripci'on del ensamble can'onico para un agujero negro neutro asint'oticamente plano, la capacidad cal'orica puede hacerse positiva si el agujero negro es puesto dentro de una caja. Las condiciones de borde deben ser especificadas, es decir, en el caso del agujero negro de Schwarzschild uno puede fijar la temperatura de la caja y su radio. Se sigue entonces que la estabilidad puede ser lograda s'olo para cavidades suficientemente peque\~nas, conduciendo a un ensamble can'onico bien definido.

Despu'es de una revisi'on sobre las condiciones para la estabildad termodin'amica, aplicaremos este an'alisis para las soluciones de inter'es.

\newpage
\subsection{Condiciones para la estabilidad termodin'amica}
\label{stabilitycond}

Para analizar la estabilidad termodin'amica de las soluciones presentadas hasta ahora, es importante distinguir entre estabilidad local y global. La primera est'a relacionada con c'omo las configuraciones de equilibrio responden bajo peque\~nas fluctuaciones en las variables termodin'amicas, mientras la 'ultima est'a relacionada con el m'aximo global en la entrop'ia (o m'inimo global en la energ'ia). Describiremos la estabilidad local, aunque para m'as detalles puede consultarse \cite{Landau} (o tambi'en \cite{Callen}).

Estamos interesados en agujeros negros cargados (con o sin pelo) est'aticos, con las cargas conservadas $M$ y $Q$, para los cuales la primera ley de la termodin'amica pueda ser escrita como
\begin{equation}
\label{masss}
dM=TdS+\Phi dQ \,\qquad \rightarrow \,\qquad T=\(\frac{\pa M}{\pa S}\)_Q ,\quad
\Phi=\(\frac{\pa M}{\pa Q}\)_S
\end{equation}
Todos los procesos irreversibles en sistemas aislado que conducen al equilibrio est'an gobernados por un incremento en la entrop'ia, y el equilibrio ser'a restablecido s'olo cuando la entrop'ia asuma su m'aximo valor. Esta es la segunda ley de la termodin'amica, $dS \geqslant 0$. Ya que primero trabajaremos en el ensamble gran can'nonico, es m'as conveniente usar el potencial termodin'amico correspondiente para estudiar la estabilidad local y no la condici'on de m'axima entrop'ia.

En el ensamble gran can'onico, para el cual el potencial electrost'atico $\Phi$ es constante, tenemos que, al equilibrio, el potencial termodin'amico $\mathcal{G}$ satisface
\begin{equation}
\label{freeenrgy2}
\mathcal{G}_e(T_e, \Phi_e)=M-T_eS-\Phi_e Q \,,\qquad 
d\mathcal{G}_e = -(SdT+Qd\Phi)_e = 0
\end{equation}
donde el sub'indice $e$ indica el valor al equilibrio termodin'amico.
Considere ahora una peque\~na desviaci'on, para las que las condiciones para el equilibrio local son
\begin{equation}
\left(\frac{\partial^2 M}{\partial Q^2}\right)_{S}\left(\frac{\partial^2 M}{\partial S^2}\right)_{Q}-\left[\left(\frac{\partial}{\partial S}\right)_{Q}\left(\frac{\partial M}{\partial Q}\right)_{S}\right]^{2}= \frac{T}{C_{Q}}
\(\frac{1}{\epsilon_{S}}-\frac{T\alpha_{Q}^{2}}{C_{Q}}\)>0
\label{mixcondM}
\end{equation}
\begin{equation}
\(\frac{\pa^2 M}{\pa S^2}\)_{Q}=\frac{T}{C_{Q}} > 0 
\end{equation}
\begin{equation}
\(\frac{\pa^2 M}{\pa Q^2}\)_{S}=\frac{1}{\epsilon_{S}}>0
\label{indcondM}
\end{equation}
Las expresiones arriba fueron escritas tambi'en en t'erminos de las cantidades termodin'amicas usuales como la capacidad cal'orica ($C_Q$), la permitividad el'ectrica ($\epsilon_S$) y $\alpha_Q \equiv (\pa \Phi/\pa T)_Q$ (vea, por ejemplo, \cite{Chamblin:1999hg}).

Ahora, queremos obtener unas relaciones similares para las fluctuaciones en el ensamble gran canon'ico y can'onico, usando segundas derivadas de los potenciales termodin'amicos correspondientes. Para ello, usaremos las siguientes relaciones entre las cantidades termodin'amicas de inter'es,
\begin{equation}
\label{relations}
C_{\Phi}=C_{Q}+T\epsilon_T\alpha_Q^2
,\qquad 
\epsilon_S=\epsilon_T-\frac{T\alpha_\Phi^2}{C_{\Phi}}
,\qquad 
\alpha_{\Phi}=-\epsilon_{T}\alpha_{Q}
\end{equation}
Es sencillo ahora reescribir las condiciones para la estabilidad local en una forma compacta
\begin{equation}
\epsilon_S>0, \qquad C_\Phi>0 \qquad \text{ensamble gran can'onico}
\end{equation}
\begin{equation}
\epsilon_T>0, \qquad C_Q>0 \qquad \,\,\,\,\,\,\,\,\,\,\,\,\,\,\,\, \text{ensamble can'onico}
\end{equation}
Estas relaciones son consistentes con el criterio de estabilidad general que establece que los potenciales termodin'amicos son funciones convexas de sus variables extensivas, y funciones c'oncavas de sus variables intensivas (vea, por ejemplo, \cite{Callen}).

\newpage
\subsection{El agujero negro de Reissner-Nordstr\"{o}m}
\label{RN}

Si un agujero negro est'a en equilibrio termodin'amico a una temperatura $T$, entonces debe estar rodeado por una ba\~no t'ermico a la misma temperatura. El agujero negro de Schwarzschild, la soluci'on est'atica y esf'ericamente sim'etrico m'as simple de las ecuaciones de Einstein, es termodin'amicamente estable. Esto se puede obtener simplemente observando que la temperatura es inversamente proporcional con respecto a la masa del agujero negro,
\begin{equation}
T=\frac{\hbar c^3}{k_B G}\frac{1}{8\pi M} \,,\,\,\, S=\frac{k_B c^3}{4\hbar G}A_h = \frac{4\pi k_B G}{\hbar c}M^2 \,\,\, \rightarrow \, C=T\(\frac{\pa S}{\pa T}\) =-\frac{8\pi k_BG}{\hbar c}M^2
\end{equation}
Ya que la capacidad cal'orica es negativa, el agujero negro se calienta mientras emite radiaci'on y pierde energ'ia. La consecuencia inmediata es que el ensamble can'onico no est'a bien definido.
Una forma de arreglar esta inestabilidad, fue propueta por York en \cite{York:1986it}, en la cual considera al agujero negro en una caja de radio $r_B$. El principio de equivalencia requiere que la temperatura medida localmente por un observador est'atico est'a ``corrida al azul"{} con respecto a la temperatura usual que es determinada en, asint'oticamente plano, infinito espacial:
\begin{equation}
\label{t}
T(r_B) = T_{\infty} \left|g_{tt}(r_B)\right|^{-1/2} =\frac{\hbar}{8\pi GM}\(1-\frac{2GM}{r_B} \)^{-1/2}
\end{equation} 
donde, como en \cite{York:1986it}, hemos usado las convenciones en las cuales $c=k_B=1$. La capacidad cal'orica a un valor fijo de $r_B$ puede ahora ser calculada usando la temperatura (\ref{t}) y el resultado es
\begin{equation}
C=T\(\frac{\pa S}{\pa T}\)
=\frac{8\pi G M^2}{\hbar} \(1-\frac{2GM}{r_B}\)
\(\frac{3GM}{r_B} - 1\)^{-1}
\end{equation}
Es claro ahora que, cuando $2M\leq r_B < 3M$, la capacidad cal'orica es positiva y entonces el ensamble can'onico es bien definido.

Una pregunta natural es, ?`es posible obtener una capacidad cal'orica positiva cuando se agrega un campo el'ectrico (carga el'ectrica) sin poner al sistema en una caja? Si bien la respuesta es positiva, la carga el'ectrica introduce otra funci'on respuesta relevante, la permitividad el'ectrica, que, como mostraremos para el agujero negro de Reissner-Nordstr\"om, no es simult'aneamente positiva junto con la capacidad cal'orica y, en consecuencia, no permite una estabildiad termodin'amica.

La m'etrica y el potencial de gauge que resuelven las ecuaciones de movimiento son las dadas en (\ref{rnmetric}) y (\ref{rnpotential}), y las cantidades termodin'amicas est'an dadas en (\ref{thermod}). La existencia de agujeros negros est'a condicionada por la desigualdad $Q\leq M $, de otra forma la soluci'on describe una singularidad desnuda. En el caso extremo, $M=Q$, la temperatura se anula y el radio del horizonte se vuelve $r_+=M=Q$. Esto va a imponer una restricci'on sobre los valores del potencial electrost'atico, $\Phi=Q/r_+ \leq 1$. 

La permitividad el'ectrica es una medida de las fluctuaciones el'ectricas. Los agujeros negros ser'an el'ectricamente inestables bajo fluctuaciones el'ectricas si la permitividad el'ectrica es negativa.
Esto ocurre si el potencial electrost'atico de sistema decrece como resultado de situar m'as cargas sobre el agujero negro. Este potencial, en configuraciones de equilibrio, deber'ia por supuesto aumentar, en un intento por hacer m'as dif'icil poner m'as carga sobre el agujero negro.

\begin{figure}[H]
	\centering
	\includegraphics[width=7 cm]{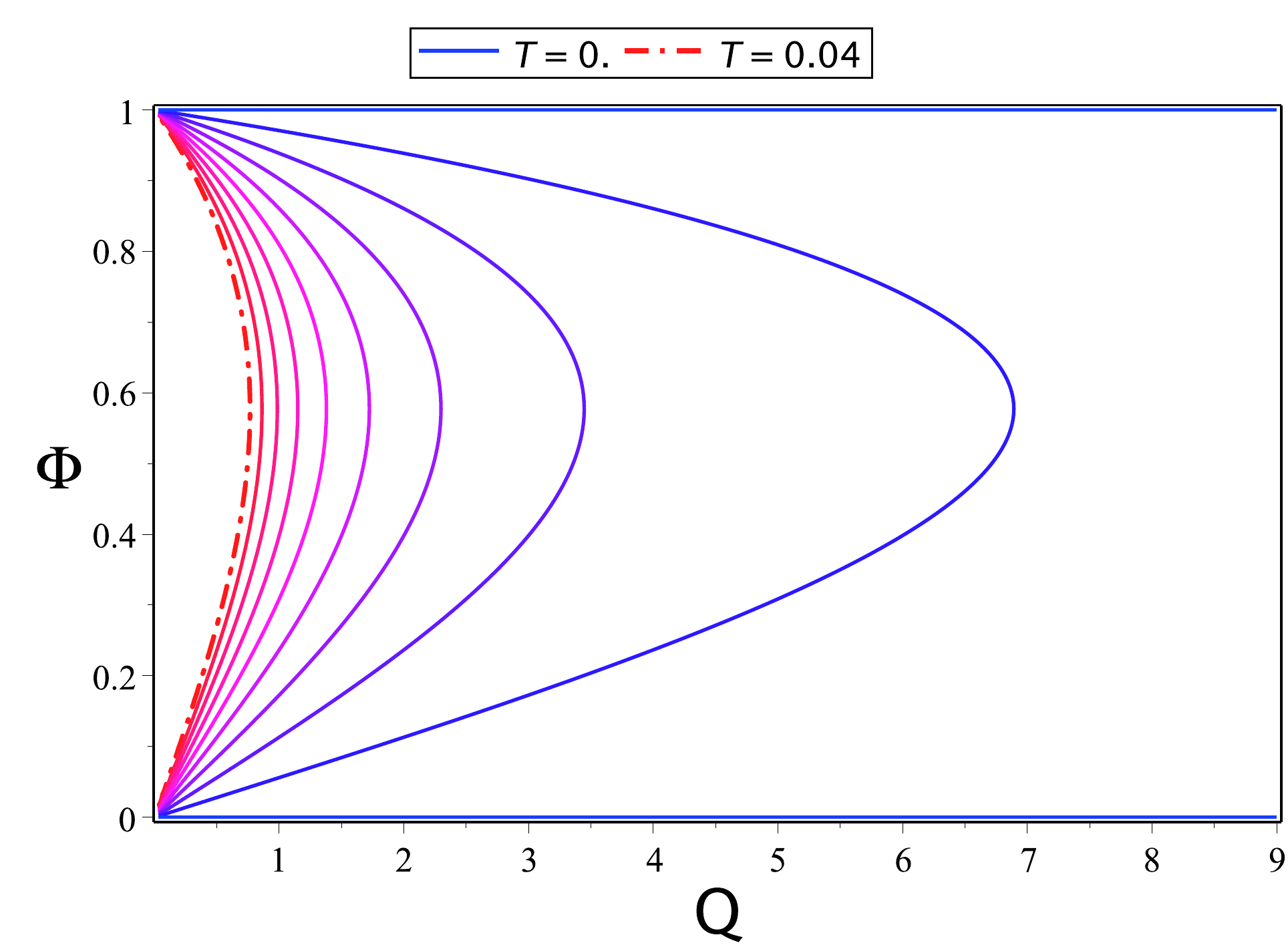} \quad
	\includegraphics[width=7 cm]{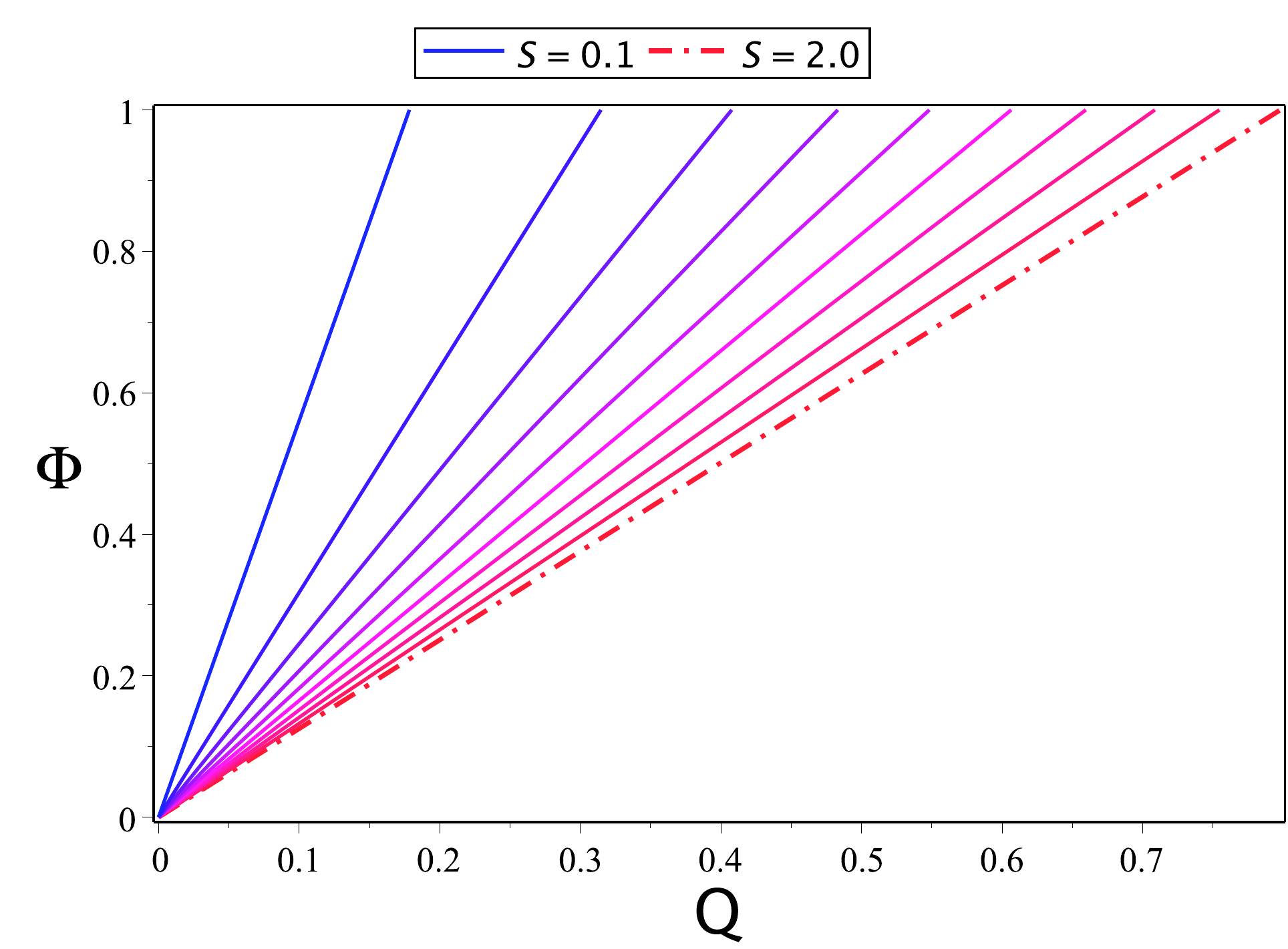}
	\caption{\small{A la izquierda: Ecuaci'on de estado para el agujero negro de RN. La l'inea horizontal (azul) ubicada en $\Phi=1$ representa el agujero negro extremo. A la derecha: $\Phi$ vs $Q$ a entrop'ia fija. Se ha toamdo $\alpha=10$.}}
	\label{RN1}
\end{figure}

La ecuaci'on de estado da informaci'on importante sobre la estabilidiad frente a fluctuaciones tanto en la carga el'ectrica $Q$ o en su potencial conjugado $\Phi$, a una temperatura fija.
Combinando la primera ley y la 'ultima ecuaci'on en (\ref{thermod}) para eliminar $r_+$, obtenemos la ecuaci'on de estado para el agujero negro de RN,
\begin{equation}
4\pi TQ+\Phi\, \left(\Phi^2-1\right)=0
\label{Eqstatern}
\end{equation}
de donde la permitividad el'ectrica a temperatura fija, $\epsilon_T$, puede ser le'ida. Tambi'en, combinando las primeras dos ecuaciones en (\ref{thermod}), se obtiene $\pi Q^2-S\Phi^2=0$,
desde donde se puede leer la permitividad el'ectrica a entrop'ia fija, $\epsilon_S$. Las ecuaciones $Q$ vs $\Phi$ a temperatura y entrop'ia fijas est'an graficadas en la Fig. \ref{RN1}. Es sencillo mostrar que s'olo agujeros negros con $\Phi<1/\sqrt{3}$ son el'ectricamente estables, puesto que diferenciando la ecuaci'on $(\ref{Eqstatern})$ con respecto a $\Phi$ a temperatura fija, se obtiene
\begin{equation}
\epsilon_T=\(\frac{\pa Q}{\pa\Phi}\)_{T}=\frac{1-3\Phi^2}{4\pi T}
\end{equation}

Ahora, presentaremos la estabilidad bajo fluctuaciones mixtas, de acuerdo con la discusi'on en la secci'on previa. En el ensamble gran can'onico, la estabilidad termodin'amica est'a garantizada por la positividad simult'anea de las funciones respuestas $C_\Phi$ y $\epsilon_S$. Tenemos,
\begin{equation}
\epsilon_S=\frac{1-\Phi^2}{4\pi T}>0
\end{equation}
ya que $\Phi \leq 1$. Por otra parte, tenemos
\begin{equation}
\mathcal{G}(r_+,\Phi)=\frac{\(1-\Phi^2\)r_+}{4},
\quad 
T(r_+,\Phi)=\frac{1-\Phi^2}{4\pi r_+}
\quad\rightarrow\quad
\mathcal{G}(T,\Phi)
=\frac{\(1-\Phi^2\)^2}{16\pi T}
\end{equation}
y entonces la capacidada cal'orica es
\begin{equation}
C_\Phi=
-\frac{1}{8\pi}\(\frac{1-\Phi^2}{T}\)^2<0
\end{equation}
En otras palabras, $\mathcal{G}(T,\Phi)$ no cambia su concavidad con respecto a la temperatura, a $\Phi$ fijo. Este resultado implica que no hay configuraciones termodin'amicamente estables en este ensamble.

En el ensamble can'onico, debemos investigar si $C_Q$ y $\epsilon_T$ pueden ser simult'aneamente poitivas. La permitividad el'ectrica a temperatura fija puede ser expreasada como
\begin{equation}
\epsilon_T=\frac{\(r_+^2-3Q^2\)r_+}{r_+^2-Q^2}
\end{equation}
Note que, puesto que $r_+^2-Q^2$ es una cantidad positiva, la regi'on donde $\epsilon_T>0$ corresponde a aquellas configuraciones que satisfacen $r_+^2-3Q^2>0$ (esto es equivalente a $\Phi<1/\sqrt{3}$).
En otras palabras, el potencial termodin'amico y la temperatura pueden expresarse como
\begin{equation}
\mathcal{F}(r_+,Q)=\frac{r_+^2+3Q^2}{4r_+},
\qquad T(r_+,Q)=\frac{r_+^2-Q^2}{4\pi r_+^3}
\end{equation}
y, por lo tanto, la capacidad cal'orica,
\begin{equation}
C_Q=-\frac{2\pi r_+^2\(r_+^2-Q^2\)}{r_+^2-3Q^2}
\end{equation}
tiene valores positivos solamente si $r_+^2-3Q^2<0$, pero entonces, $C_Q$ y $\epsilon_T$ no son ambas positivas a la vez, para un mismo valores en las cantidades termodin'amicas. Esto confirma que no hay configuraciones termodin'amicamente estables en el ensamble can'onico.

\newpage
\section{Agujeros negros con pelo}
\label{sec:therm2}

En esta secci'on, obtendremos el principal resultado de este cap'itulo, que los agujeros negros cargados con pelo escalar asint'oticamente planos en teor'ias con una interacci'on no trivial para el campo escalar pueden ser termodin'amicamente estables, y entonces ponerlos en una caja no es necesario. Quisi'eramos comentar que en ausencia de la auto-interacci'on, tambi'en existen soluciones regulares asint'oticamente planas de agujeros negros, pero no son termodin'amicamente estables.
El caso de inter'es es cuando el potencial dilat'onico es no trivial, pero, por completitud, presentamos el an'alisis del caso sin potencial en el ap'endice \ref{apeC}. Por simplicidad, vamos a influir en esta secci'on s'olo el caso $\gamma=1$ para ambas ramas de la soluci'on, y dejaremos para el mismo ap'endice \ref{apeC2}, el estudio del caso $\gamma=\sqrt{3}$, para el cual las conclusiones sobre estabilidad son las mismas que para $\gamma=1$. La principal observaci'on es que la auto-interacci'on del campo escalar juega un rol similar al que juega la ``caja"{} para agujeros negros sin pelo, como el caso discutido inicialmente en este cap'itulo, y, por lo tanto, es un ingrediente clave para la estabilidad termodin'amica de agujeros negros asint'oticamente planos.

\subsection{La rama positiva}
\label{main}

En esta subsecci'on, investigamos la estabilidad termodin'amica del agujero negro presentado en la secci'on previa \ref{sec:sols}, para la rama positiva, es decir, para valores positivos del campo escalar o, equivalentemente, $x\in (1, +\infty)$. La teor'ia considerada corresponde a $\gamma=1$, y tomaremos el caso $\alpha>0$.

\subsubsection{\textit{Ensamble gran can'onico}}
\label{gcpb}

\begin{figure}[h]
	\centering
	\includegraphics[width=7 cm]{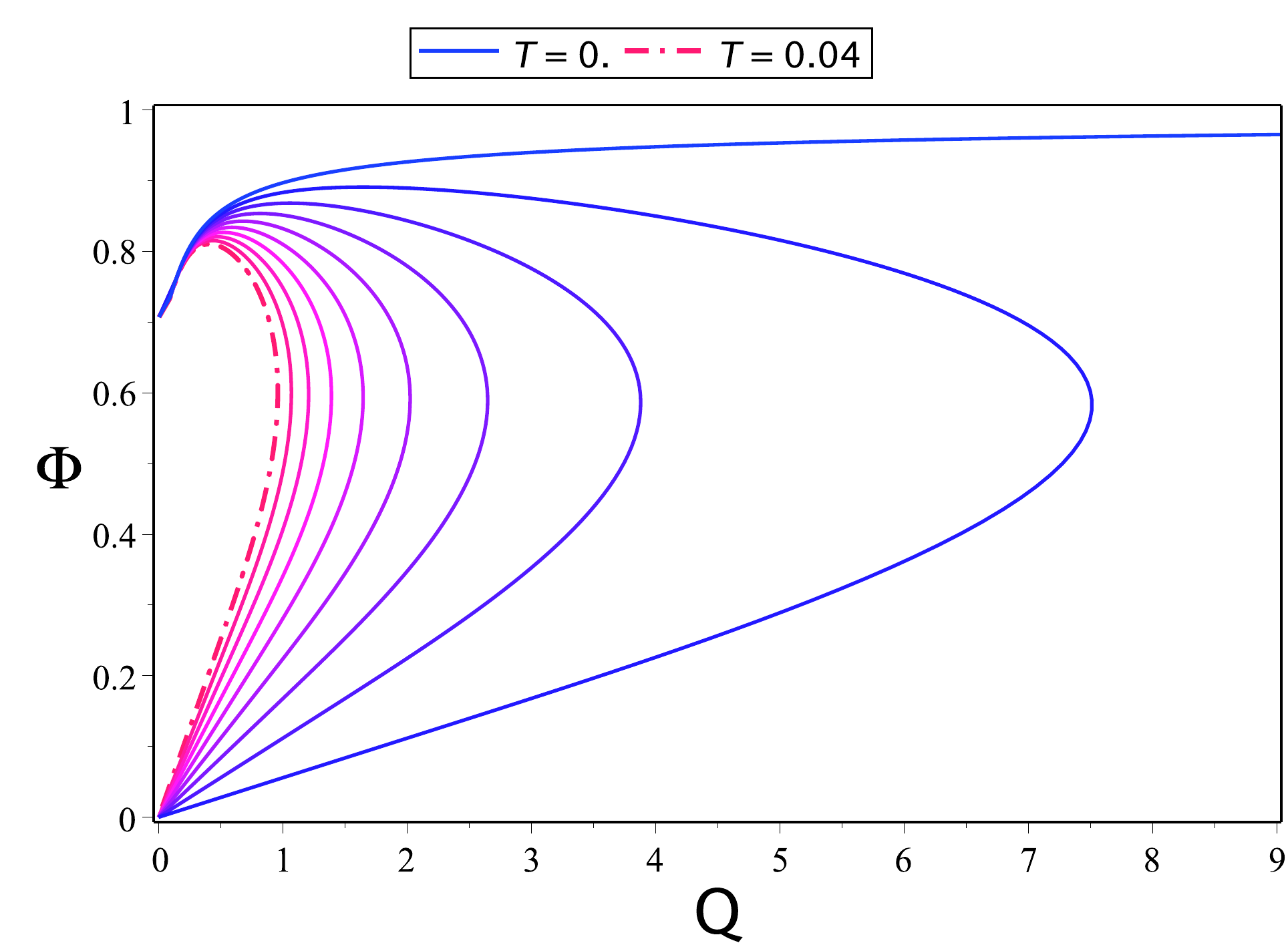}
	\includegraphics[width=7 cm]{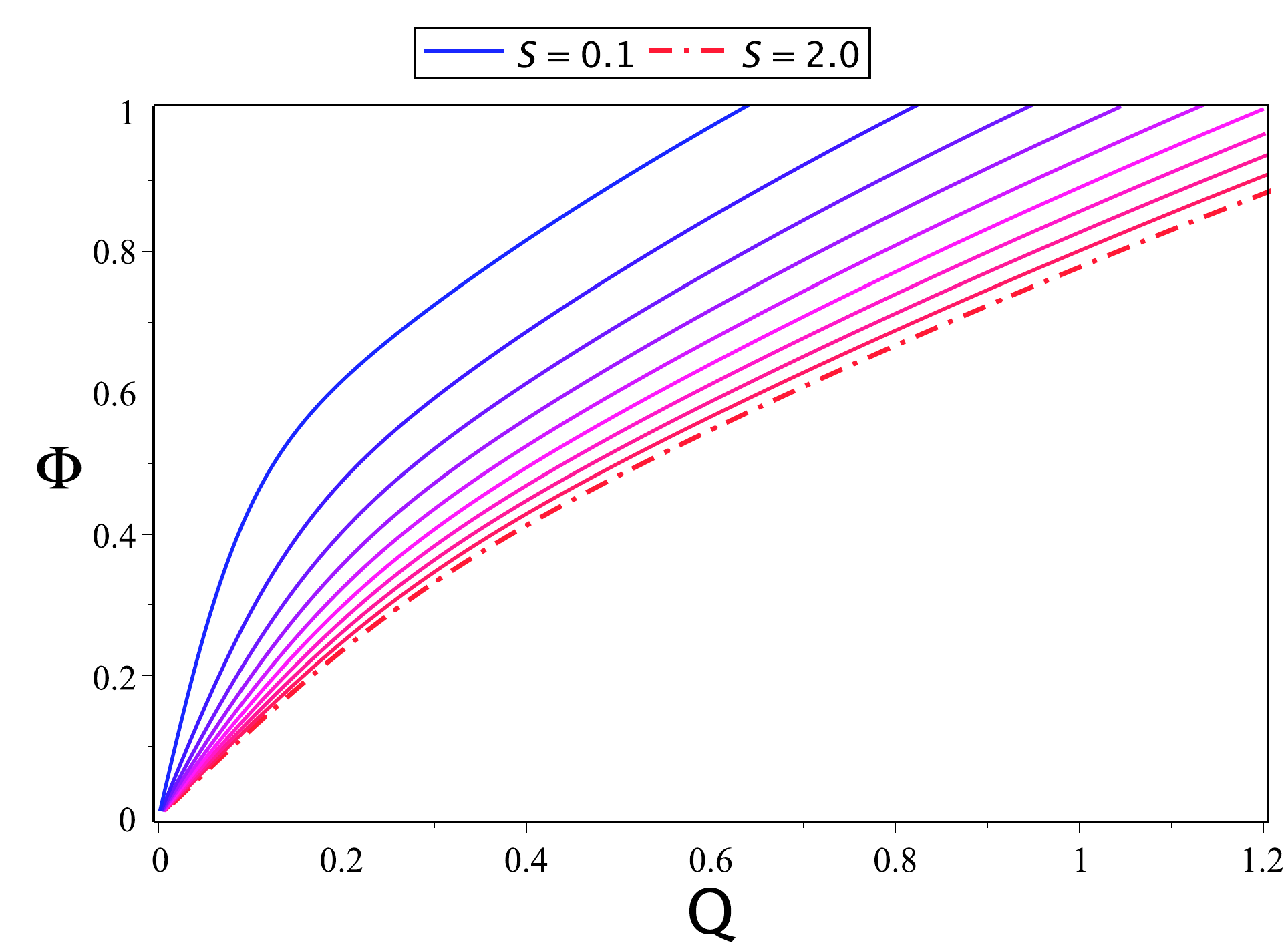}
	\\\caption{\small{Izquierda: Isotermas $\Phi$ vs $Q$. Derecha: $\Phi$ vs $Q$ a entrop'ia fija, para $\alpha=10$}.}
	\label{state_hairy1}
\end{figure}

\begin{figure}[h]
	\includegraphics[width=7 cm]{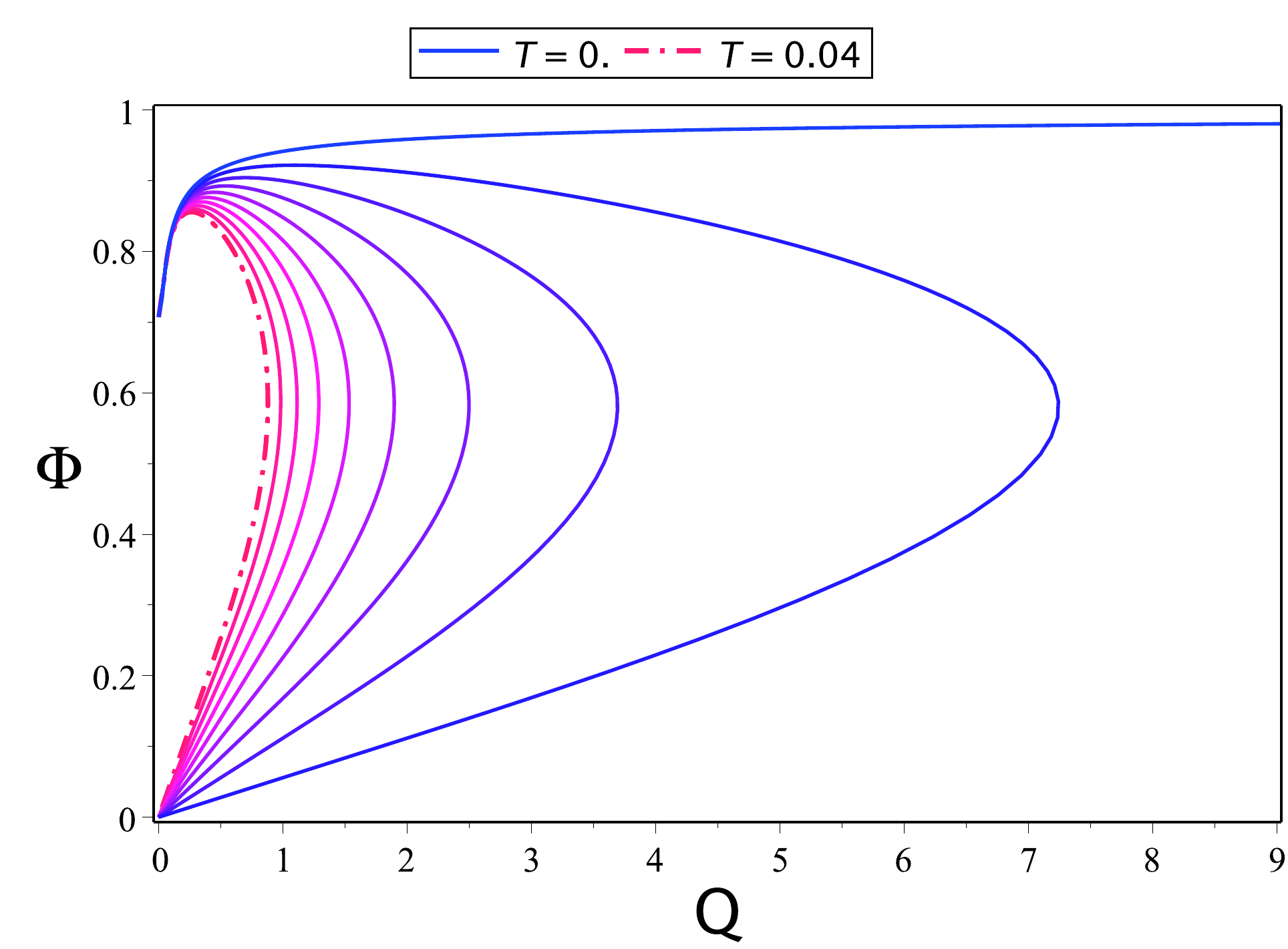}
	\includegraphics[width=7 cm]{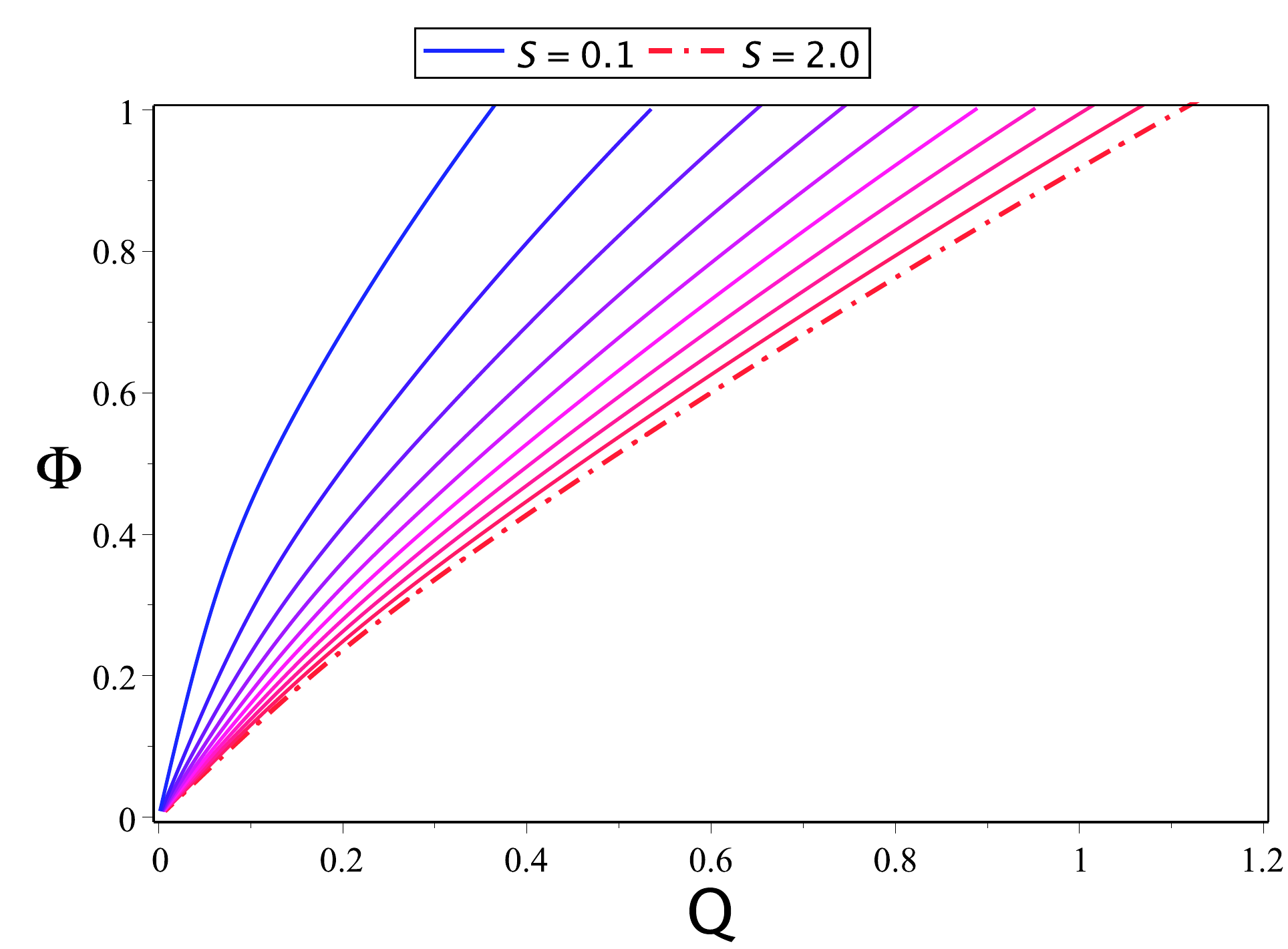}
	\caption{\small{Isotermas y curvas isoentr'opicas, $\Phi$ vs $Q$, para $\alpha=100$}.}
	\label{state_hairy2}
\end{figure}

Como es claro mirando la ecuaci'on (\ref{1gamma}), no es posible despejar $x_+$ de la ecuaci'on del horizonte, $f(x_+)=0$. Por lo tanto, deberemos trabajar con ecuaciones param'etricas. Para obtener la ecuaci'on de estado, es necesario expresar tanto la carga el'ectrica $Q$ como su potencial conjugado $\Phi$ como funciones de $(x_+,T)$. Para concretar este plan, primer despejamos la ra'iz negativa\footnote{De esta manera, estamos trabajando con $Q>0$ y $\Phi>0$.} de $q$ desde la ecuaci'on del horizonte,
\begin{equation}
q=-\frac{\sqrt{x_{+}\(2\,\eta^2x_+^2
		-2\alpha x_{+}\ln x_{+}+\alpha{x_+}^2
		-4\,\eta^2x_{+}+2\eta^2-\alpha\)}}
{2\eta\(x_{+}-1\)^{3/2}}
\label{q1}
\end{equation}
Luego, reemplazamos $q$ en la expresi'on para la temperatura (\ref{quant1}), para obtener
\begin{equation}
\label{tempbun}
T(x_+,\eta)=\frac{\(x_+-1\)\eta}{4\pi x_+}
-\frac{\(2x_+^2\ln{x_+}+4x_+\ln{x_+}-5x_+^2+4x_++1\)\alpha}
{8\pi x_{+}\(x_{+}-1\)\eta}
\end{equation}
de donde depejamos la ra'iz positiva de $\eta$,
\begin{equation}
\eta(x_+,T)=\frac{4\pi T x_{+}+ w(x_+,T)}
{2(1-x_+)}
\label{eta1}
\end{equation}
donde $w(x_+,T)=\sqrt{ 16\pi^2T^2x_+^2
+\alpha(4x_+^2\ln{x_+}+8x_+\ln{x_+}-10x_+^2+8x_{+}+2)}$. Finalmente, reemplazamos (\ref{eta1}) de vuelta en (\ref{q1}) para obtener $q=q(x_+,T)$. Es sencillo ahora encontrar expresiones concretas para $Q=Q(x_+,T)$ y $\Phi=\Phi(x_+,T)$, usando (\ref{charge1}) y (\ref{quant1}). 
Obtenemos los siguientes resultados,
\begin{align}
\Phi&=\frac{
\sqrt{\(x_+-1\)\(32\pi^2T^2x_+^2+ 16\pi T x_+w-8\alpha x_+\ln{x_+}+4\alpha x_+^2+2w^2-4\alpha\)}}
{2\sqrt{x_+}\(4\pi x_+T+ w\)}
\label{hairychem1} \\
Q&=\frac{
\sqrt{x_+(x_+-1)\(32\pi^2T^2x_+^2+ 16\pi T x_+w-8\alpha x_+\ln{x_+}+4\alpha x_+^2+2w^2-4\alpha\)}}
{\(4\pi x_+T+ w\)^2}
\label{hairycharg1}
\end{align}
Un procedimiento similar es usado para obtener expresiones para $\Phi(x_+,S)$ y $Q(x_+,S)$,
\begin{align}
\Phi&=\frac{
\sqrt{\({x_+}-1\)\[\alpha{S}\(x_+^2-1-2x_+\ln{x_+}\)+2\pi x_+\]}}
{2x_+\sqrt{\pi}}
\\
Q&=\frac{
\sqrt{\({x_+}-1\)\[\alpha{S}\(x_+^2-1-2x_+\ln{x_+}\)+2\pi x_+\]S}}
{2\pi\sqrt{x_+}}
\end{align}
En las Figs. \ref{state_hairy1} y \ref{state_hairy2} se muestra $\Phi$ vs $Q$ a temperatura y entrop'ia fijas, para diferentes 'ordenes de magnitud de $\alpha$.
A partir de la representaci'on gr'afica, uno puede concluuir que la permitividad el'ectrico isoentr'opica es positiva, $\epsilon_S>0$.  SIn embargo, este resultado tambi'en puede ser obtenido anal'iticamente, puesto que
\begin{equation}
\epsilon_S=\frac{\alpha S
	\(2x_+^3-2x_+^2\ln{x_+}-3x_+^2+2{x_+}-1\)
	+2\pi x_+^2}
{\alpha S\(x_+^3-2x_+\ln{x_+}-2x_+^2+3{x_+}-2\)
	+2\pi x_+} \(\frac{Sx_+}{\pi}\)^{\frac{1}{2}}
\end{equation}
donde, como puede mostrarse, tanto el numerador como el denominador son funciones definidas positivas en la rama positiva.

La posibilidad de esta funci'on respuesta est'a directamente relacionada con la estabilidad en el ensamble gran can'onico. Si algunas configuraciones con $\epsilon_S>0$ estuvieran tambi'en caracterizadas por tener $C_\Phi>0$, entonces, representar'ian agujeros negros termodin'amicamente estables, de acuerdo con la discusi'on en la secci'on previa.

\begin{figure}[h]
	\centering
	\includegraphics[width=11 cm]{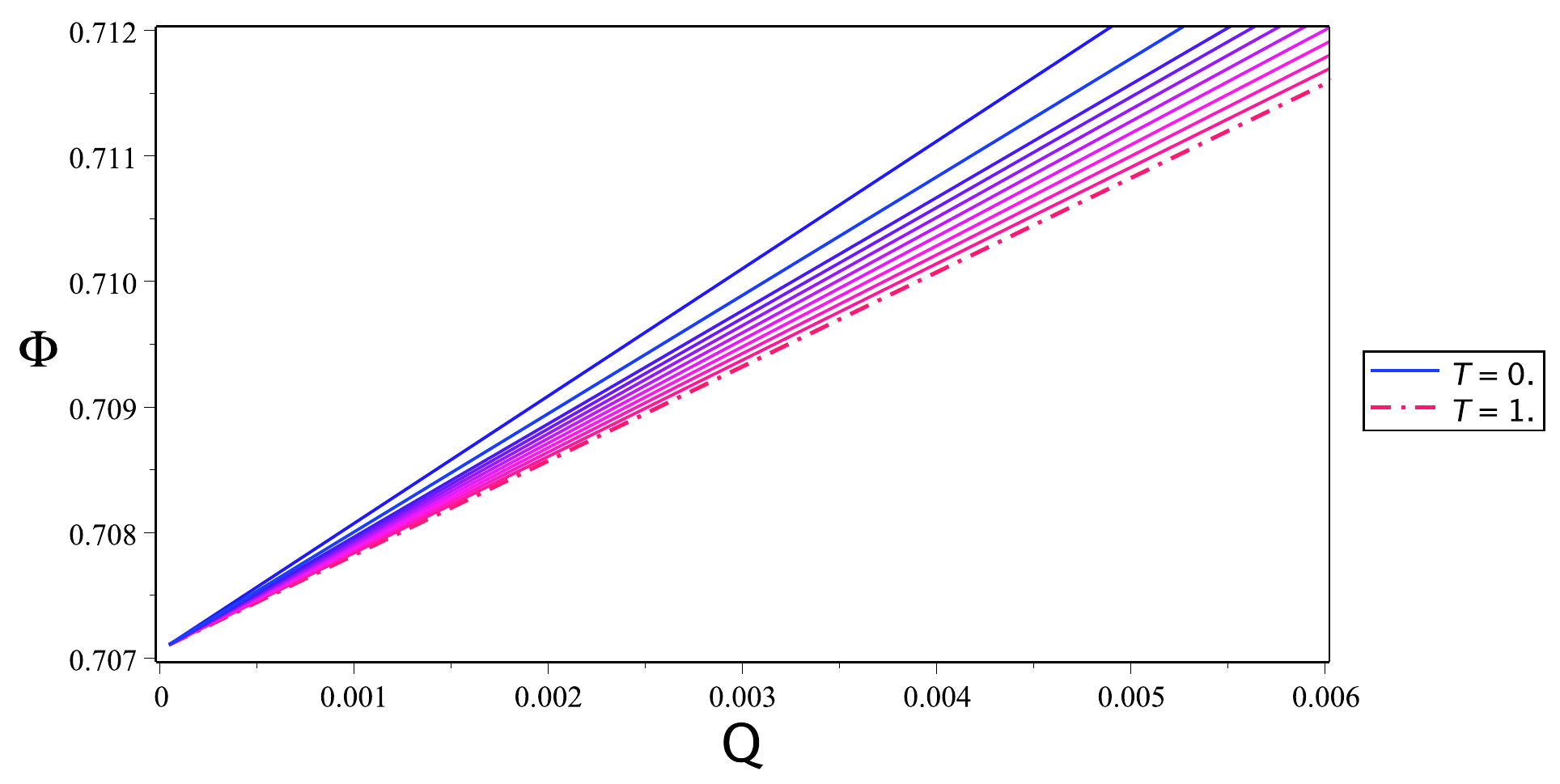}
	\caption{\small{Un acercamiento (zoom) para la ecuaci'on de estado, en las cercan'ias de $Q=0$ y $\Phi=-1/2q$. La temperatura se tom'o desde $T=0$ hasta $T=1$}. }
	\label{zoomeqestate}
\end{figure}

La ecuaci'on de estado, mostrada en el lado izquierdo en las Figs. \ref{state_hairy1} y \ref{state_hairy2}, revela la existencia de dos regiones separadas donde $\epsilon_T>0$. Tambi'en, note que los agujeros negros extremos son el'ectricamente estables en este caso. Estas dos caracter'isticas no est'an presentes en el agujero negro de Reissner-Nordstr\"om. Otro aspecto interesante de la ecuaci'on de estado es que todas las isotermas comienzan y terminan en los puntos $Q=0$, $\Phi=0$ (justo como el agujero negro de RN) y en $Q=0$, $\Phi=1/\sqrt{2}<1$. Para entender porqu'e, 
observe que la carga f'isica es $Q=-q/\eta$ (estamos considerando $q$ negativo en nuestro an'alisis) y $Q=0$ puede alcanzarse en los l'imites $q=0$ y $\eta\rightarrow\infty$\footnote{Aunque es importante notar que esta rama no contiene agujeros negros cuando $Q=0$, puesto que en el l'imite $Q\rightarrow 0$, $x_+\rightarrow 1$.}, 'ultimo caso en el cual tenemos que, de la ecuaci'on del horizonte,
\begin{equation}
\frac{\eta^{2}(x_+ - 1)^2}{x_+}\left[1-\frac{2q^2(x_+ - 1)}{x_+}\right] = 0 \quad\longrightarrow \quad 
\frac{x_+}{x_+ - 1} =2q^2
\end{equation}
de donde se sigue que, reemplazando en (\ref{quant1}),
\begin{equation}
\Phi=-\frac{Q\eta (x_{+}-1)}{x_{+}} \,\,\,\Longrightarrow \,\,\, \Phi=-\frac{1}{2q}
\label{limq}
\end{equation}
Por un lado, entonces, tenemos que en el l'imite $Q=0$, el potencial conjugado adquiere dos valores distintos, $\Phi=0$ y $\Phi=-1/2q$. Observe que, en el l'imite $\eta\rightarrow\infty$, $q=-1/\sqrt{2}$ cuando $x_+\rightarrow\infty$ (vea la ecuaci'on (\ref{limq})). En este caso, $\Phi\rightarrow 1/\sqrt{2}$, como se puede observar en las Figs. \ref{state_hairy1}, \ref{state_hairy2}, y \ref{zoomeqestate}, donde se ha hecho un zoom,

Examinemos la otra funciones respuesta relevante, la capacidad cal'orica $C_\Phi$, por medio del potencial termodin'amica. Resolviendo $\eta=\eta(x_+,\Phi)$ de la ecuaci'on del horizonte, encontramos
\begin{align}
M&=-\frac{\alpha}{12\eta^3}
+\frac{x_+^2\Phi^2}{\eta(x_{+}-1)^2}, \quad
Q=\frac{x_{+}\Phi}{\eta(x_{+}-1)}, \quad
S=\frac{\pi x_+}{\eta^2(x_{+}-1)^2},\\
T&=\frac{(x_{+}-1)^2}{8\pi\eta\,x_+}
\[-\alpha+4\Phi^2\eta^2x_+\,\frac{x_{+}+2}{(x_{+}-1)^2}
-2\eta^2\,\frac{x_{+}+1}{x_+-1}\]
\end{align}
y el potencial termodin'amico es
\begin{equation}
\mathcal{G}(x_+,\Phi)=\frac{\alpha}{24\eta^3}
-\frac{\Phi^2x_+^2}{2\eta(x_+-1)^2}
+\frac{x_{+}+1}{4\eta(x_+-1)}
\end{equation}
La capacidad cal'orica $C_\Phi$ puede ahora ser directamente calculada, obteniendo la segunda derivada del potencial termodin'amico. Concretamente, las condiciones para la estabilidad local en el gran can'onico son
\begin{align}
-\(\frac{\pa^2\mathcal{G}}{\pa T^2}\)_\Phi&>0
\label{firstcondition}
\\
\(\frac{\pa^2\mathcal{G}}{\pa T^2}\)_\Phi
\(\frac{\pa^2\mathcal{G}}{\pa\Phi^2}\)_T
-\[\(\frac{\pa}{\pa\Phi}\)_T
\(\frac{\pa\mathcal{G}}{\pa T}\)_\Phi\]^2&>0
\label{secondcondition}
\\
-\(\frac{\pa^2\mathcal{G}}{\pa T^2}\)_\Phi
\left\{\(\frac{\pa^2\mathcal{G}}{\pa T^2}\)_\Phi
\(\frac{\pa^2\mathcal{G}}{\pa\Phi^2}\)_T
-\[\(\frac{\pa}{\pa\Phi}\)_T
\(\frac{\pa\mathcal{G}}{\pa T}\)_\Phi\]^2\right\}&>0
\label{thirdcondition}
\end{align}
De acuerdo a la discusi'on en la secci'on (\ref{stabilitycond}), o como puede directamente verse de estas condiciones, es consistente que s'olo dos de ellas se satifagan, ya que la restante se satisface autom'aticamente. Para ser m'as expl'icito, las desigualdades (\ref{firstcondition}) y (\ref{thirdcondition}) son equivalentes a las condiciones $C_\Phi>0$ y $\epsilon_S>0$, respectivamente, mientras que (\ref{secondcondition}) es equivalente a $C_\Phi\epsilon_S>0$. 

\begin{figure}[H]
	\centering
	\includegraphics[width=4.6 cm]{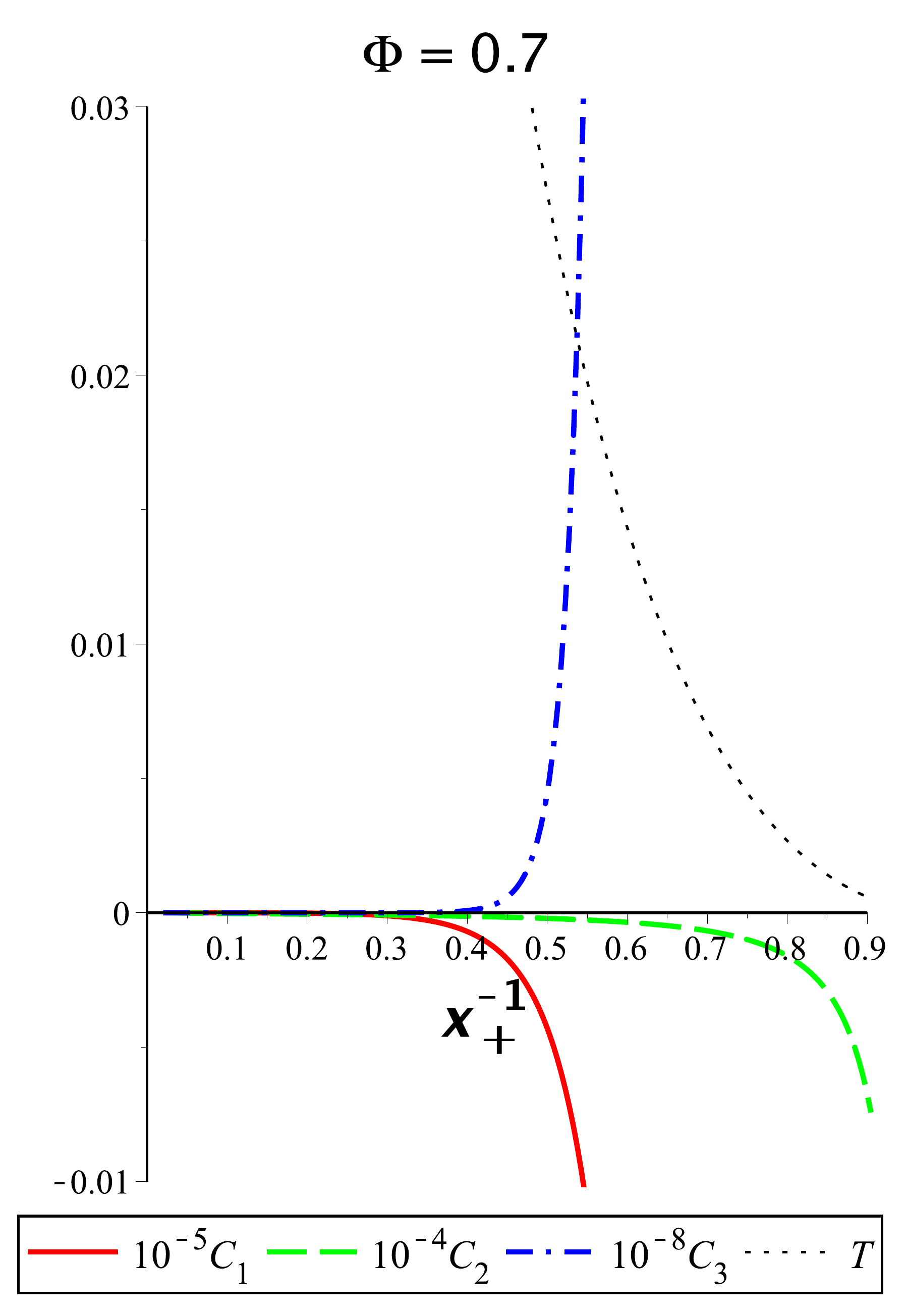}
	\quad
	\includegraphics[width=4.6 cm]{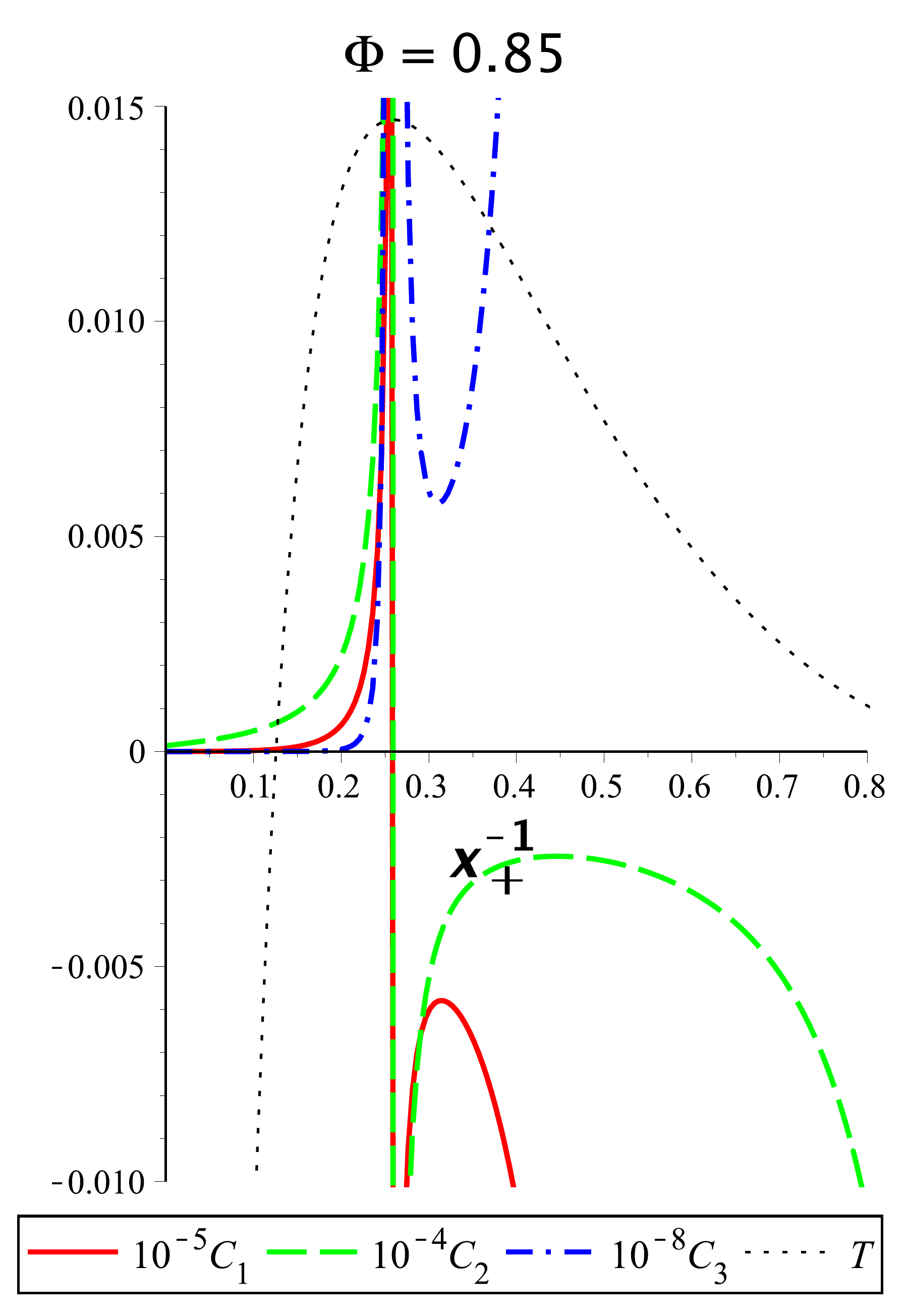}
	\quad
	\includegraphics[width=4.6 cm]{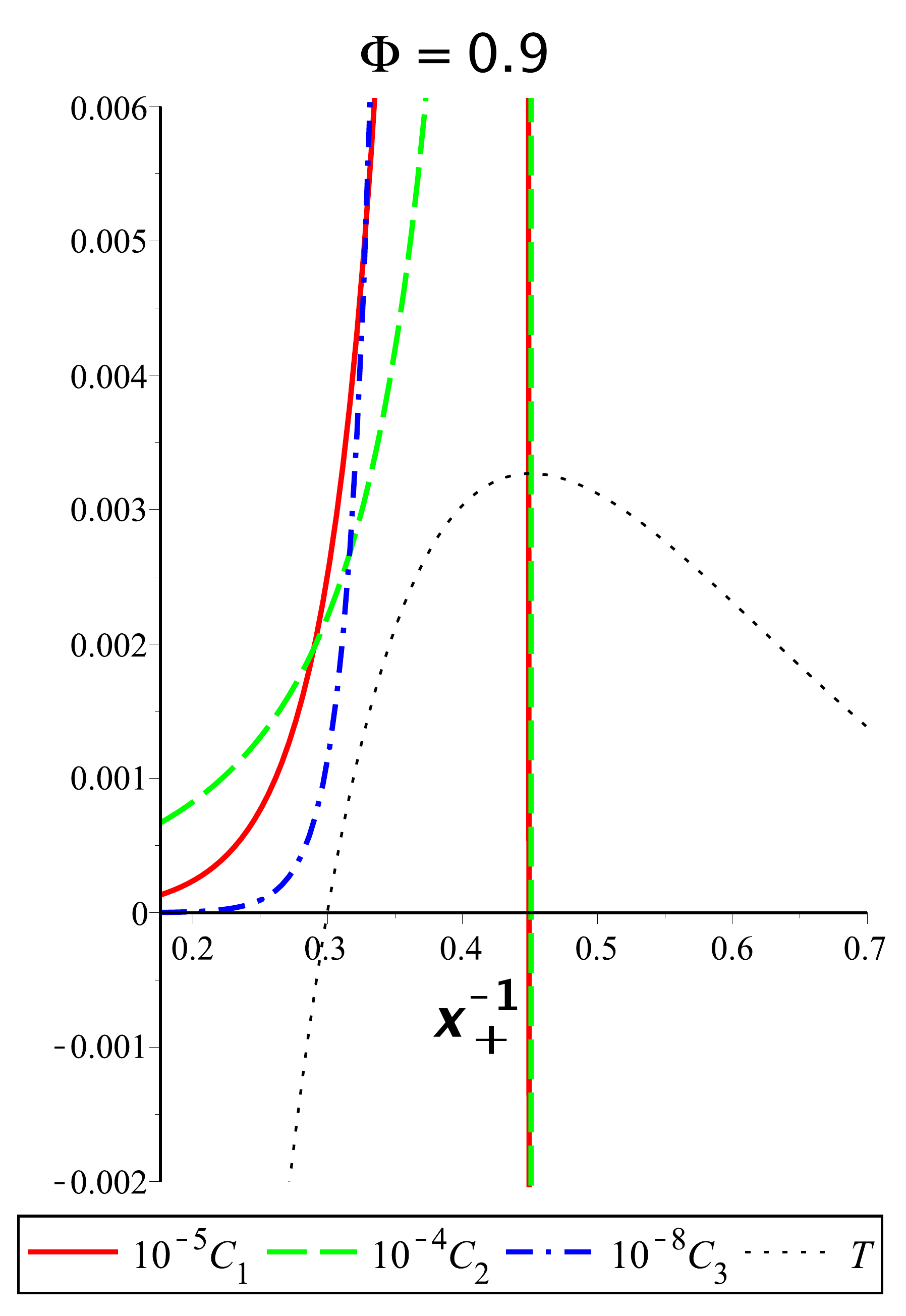}
	\caption{\small{Funciones repuesta en t'erminos de las segundas derivadas del potencial termodin'amico, para $\alpha=10$. Tres diferentes valores del potencial conjugado $\Phi$ se han mostrado. La positividad de $C_1$ (en rojo) corresponde a $C_\Phi>0$, mientras que la positividad de $C_3$ (azul) corresponde a $\epsilon_S>0$. Si ambas son positivas, se sigue que $C_2>0$ (en verde).}}
	\label{resp10}
\end{figure}

En la Fig. \ref{resp10}, se muestra que el criterio para la estabilidad se cumple para un conjunto de configuraciones de agujeros negros. Concretamente, hemos graficado las siguientes cantidades (amplificadas por un factor constante, por conveniencia)
\begin{align}
C_1
&:=-\(\frac{\pa^2\mathcal{G}}{\pa T^2}\)_\Phi ,\qquad
C_2
:=-\(\frac{\pa^2\mathcal{G}}{\pa\Phi^2}\)_T
\label{cc2}\\
C_3
&:=-\(\frac{\pa^2\mathcal{G}}{\pa T^2}\)_\Phi
\left\{\(\frac{\pa^2\mathcal{G}}{\pa T^2}\)_\Phi
\(\frac{\pa^2\mathcal{G}}{\pa\Phi^2}\)_T
-\[\(\frac{\pa}{\pa\Phi}\)_T
\(\frac{\pa\mathcal{G}}{\pa T}\)_\Phi\]^2\right\}
\label{cc3}
\end{align}
Como se coment'o, $C_1>0$ y $C_3>0$ son suficientes para determinar una estabildiad completa, sin embargo, por completitud, hemos tambi'en graficado $C_2$ para mostrar expl'icitamente que todo es consistente.

\begin{figure}[h]
	\centering
	\includegraphics[width=13 cm]{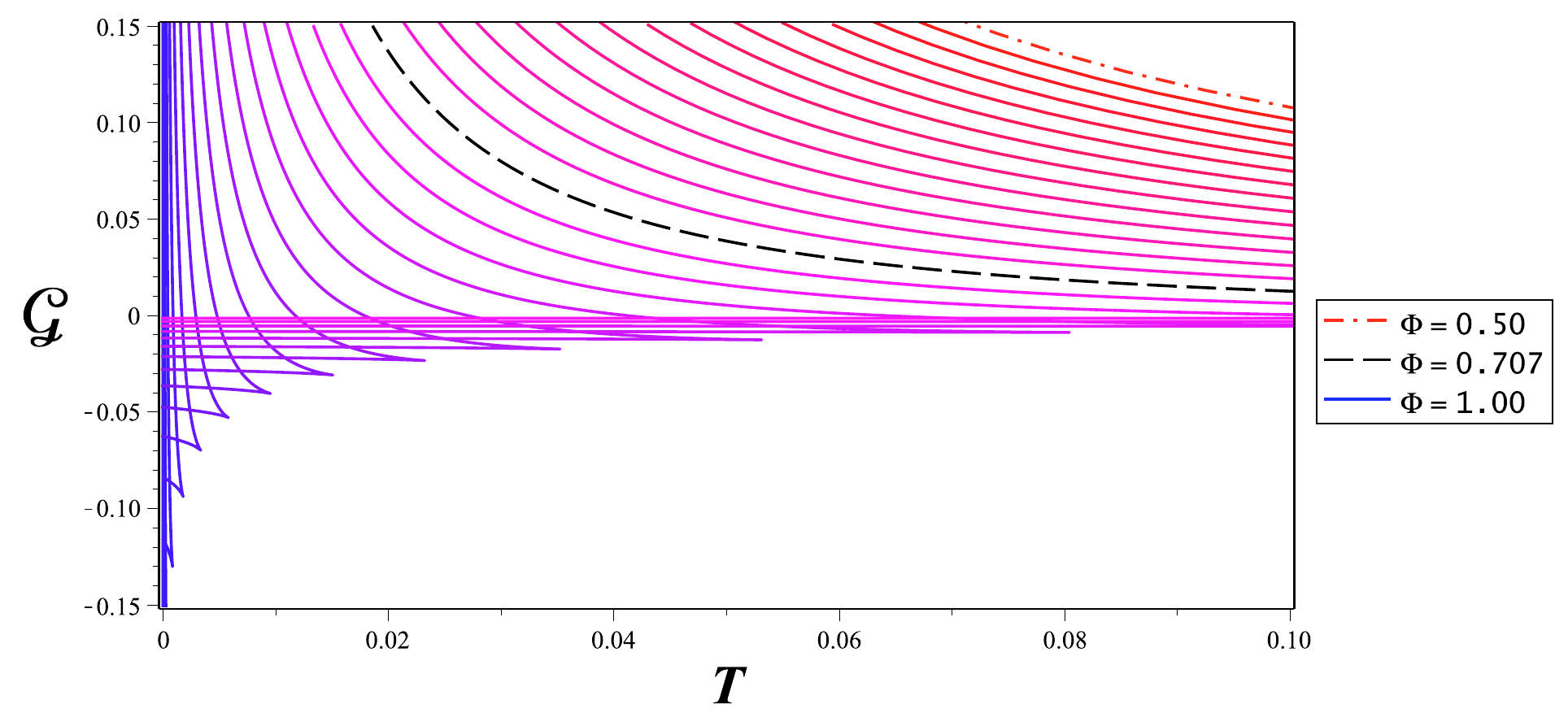}
	\caption{\small{Potencial termodin'amico $\mathcal{G}$ vs $T$ para la rama positiva, para $\alpha=10$. Para $\Phi>1/\sqrt{2}$, el potencial desarrolla un sector con concavidad negativa, esto es, $C_\Phi>0$.}}
	\label{GT}
\end{figure}

En la Fig. \ref{resp10}, observamos que, para $\Phi>{1}/{\sqrt{2}}$ (segundo y tercer gr'aficos), tanto $C_1$, $C_2$ como $C_3$ desarrollan una divergencia y resultan simult'aneamente positivas dentro de una regi'on f'isica que comienza en $T=0$ y se extiende hasta $T=T_{max}$ (donde se localizan estas divergencias). Esta regi'on nueva (que no existe en el agujero de RN) est'a caracterizada por $\epsilon_S>0$ y $C_\Phi>0$. En la Fig. \ref{GT}, m'as abajo, se puede observar que estos agujeros negros estables aparecen siempre que $\Phi>1/\sqrt{2}$ y est'an caracterizadas tambi'en por $\mathcal{G}<0$ y por $\left(\pa^2\mathcal{G}/\pa{T}^2\right)_\Phi<0$, como es esperado, puesto que $C_\Phi>0$. Quisi'eramos enfatizar que, en el caso del agujero negro de RN, no hay configuraciones f'isicas con capacidad cal'orica positiva para un $\Phi$ fijo.

Una discusi'on detallada sobre la interpretaci'on f'isica de etos reultados, y una comparaci'on con el agujero negro estable en AdS ser'a presentada en la secci'on \ref{disc}.

\subsubsection{\textit{Ensamble can'onico}}
\label{cpb}

\begin{figure}[h]
	\centering
	\includegraphics[width=5.6 cm]{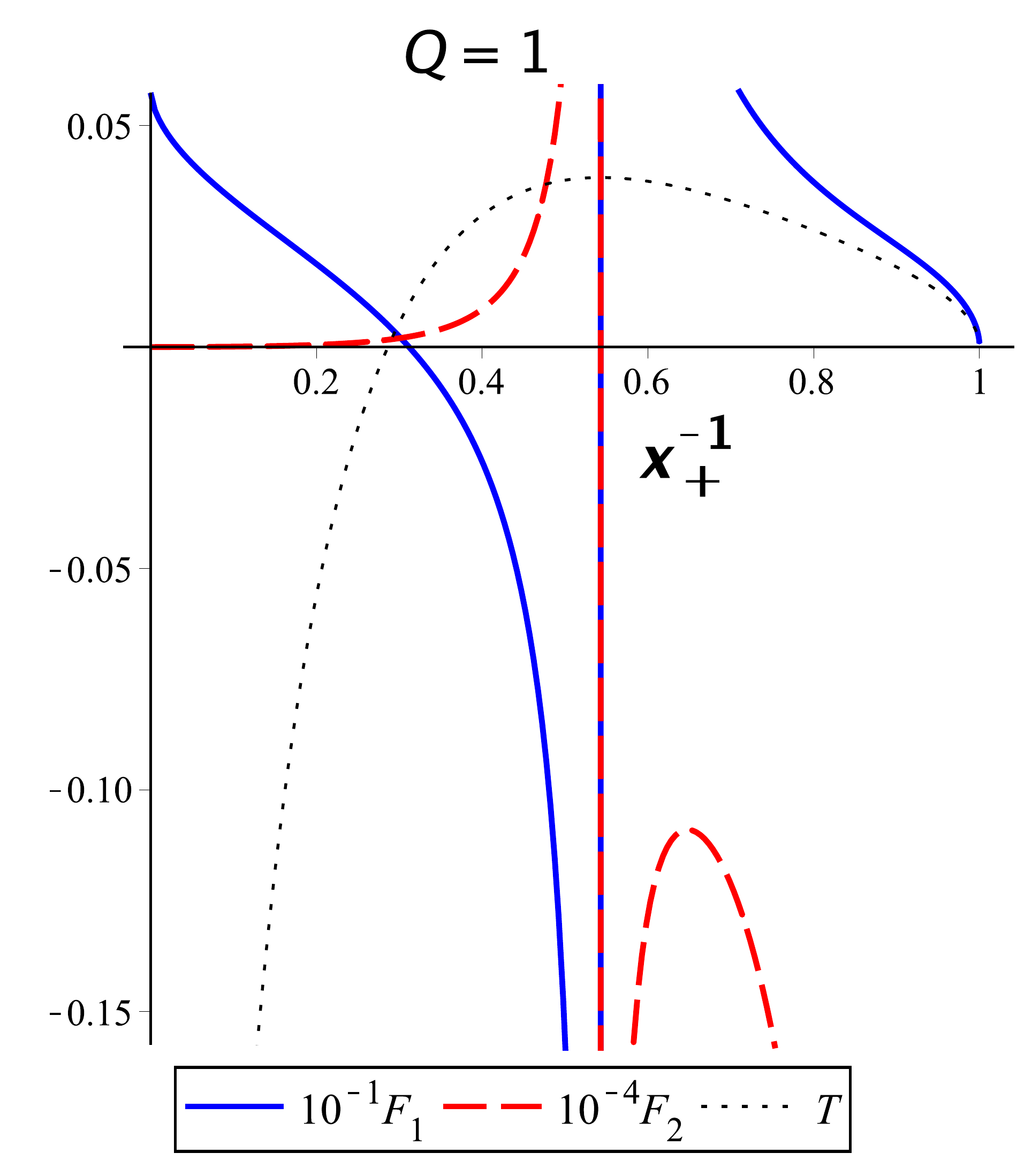} \qquad
	\includegraphics[width=5.6 cm]{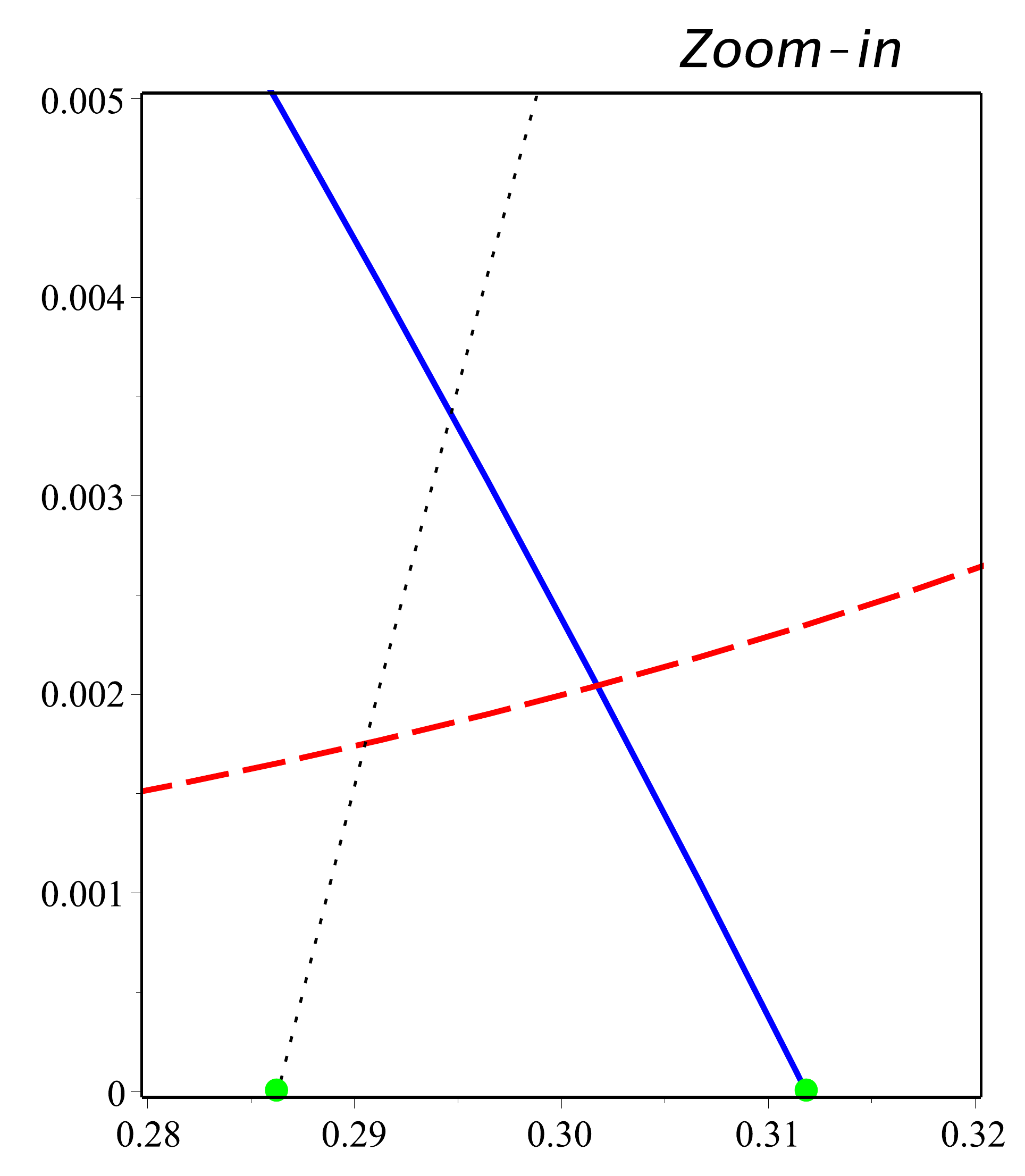}
	\caption{\small{Segundas derivadas del potencial termodin'amico, dadas por $F_1$ y $F_2$, para $\alpha=10$ y $Q=1$. Existe una regi'on donde ambas son positivas, entre $T=0$ y donde $F_1$ desarrolla un cero ($\epsilon_T=0$). En el gr'afico a la derecha se ha hecho un zoom sobre dicha regi'on y se ha marcado el intervalo (en el eje $x_+$) de estabilidad, mediante puntos verdes.}}
	\label{resp4}
\end{figure}

\begin{figure}[H]
	\centering
	\includegraphics[width=5.6 cm]{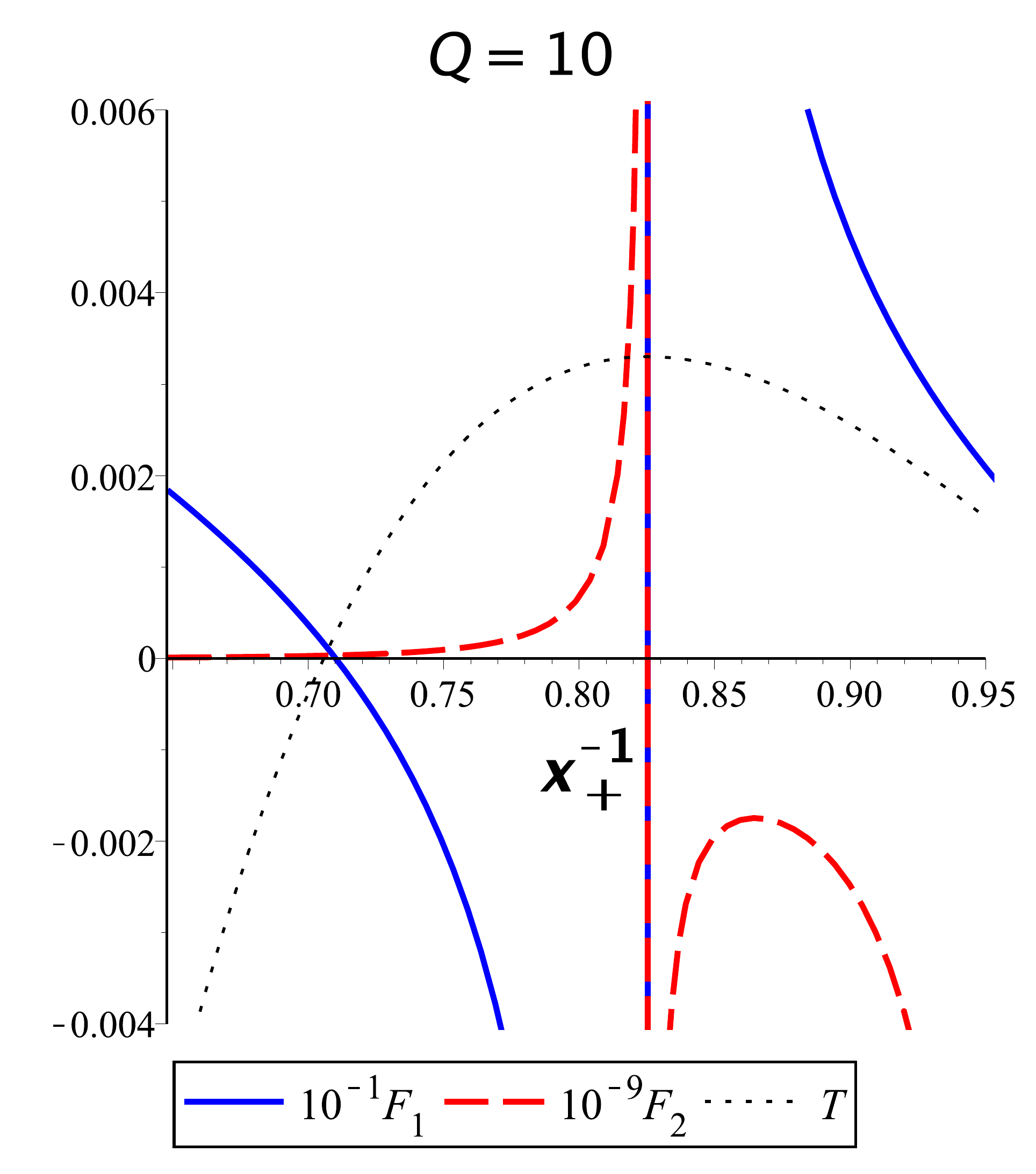} \qquad
	\includegraphics[width=5.6 cm]{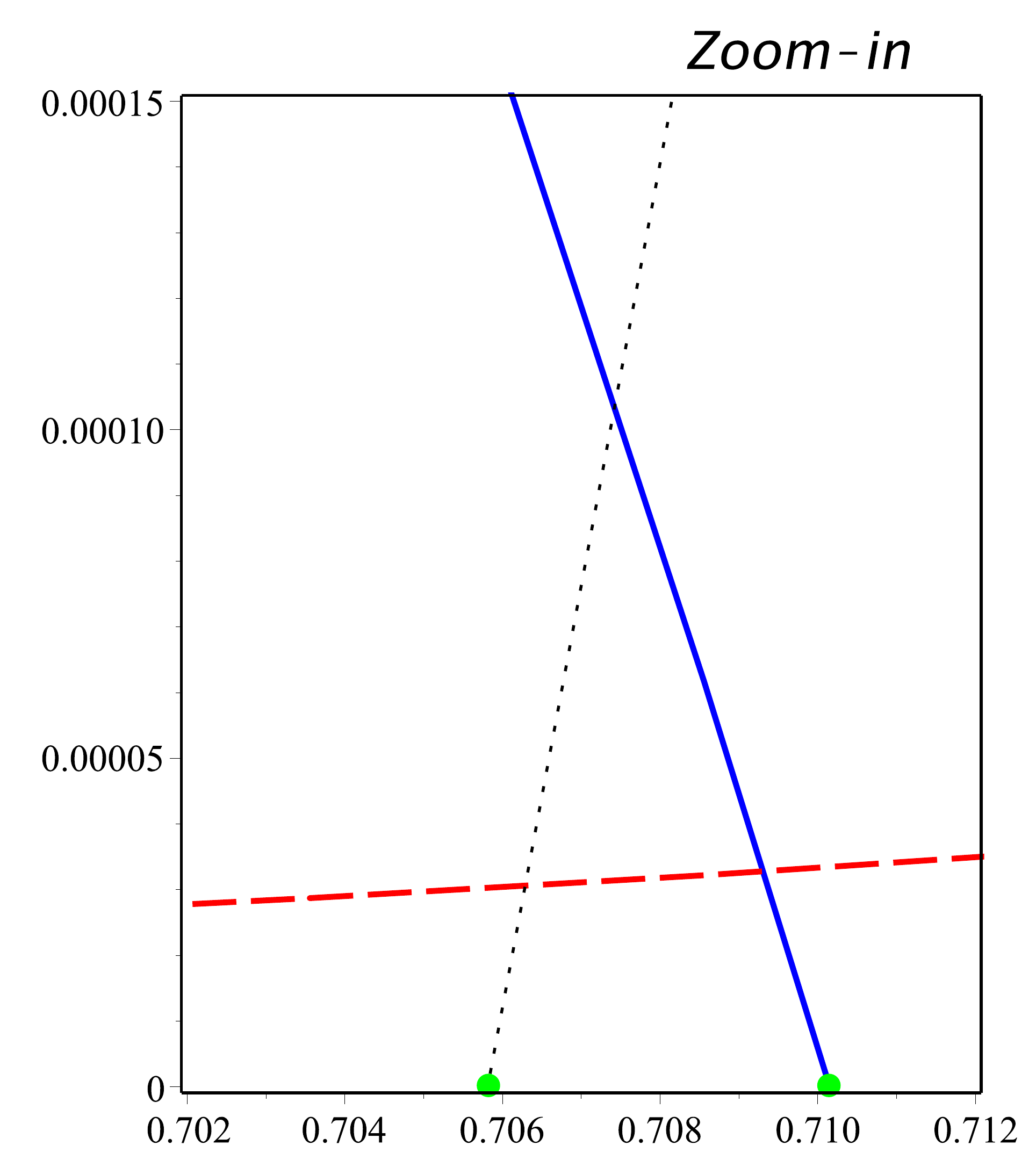}
	\caption{\small{Funciones respuestas, para $\alpha=10$ y $Q=10$. Existe la misma regi'on de estabilidad mostrada en la figura previa.}}
	\label{resp44}
\end{figure}

\begin{figure}[H]
	\centering
	\includegraphics[width=13 cm]{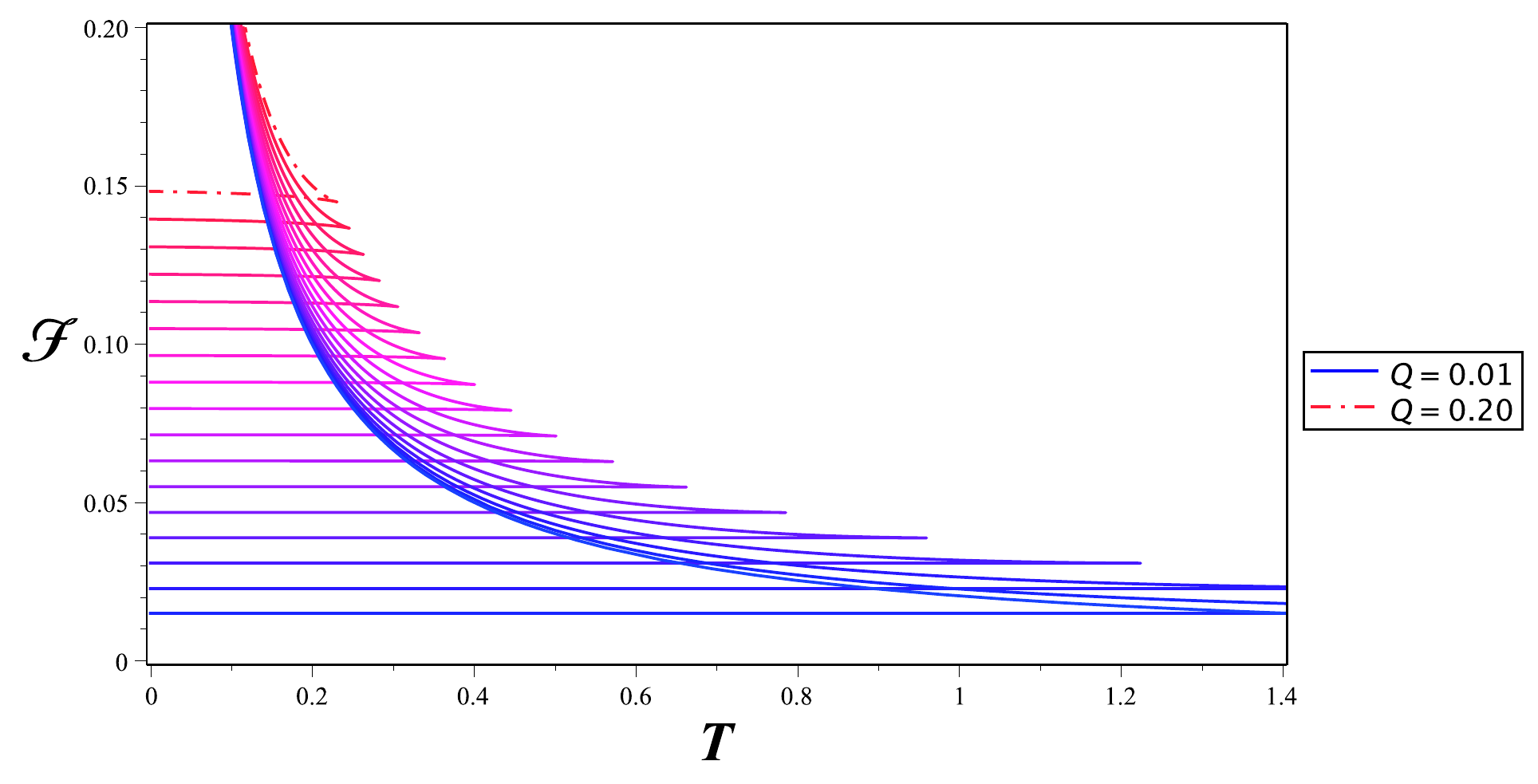}
	\caption{\small{Potencial termodin'amico $\mathcal{F}$ vs $T$ para $\alpha=10$. El sector con concavidad negativa existe para cualquier valor de $Q$}}
	\label{FT}
\end{figure}

Ahora presentamos un an'alisis similar para el ensamble can'onico. Para hacer expl'icita la dependencia de las cantidades termodin'amicas relevantes en $Q$, usaremos la ecuaci'on $q=-\eta Q$ para eliminar $q$ de la ecuaci'on del horizonte, $f(x_+)=0$, y entonces resolvemos la ra'iz positiva $\eta=\eta(x_+,Q)$.
%
%
Esto permite escribir las siguientes expresiones param'etricas:
\begin{align}
M&=\frac{12\eta^4Q^2-\alpha}{12\eta^3} ,\quad 
\Phi=\frac{\eta\,(x_{+}-1)Q}{x_+},\quad
S=\frac{\pi x_+}{\eta^2(x_{+}-1)^2}, \\
T&=\frac{\(x_{+}-1\)^2}{8\pi\eta x_+}
\(-\alpha-2\eta^2\,\frac{x_{+}+1}{x_{+}-1}
+4\eta^4Q^2\,\frac{x_{+}+2}{x_+}\)
\end{align}
Consecuentemente, el correspondiente potencial termodin'amico puede ser expresado param'etricamente en la siguiente forma compacta:
\begin{equation}
\mathcal{F}(x_+,Q)=
\frac{\alpha}{24\eta^3}
+Q^2\eta\,\frac{x_{+}-2}{2x_{+}}
+\frac{1}{4\eta}\,\frac{x_{+}+1}{x_{+}-1}
\end{equation}
Para investigar la estabilidad termodin'amica local, debemos verificar las desigualdades $\epsilon_T>0$ y $C_Q>0$, las cuales, en t'erminos de segundas derivadas de $\mathcal{F}$, son equivalentes a las condiciones
\begin{equation}
F_1:=(\partial^2\mathcal{F}/\partial Q^2)_T>0,
\qquad
F_2:=-(\partial^2\mathcal{F}/\pa T^2)_Q>0
\end{equation}
respectivamente.\footnote{La tercera condici'on para estabilidad bajo fluctuaciones mixtas, $\epsilon_TC_Q>0$, sigue de $\epsilon_T>0$ y $C_Q>0$ y entonces no necesita ser impuesta como una condici'on independiente.} Los gr'aficos en el lado derecho de las Figs. \ref{resp4} y \ref{resp44} muetran un zoom hecho en el espacio de parametros para los cuales los agujeros negros son estables. La estabilidiad termodin'amica ocurre en el sector $\epsilon_T>0$, localizado all'i donde $\Phi>\frac{1}{\sqrt{2}}$ (vea la Fig. \ref{state_hairy1}), y, por lo tanto, es consistentes con nuestros resultados en el ensamble gran can'onico.
En la Fig \ref{FT}, se ha graficado el potencial termodin'amico como funci'on de la temperatura, indicando que hay agujeros negros con $C_Q>0$ para cualquier $Q$. 
Los agujeros negros estables exiten siempre que $\epsilon_T>0$ y $C_Q>0$, pero entonces no todos aquellos con $C_Q>0$ en la Fig. \ref{FT} son termodin'amicamente estables, sino una relativamente peque\~na parte (vea la Fig. \ref{resp44}).

\subsection{La rama negativa}

Ahora investigamos la estabildiad en la rama negativa, esto es, cuando $\phi<0$ o, equivalentemente, $x\in (0,1)$, para la teor'ia con $\gamma=1$ y $\alpha>0$, presentada en la secci'on (\ref{sec:sols}). Como motraremos, aunque sin presentar detalles (puesto que los pasos son b'asicamente los mismos que aquellos mostrados para la rama positiva), no existen agujeros negros estables en este caso.

\subsubsection{Ensamble gran can'onico}
\label{gcnb}

\begin{figure}[H]
	\centering
	\includegraphics[width=7.2 cm]{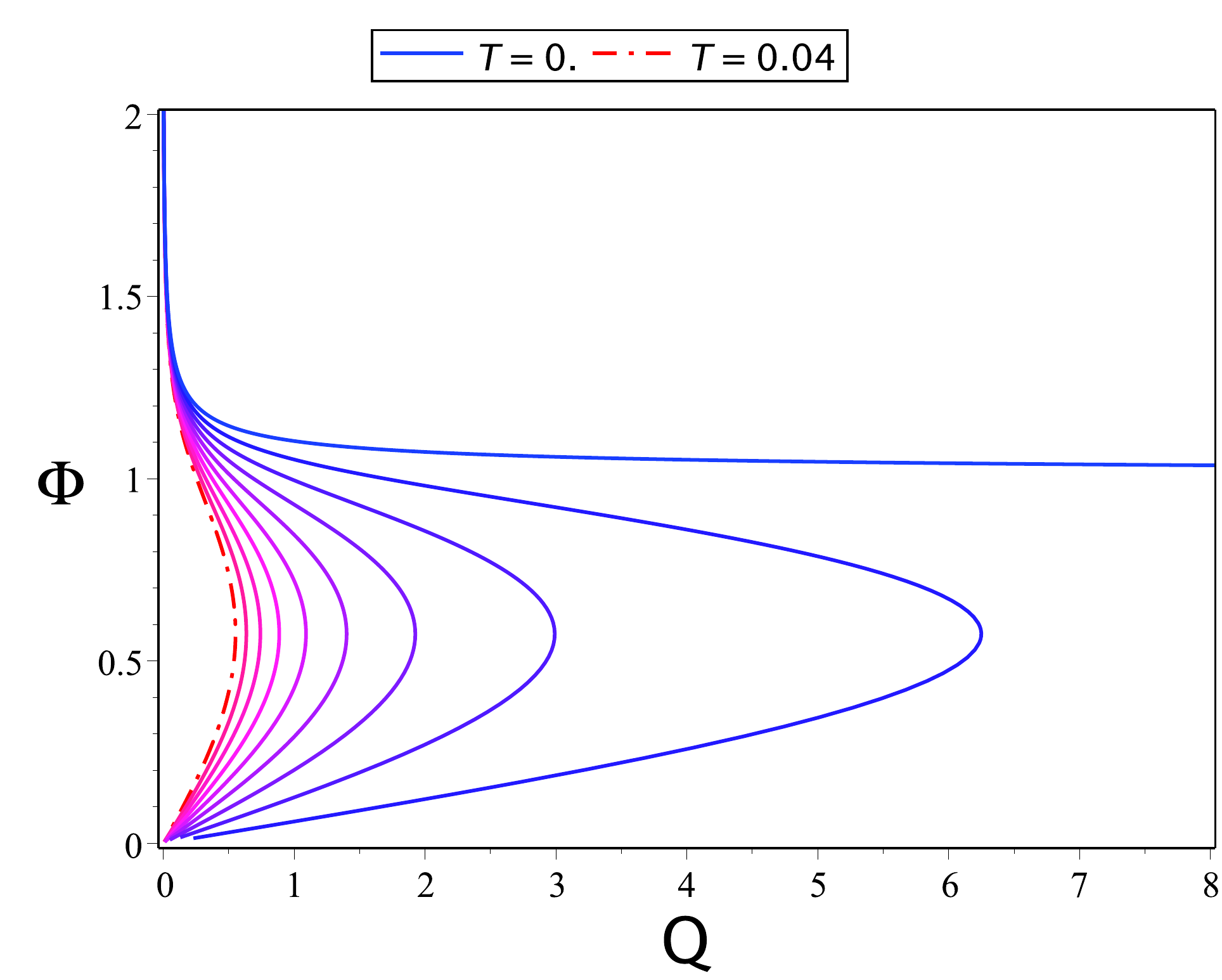}\quad
	\includegraphics[width=7.2 cm]{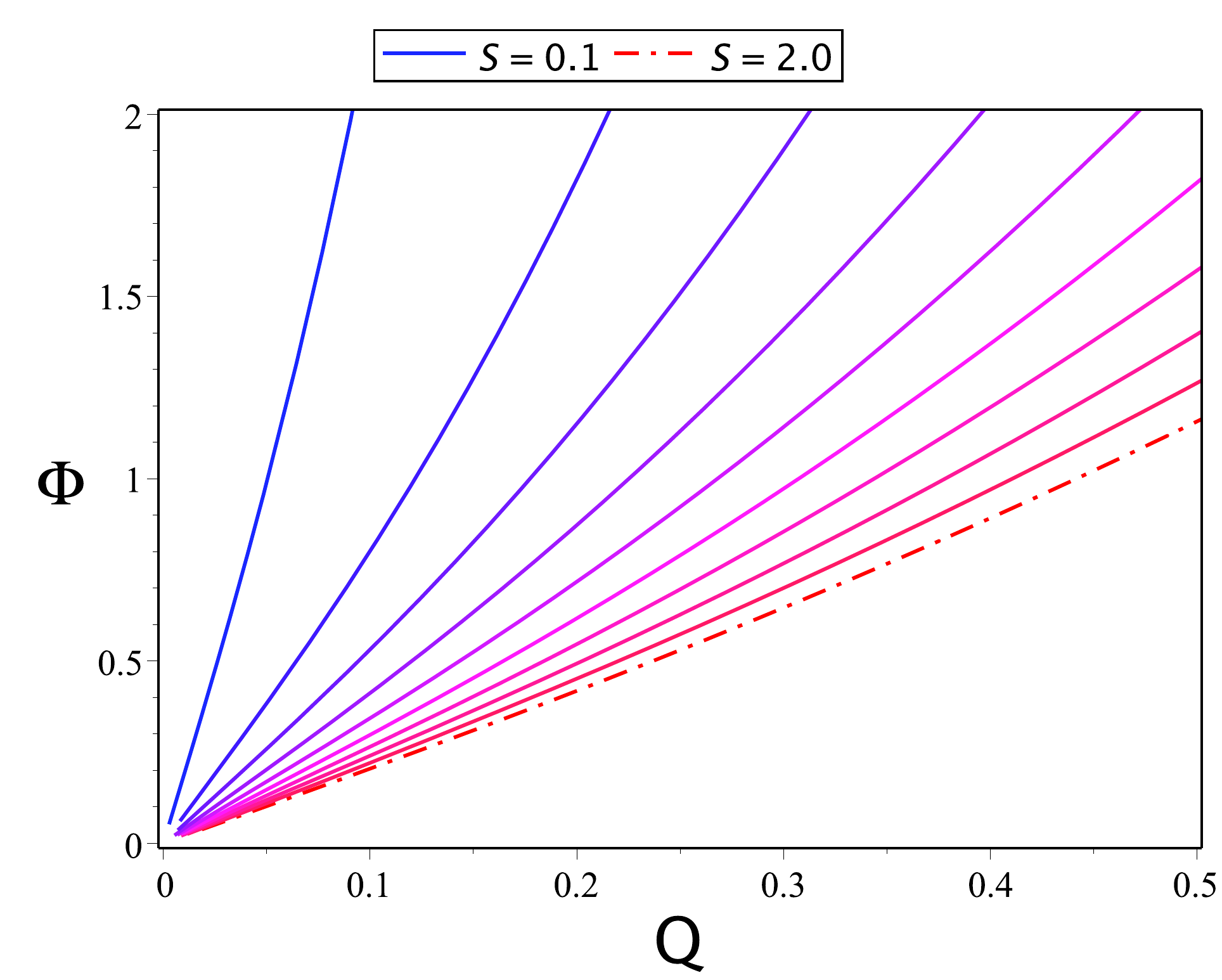}
	\caption{\small{Izquierda: Ecuaci'on de estado en la rama negativa, para $\alpha=10$. Derecha: $\Phi$ vs $Q$ para entrop'ia fija.}}
	\label{state1}
\end{figure}

En la Fig. \ref{state1} se muestra $\Phi$ vs $Q$, a temperatura (ecuaci'on de estado) y entrop'ia fija, respectively. Comparado con la ecuaci'on de estado en la rama positiva, la parte superior del gr'afico no contiene una regi'on donde $\epsilon_T>0$. Esto representa una importante diferencia con el caso estudiado previamente.

\begin{figure}[h]
	\centering
	\includegraphics[width=5.3 cm]{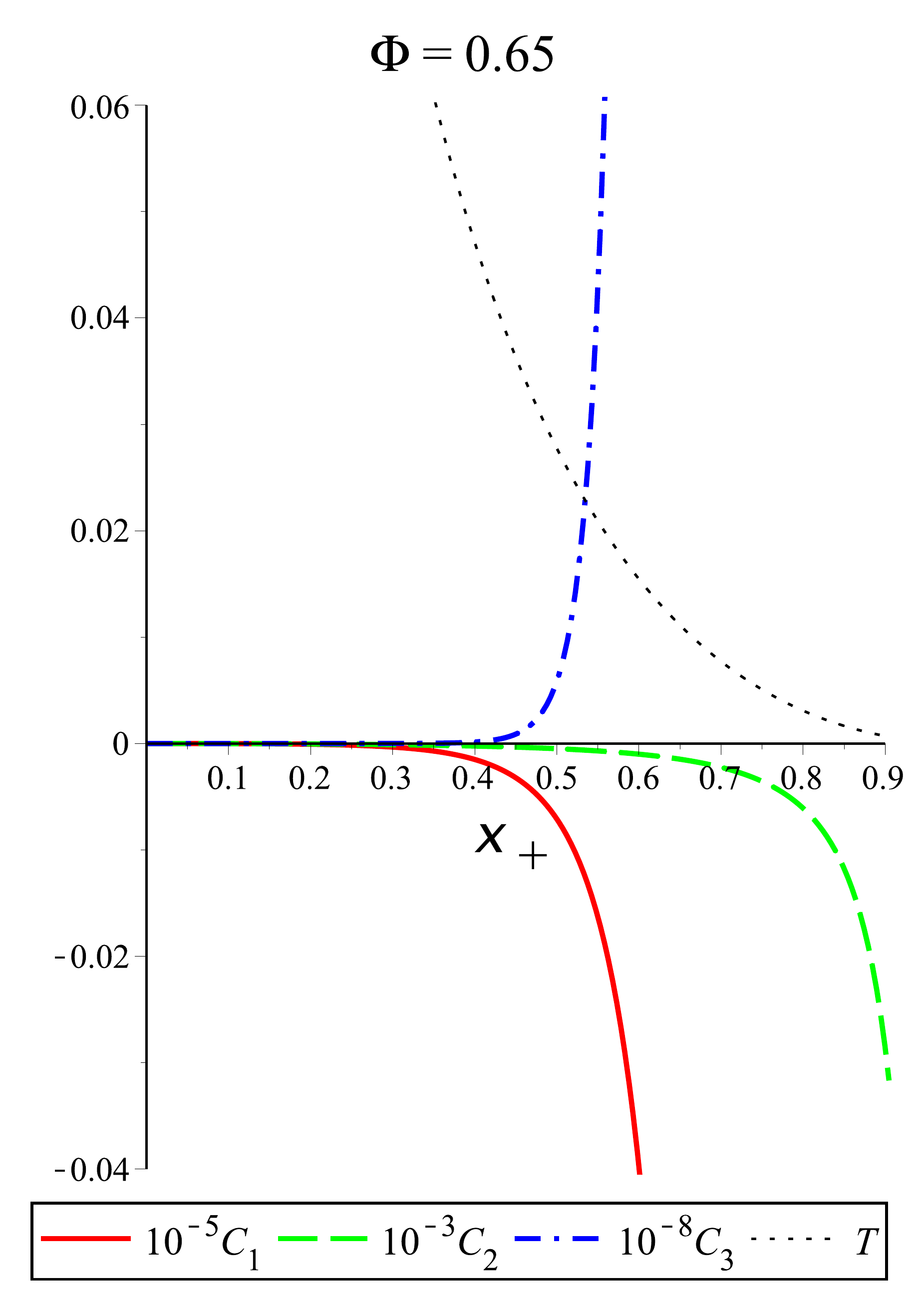}
	\includegraphics[width=5.3 cm]{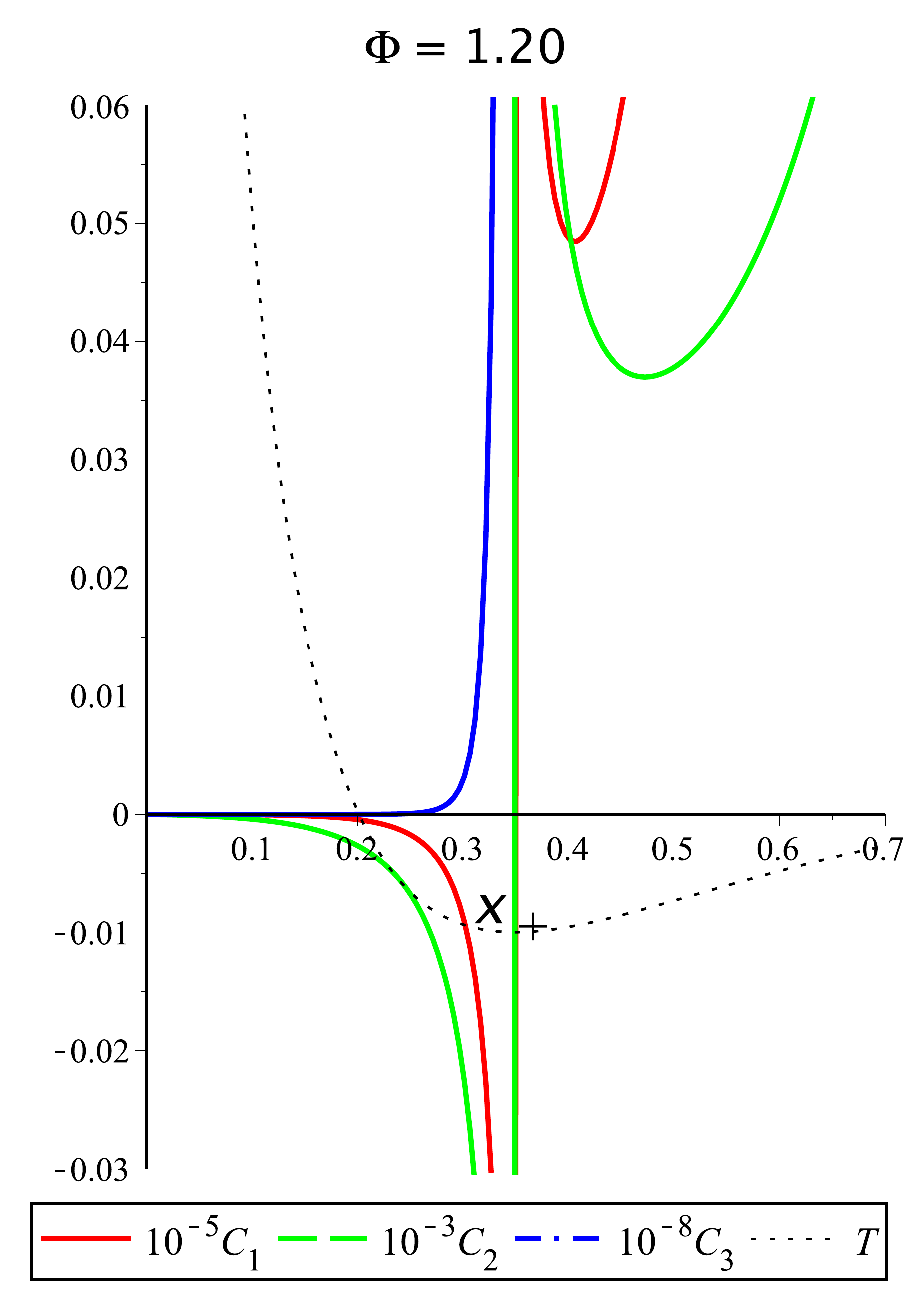}
	\caption{\small{Funciones respuesta en t'erminos de la segundas derivadas de $\mathcal{G}$, para $\alpha=10$. La positividad simult'anea de $C_1$ (en rojo) y $C_3$ (en azul) indica estabilidad.}}
	\label{resp1}
\end{figure}

La ecuaci'on de estado es bastante similar con el agujero negro de RN, en el sentido que s'olo contiene una regi'on con $\epsilon_T>0$ correspondiente a la parte baja del gr'afico. Por otra parte, uno expl'icitamente mostrar que, como en el caso previo, $\epsilon_S>0$. Resulta, sin embargo, que las funciones respuesta relevantes no comparten valores positivos en ninguna regi'on f'isica, esto es, $C_\Phi<0$, como se muestra en la Fig. \ref{resp1} --- hemos usado las mismas convenciones que en caso del ensamble gran can'onico en la rama positiva para las definiciones (\ref{cc2}) y (\ref{cc3}).
A pesar de que $\epsilon_S>0$ (en azul), como puede verse en Fig. \ref{state1}, no existe una regi'on f'isica donde $C_\Phi>0$. Por lo tanto, no existen agujeros negros estables en la rama negativa.
%
%
\begin{figure}[H]
	\centering
	\includegraphics[width=12 cm]{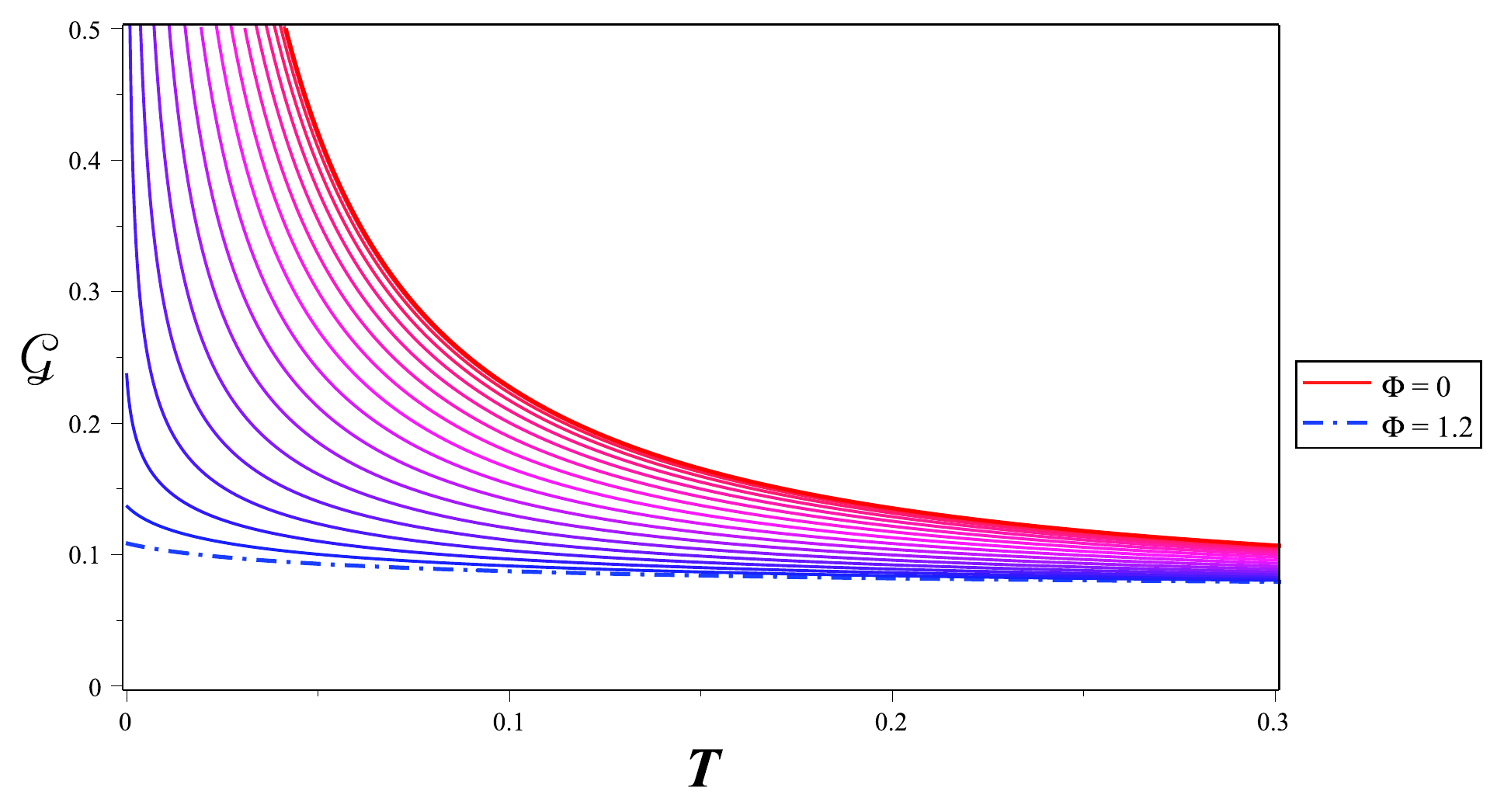}
	\caption{\small{La concavidad del potencial termodin'amico es definida positiva y entonces $C_\Phi<0$. Se tom'o $\alpha=10$.}}
	\label{gtcan}
\end{figure}

\subsubsection{\textit{Ensamble can'onico}}
\label{cnb}

\begin{figure}[h]
	\centering
	\includegraphics[width=6 cm]{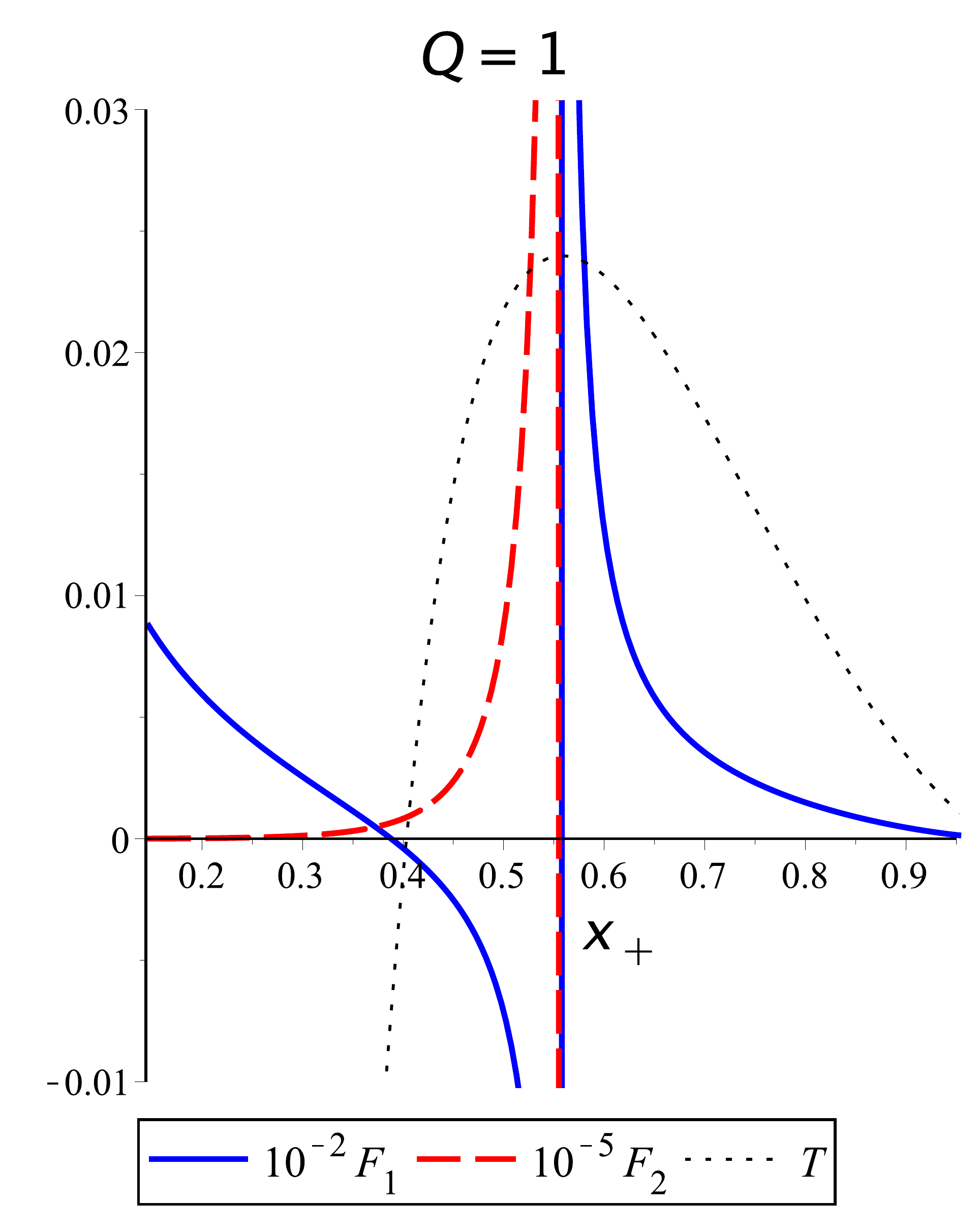} \quad
	\includegraphics[width=6 cm]{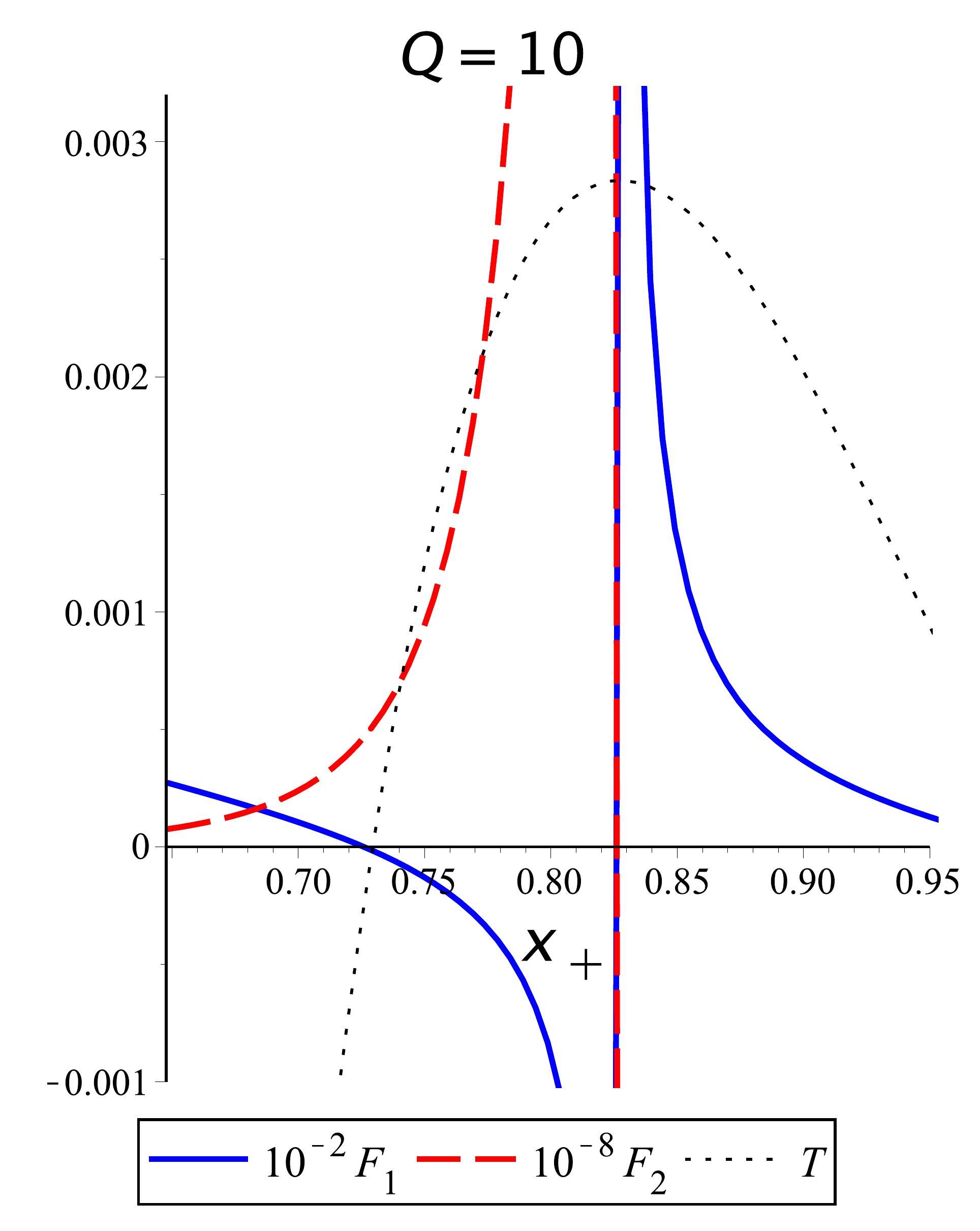}
	\caption{\small{Segundas derivadas del potencial termodin'amico, para $\alpha=10$.}}
	\label{resp3}
\end{figure}

Las funciones respuestas est'an graficadas en la Fig. \ref{resp3}, donde se observa que, en sinton'ia con los resultados en el ensamble gran can'onico, no existen una regi'on donde $\epsilon_T>0$ y $C_Q>0$, simult'aneamente. Adem'as, el producto $C_Q\epsilon_T$ es negativo para $T>0$.

No existen configuraciones estables en el ensamble can'onico para la rama negativa. El cero de $F_1$ est'a localizado en la regi'on $T<0$ o, en otras palabras, el agujero negro extremo es instable (vea Fig. \ref{state1}) y, por lo tanto, $\epsilon_T$ no toma valores positivos en regiones donde $C_Q>0$.


\newpage
\section{Discusi'on}
\label{disc}
Los campos escalares juegan un rol central en cosmolog'ia y f'isica de part'iculas y aparecen naturalmente en teor'ia de unificaci'on en f'isica de altas energ'ias. Es entonces importante entender propiedades generales de las teor'ias de gravedad acopladas a escalares (y otros campos de materia), particularmente el rol que juegan en la f'isica de agujeros negros.\footnote{Algunas aplicaciones recientes interesantes pueden ser encontradas en \cite{Hirschmann:2017psw,Jai-akson:2017ldo,Cardenas:2017chu,McCarthy:2018zze,Brihaye:2018woc, Bzowski:2018aiq}.} En este trabajo, hemos considerado propiedades termodin'amicas de una familia de agujeros negros con pelo asin'oticamente planos, con el objetivo de arrojar luz sobre su estabilidad termodin'amica. 

En nuestra investigaci'on, hemos sido directamente motivados por los resultado de \cite{Nucamendi:1995ex}, donde fue conjeturada la existencia de tales soluciones de agujeros negros en teor'ias con un potencial no trivial que se anula en el borde\footnote{El potencial escalar diverge en la singularidad, $x=\{0,\infty\}$, pero esto no es materia de atenci'on puesto que la singularidad est'a protegida por el horizonte. En la singularidad uno espera que los efectos de la gravedad cu'antica se vuelven relevantes y, entonces, la teor'ia que estamos considerando debe ser interpretada como una teor'ia efectiva.} y en \cite{Anabalon:2013qua}, donde soluciones exactas regulares de agujeros negros con pelo fueron obtenidas.

Debido a su 'intima conexi'on con la funci'on de partici'on, el formalismo (Euclidiano) de la integral de caminos de la gravedad cu'antica es ampliamente usado cuando se estudia la termodin'amica de agujero negro. Hemos presentado un an'alisis completo usando los t'erminos de borde requeridos en la acci'on de la relatividad general y hemos probado que, para algunos valores de los par'ametros, estos agujeros negros son termodin'amicamente estables en ambos, en los ensambles can'onico y gran can'onico. Este resultado podr'ia venir como sorpresa ya que, generalmente, en espaciotiempo asint'oticamente planos, para varias dimensiones, se sabe que los agujeros negros no son termodin'amicamente estables\cite{Hawking:1976de,Monteiro:2009tc, Dias:2010eu, Astefanesei:2010bm}.\footnote{Sin embargo, existen ejemplos de agujeros negros termodin'amicamente estables en teor'ias que incluyen t'erminos de derivadas de 'orden superior\cite{Myers:1988ze, Bueno:2016lrh, Bueno:2017qce}.} Es posible construir agujeros negros termodin'amicamente estables introduciendo una constante cosmol'ogica negativa $\Lambda$ y considerando agujeros negros asint'oticamente AdS, o poni'endolos dentro de una cavidad finita (como hemos comentado). Sin embargo, cuando exiten campos escalares en la teor'ia, su auto-interacci'on es el ingrediente clave para la estabilidad terodin'amica.

Se sabe que, cuando el potencial dilat'onico se anula, uno puede tambi'en variar el valor asint'otico del campo escalar, $\phi_\infty$. En este caso, se consider'o inicialmente que la primera ley de la termodin'amica deber'ia ser modificada mediante la adici'on  de una contribuci'on debido a la carga escalar \cite{Gibbons:1996af} que puede ser expl'icitamente verificada para soluciones exactas de agujero negro con pelo \cite{Gibbons:1987ps,Strominger:1991,Kallosh:1992ii}. Como mostramos en los cap'itulos anteriores, la consideraci'on del principio variacional correcto implica que la energ'ia cuasilocal no coincide con la masa $ADM$ (obtenida de la expansi'on de la componente $g_{tt}$ de la m'etrica. 
Usando una definici'on correcta para la energ'ia (cuasilocal) gravitacional, mostramos que la primera ley preserva su forma usual sin incluir ninguna contribuci'on extra viniendo de la variaci'on del dilat'on
Debemos contrastar este caso con los agujeros negros con pelo en teor'ias con un potencial dilat'onico. Particularmente, para obtener un espaciotiempo asint'oticamente plano, deber'ia ser impuesta una restricci'on importante obre el potencial, que se anule en el borde. Esto es, el valor asint'otico del dilat'on deber'ia estar fijo, de otra manera su variaci'on cambiar'ia el comportamiento asint'otico del espaciotiempo. Por lo tanto, el desaf'io de la aparici'on de la carga escalar no aparece en los casos analizados en este cap'itulo.

En el trabajo de Brown y York \cite{Brown:1992br}, fue mostrado que el tensor de estr'es cuasilocal es covariantemente conservado solamente si el comportamiento asint'otico de los campos de materia es tal que decaen lo suficientemente r'apido, lo cual es tambi'en nuestro caso: en el borde cuando $x=1$, el campo escalar es $\phi\rightarrow 0$, lo cual implica que el potencial se anula. Por ejemplo, el agujero negro en la teor'ia con $\gamma=\sqrt{3}$ tiene el siguiente tensor de estr'es cuasilocal
\begin{align}
\tau_{tt}
&=-\frac{8\alpha+3\(1-2q^{2}\)\eta^{2}}{3\eta\kappa}\,(x-1)^{2}
+\mathcal{O}\[(x-1)^{3}\] \\
\tau_{\theta\theta}
&=\frac{\tau_{\phi\phi}}{\sin^2\theta}
=\frac{\[8\alpha+3\(1-2q^{2}\)\eta^2\]^{2}-9\eta^{4}(4q^{2}-3)}{72\kappa\eta^{5}}\,(x-1)+\mathcal{O}[(x-1)^{2}] 
\end{align}
el cual es, de hecho, covariantemente conservado. Esto contraste con la situaci'on cuando la soluci'on no es regular debido a la presencia de singularidades c'onicas en el borde \cite{Astefanesei:2009mc}. 

Otra importante observaci'on es sobre el estado fundamental de la teor'ia. El m'etodo de contra'terminos, remarcablemente, provee un m'etodo de regularizaci'on que produce una definic'on intr'inseca de la acci'on sin necesidad de usar un background de referencia. Sin embargo, es importante obtener soluciones tipo solit'on (a la temperatura cero). Dejamos el an'alisis detallado de este punto para un trabajo futuro, pero, como en el caso de agujeros negros cargados en AdS \cite{Chamblin:1999tk}, uno puede, en principio, considerar la existencia de agujeros negros extremos en el ensamble can'onico (a carga $Q$ fija)
como el background de referencia. El problema que aparece para agujeros negros con pelo es que el l'imite extremo no es siempre bien definido como en el caso del agujero negro de RN. Esto est'a relacionado con el mecanismo atractor \cite{Ferrara:1995ih, Strominger:1996kf, Ferrara:1996dd} y quisi'eramos comentar ahora sobre este aspecto sutil de la teor'ia. Existen dos m'etodos diferentes para estudiar la informaci'on cerca del horizonte (near horizon data) de agujeros negros extremos, el potencial efectivo\cite{Goldstein:2005hq} y el m'etodo de la funci'on entrop'ia\cite{Sen:2005wa, Astefanesei:2006dd}. Cuando el potencial dilat'onico se anula, en teor'ias con un campo el'ectrico, el potencial efectivo no puede tener un extremo al horizonte, lo cual indica que los agujeros negros extremos no existen. Esto puede ser obtenido directamente calculando los invariantes geom'etricos en el horizonte interno y probando que algunos de ellos divergen. Sin embargo, hay un cambio dr'astico cuando el potencial dilat'onico es no trivial. Esto es, debido a la competici'on entre el potencial efectivo y el potencial dilat'onico, puede existir un l'imite extremo bien definido (en la secci'on Lorentziana) para algunos valores de los par'ametros del potencial dilat'onico. En este caso, el m'etodo del potencial efectivo deja de funcionar, pero, en cambio, uno puede usar el formalismo de la funci'on entrop'ia. Un an'alisis del caso en que estamos interesados fue hecho en \cite{Anabalon:2013qua} (vea, tambi'en, \cite{Anabalon:2013sra}) y, de hecho, existen soluciones de agujeros negros extremos regulares y el ensamble can'onico es bien definido.
 \bigskip

\begin{figure}[h]
	\centering
	\includegraphics[width=7 cm]{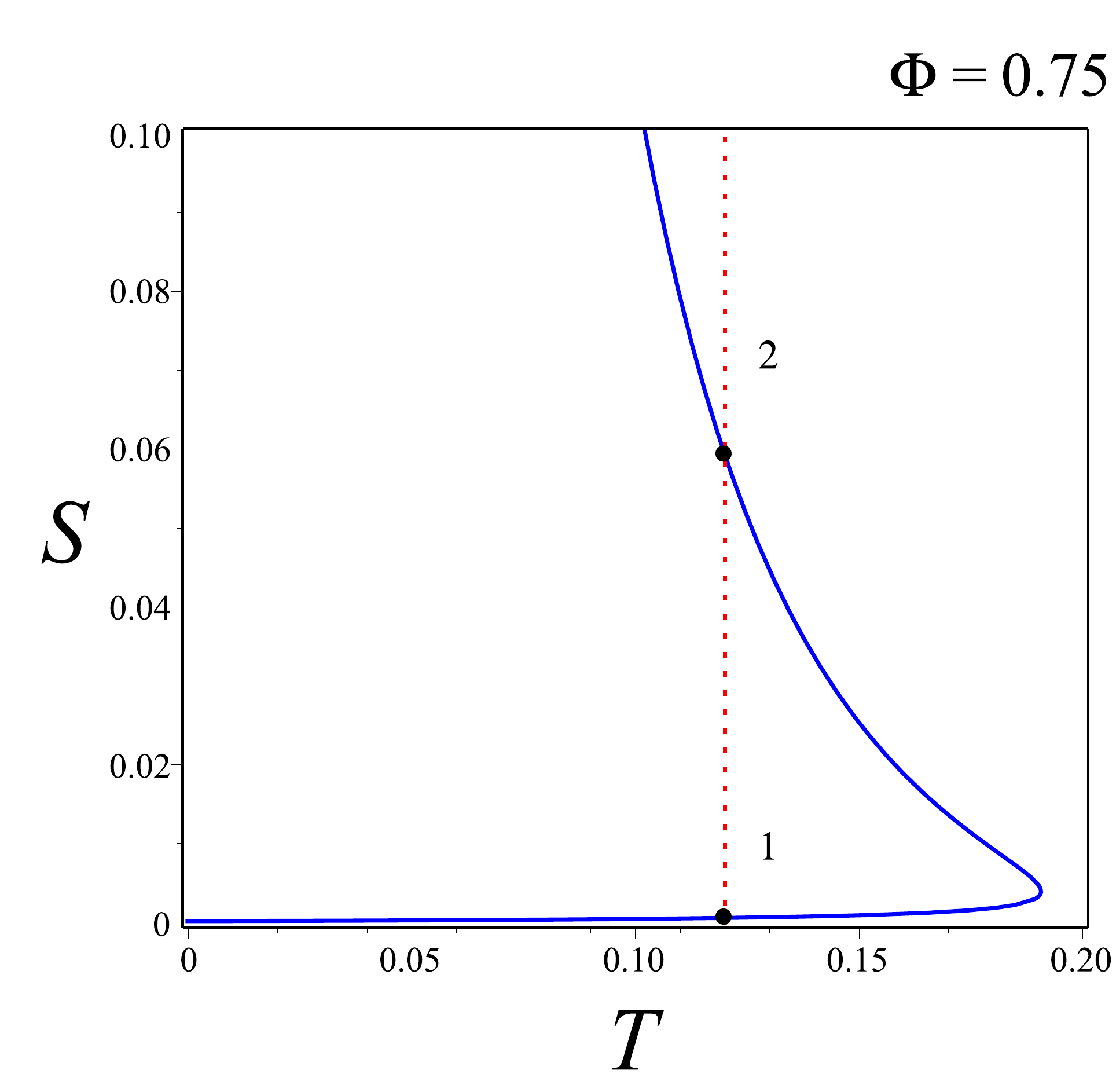}\quad
	\includegraphics[width=7 cm]{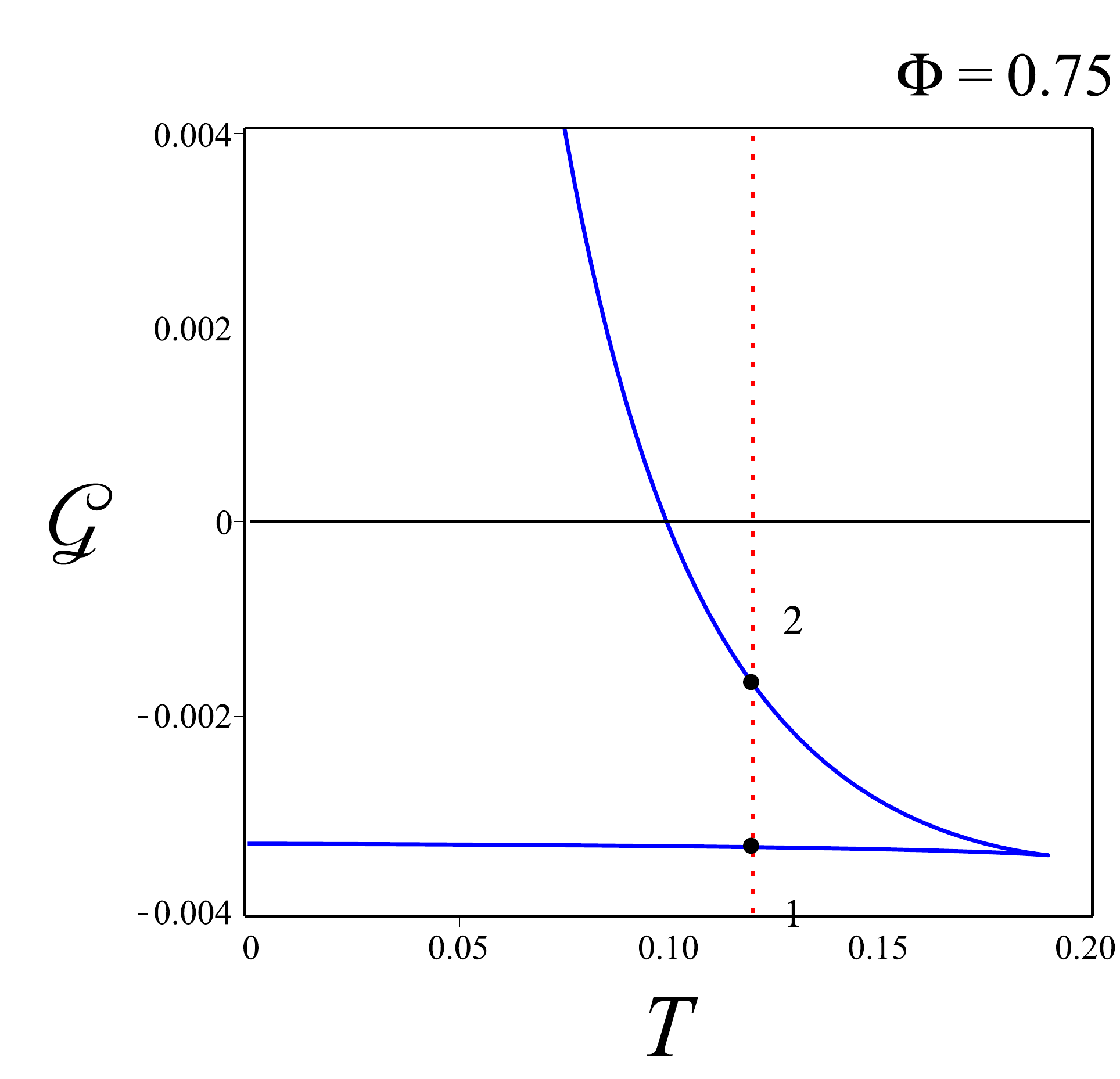}
	\caption{\small{Identificaci'on de la configuraci'on estable. Se ha fijado $\Phi=0.75$ y la l'inea roja vertical corresponde a un valor particular $T=0.06$. La configuraci'on marcada con el n'umero 1 es la estable.}}
	\label{stable1}
\end{figure}

Ahora, quisi'eramos discutir en m'as detalle nuestro principal resultado, presentados en la secci'on (\ref{main}). Comparemos los agujeros negros termodin'amicamente estables, que exiten para $\Phi>\frac{1}{\sqrt{2}}$ (el valor del potencial conjugado es menor que $1$ para los agujeros negros extremos, pero, cuando $\Phi \rightarrow 1$, el agujero negro de RN extremo es recuperado), con los agujeros negros estables en $AdS$. En la Fig. \ref{stable1}, graficamos $S$ vs $T$ y $\mathcal{G}$ vs $T$ para los agujeros negros con pelo para identificar, a una temperatura dada, cu'al configuraci'on es favorable.

\newpage
Del primer gr'afico en la Fig. \ref{stable1}, observamos que los agujeros negros estables, para los cuales $C_\Phi=T(\pa S/\pa T)_\Phi>0$, corresponden a la configuraci'on indicada con el n'umero $1$. Este agujero negro tiene menor entrop'ia que la otra configuraci'on a la misma temperatura, por lo tanto, ya que
$
S=-({\pa\mathcal{G}}/{\pa T})_\Phi
$,
este puede ser identificado en el segundo gr'afico como aquel con menor pendiente (indicado el n'umero $1$, tambi'en). Otra manera de entender esto es investigando la segunda derivada del potencial termodin'amico. Puesto que
\begin{equation}
\label{secderiv}
\(\frac{\pa S}{\pa T}\)_\Phi=-\(\frac{\pa^2\mathcal{G}}{\pa T^2}\)_\Phi
\end{equation}
los agujeros negros estables, para los cuales $(\pa S/\pa T)_\Phi>0$, deber'ian aparecer en el primer gr'afico como $({\pa^2\mathcal{G}}/{\pa T^2})_\Phi<0$, los que corresponden a la configuraci'on 1, porque tiene concavidad negativa. Con esto, la identificaci'on est'a completa.

\begin{figure}[H]
	\centering
	\includegraphics[width=7 cm]{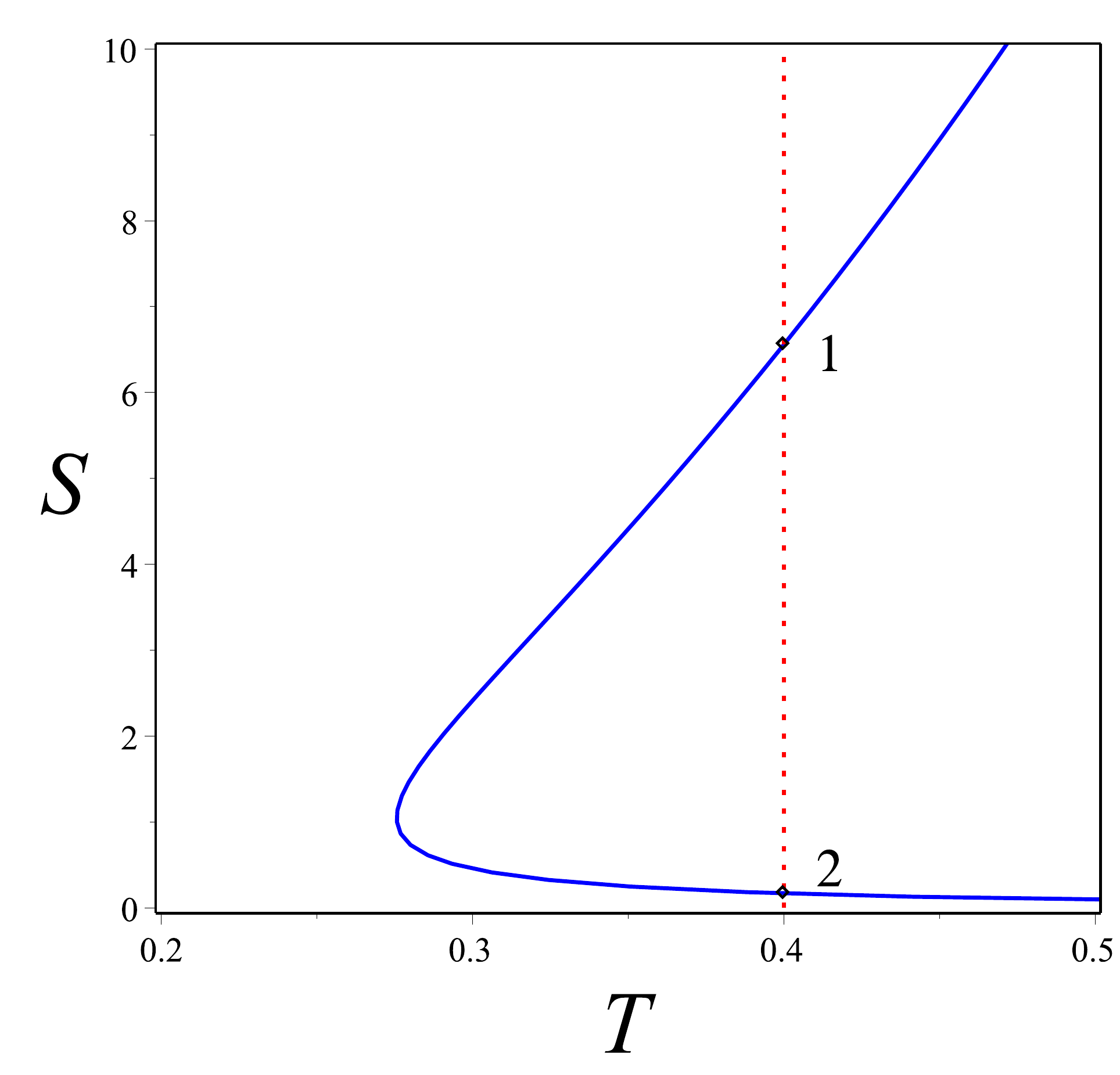} \quad
	\includegraphics[width=7 cm]{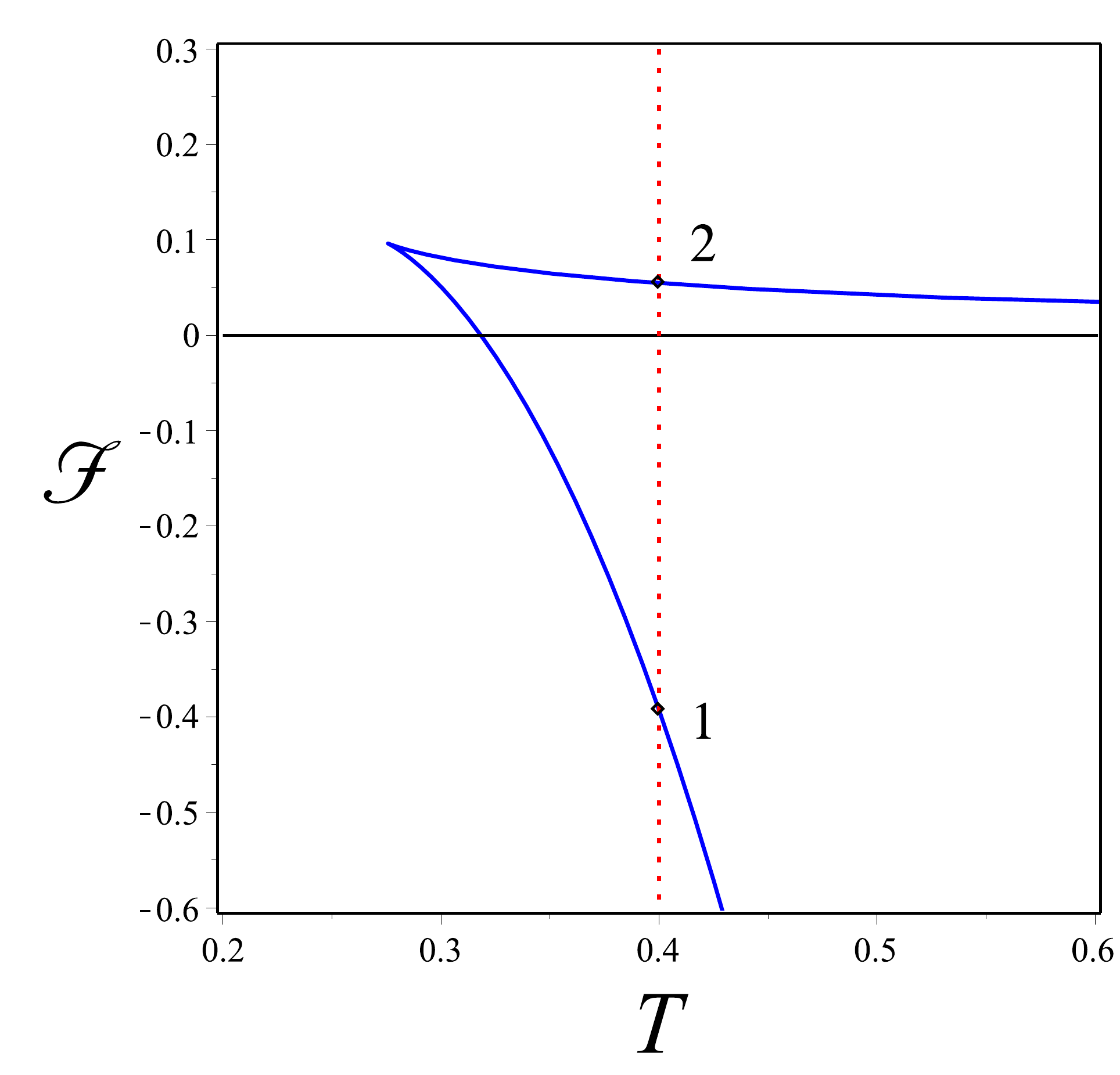}
	\caption{\small{Identificaci'on de la configuraci'on estable (n'uermo 1) para el agujero negro de Schwarzschild-AdS.}}
	\label{stable2}
\end{figure}

Ahora pondremos nuestra atenci'on en la soluci'on Schwarzschild-AdS, donde tambi'en existen dos agujeros negros a la misma temperatura, tal que podemos comparar con nuestros resultados. En la Fig. \ref{stable2}, mostramos el potencial termodin'amico correspondiente al ensamble can'onico $\mathcal{F}=M-TS$ vs $T$ y la entrop'ia $S$ vs $T$.
Para identificar los agujeros negros estables, n'otese que, en el gr'afico de $S$ como funci'on de la temperatura, la pendiente positiva corresponde a la configuraci'on $1$ a temperatura fija. De la ecuaci'on (\ref{secderiv}), esta deber'ia ser la 'unica que tiene concavidad negativa para el potencial termodin'amico. Por lo tanto, en el segundo gr'afico en la Fig. \ref{stable2}, corresponde a aquella con un vapor menor del potencial termodin'amico, indicado como la configuraci'on $1$, tambi'en. A primera vita, podr'ia parecer extra\~no que en AdS los agujeros negros grandes sean lo estables, mientras que en el espacio asint'oticamente plano los estables sean los m'as peque\~nos\footnote{Puesto que en la soluci'on con pelo asint'oticamente plana no contiene un par'ametro de longitud, como AdS, agujeros negros peque\~nos deben entenderse como $S\ll Q^2$, para $\alpha$ fijo.} (comparando los agujeros negros a la misma temperatura). Sin embargo, hay una interpretaci'on simple para este resultado. Es actualmente bien sabido
\cite{Hawking:1982dh} que los espaciotiempos AdS actu'an como una caja y, entonces, cuando el horizonte del agujero negro es comparable con el radio de AdS, $L$, ellos pueden estar en equilibrio t'ermico estable. Para los agujeros negros con pelo en espaciotiempo asint'oticamente planos, la auto-interacci'on del campo escalar juega un papel de `cavidad'{}. Cuando el radio del horizonte es grande, el potencial del campo escalar toma valores menores (se anula en el borde) y entonces los agujeros negros grandes no son estables, mientras que para los peque\~nos, la auto-interacci'on se vuelve relevante, actuando como una cavidad que permite configuraciones en equilibrio t'ermico estable.

\begin{figure}[H]
	\centering
	\includegraphics[width=8 cm]{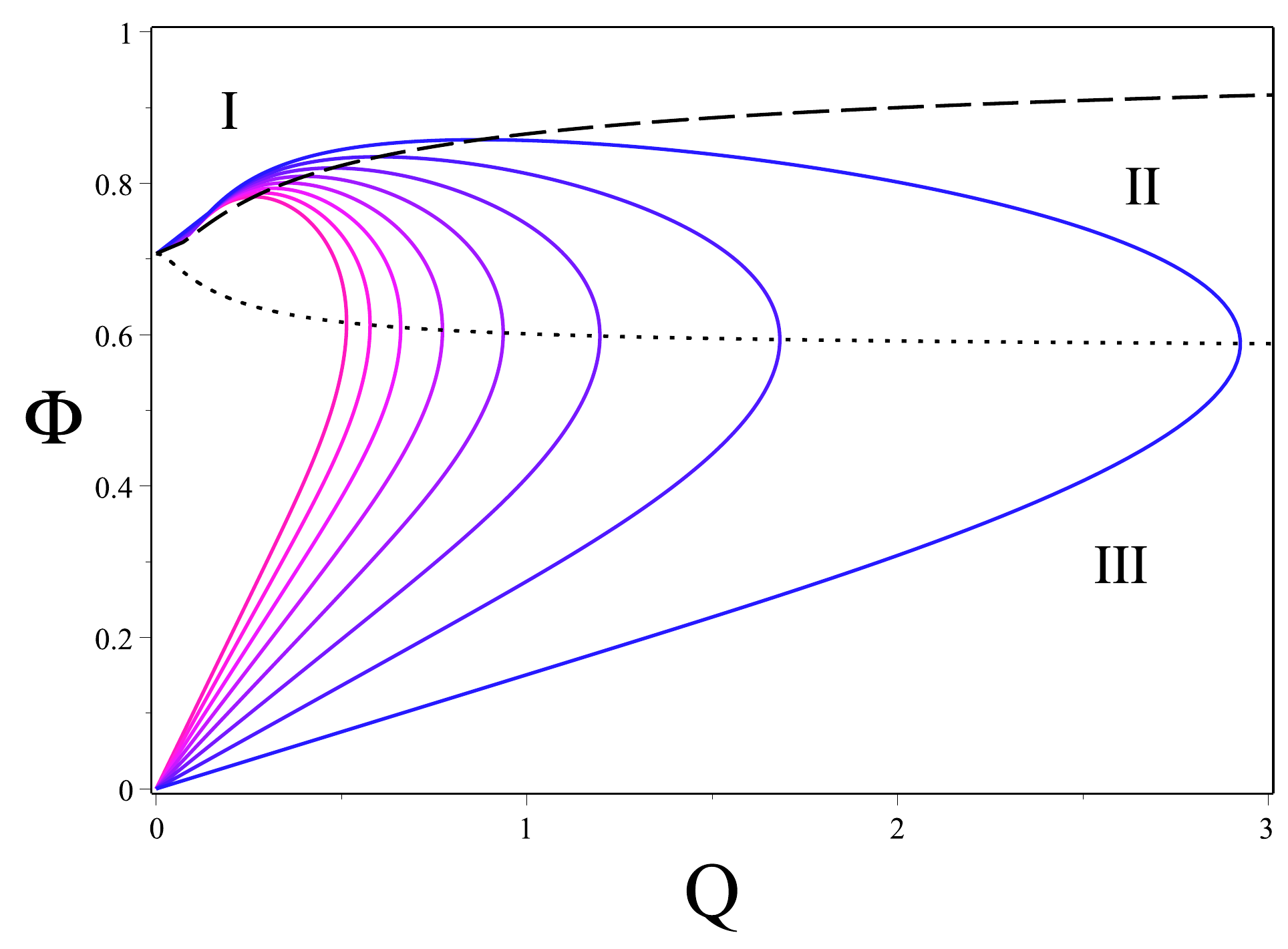}
	\caption{\small{\textbf{Regi'on I:} Esta es la regi'on donde se encuentran los agujeros negros estables. Corresponden al intervalo $\Phi(\epsilon_T=0)<\Phi<\Phi(T=0)$. En el gr'afico, la l'inea con guiones corresponde a $\epsilon_T=0$, y la l'inea punteada corresponde a $\epsilon_T\rightarrow\infty$. \textbf{Regi'on II:} En esta regi'on los agujeros negros son el'ectricamente inestables, aunque $C_Q>0$. \textbf{Regi'on III:} Los agujeros negros son el'ectricamente estables, sin embargo, t'ermicamente inestables $C_Q<0$. Se ha fijado $\alpha=10$.}}
	\label{StabilityRegions}
\end{figure}

Considere la ecuaci'on de estado, mostrada una vez m'as en la Fig. \ref{StabilityRegions} y la condici'on de estabilidad el'ectrica en el ensamble can'onico, $\epsilon_T>0$. Ahora, distinguimos las regiones relevantes, como se detalla abajo.
El resultado nuevo, comparado con el agujero negro de RN, es la existencia de la regi'on $I$, donde $\epsilon_T>0$.
Para ser m'as espec'ifico, comparemos el agujero negro con pelo en la rama positiva con el equivalente en la rama negativa y con el agujero negro de RN (asint'oticamente plano, tambi'en). La regi'on II en todos estos casos es caracter'izada por $\epsilon_T<0$ y $C_Q>0$. Sin embargo, solamente para la rama positiva, la ecuaci'on de estado desarrolla la nueva regi'on I, donde la permitividad el'ectrica cambia su signo, mientras que $C_Q$ preserva la positividad. Es el cambio en el signo de $\epsilon_T$ lo que da lugar a agujeros negros termodin'amicamente estables. Sin embargo, en la perspectiva del ensamble gran can'onico, donde $\Phi$ es fijo, la estabilidad se abre paso gracias a que $C_\Phi$ cambia su signo a positivo, en la regi'on donde $\epsilon_S$ era ya positiva.

\begin{figure}[H]
	\centering
	\includegraphics[width=6 cm]{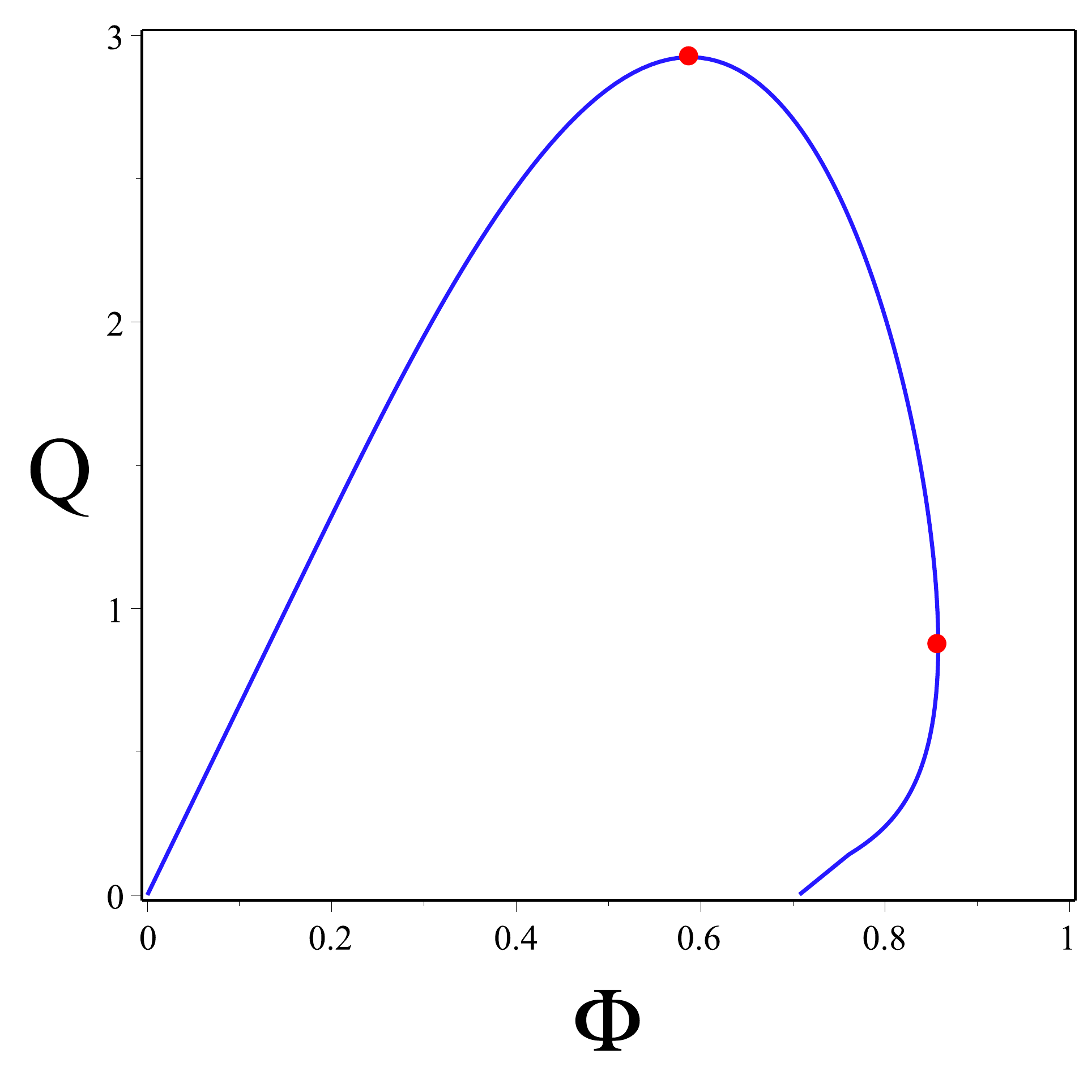}\quad
	\includegraphics[width=6 cm]{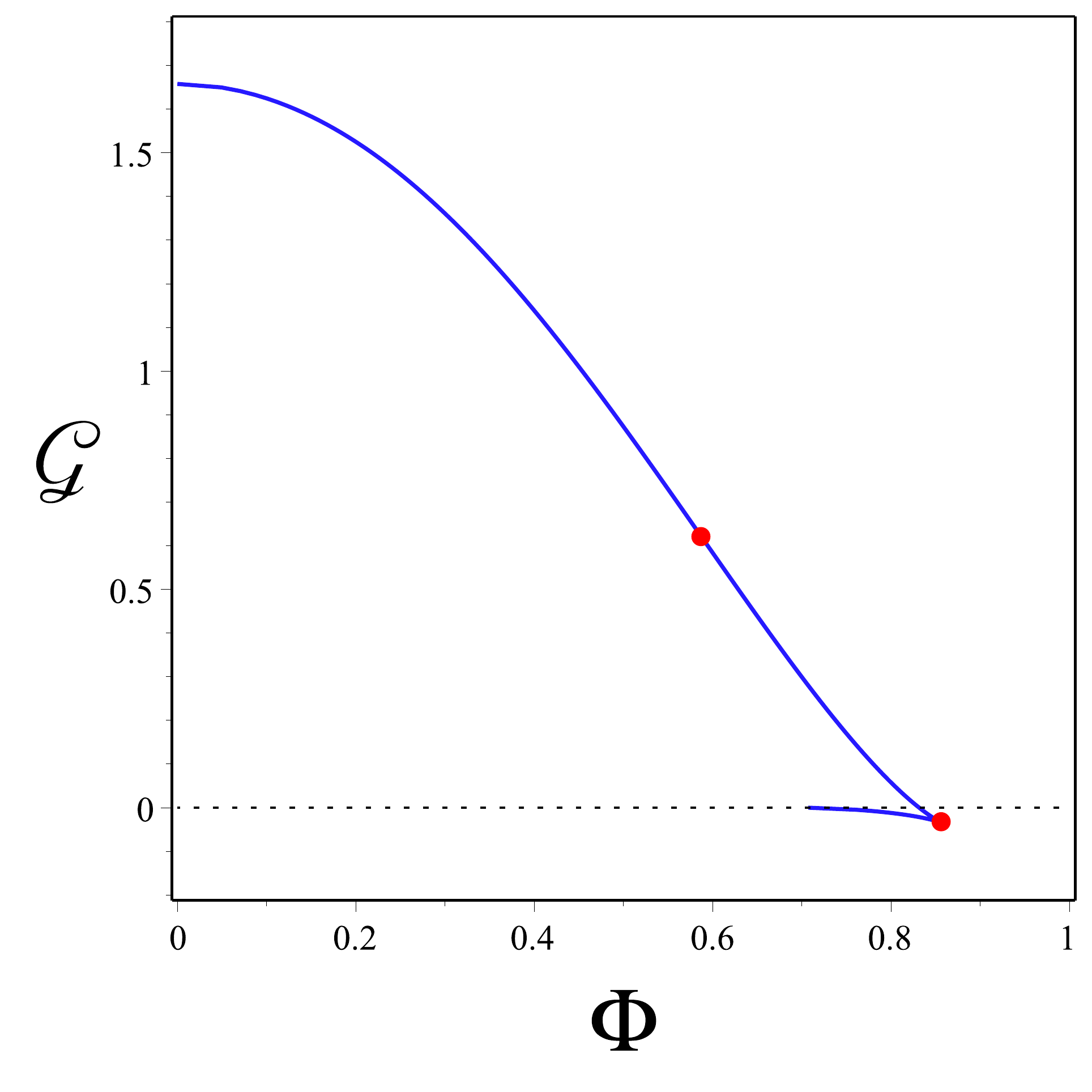}
	\caption{\small{Izquierda: Gr'afico $Q$ vs $\Phi$. Derecha: $\mathcal{G}$ vs $\Phi$, para la isoterma $T=0.012$. Los puntos rojos indican $\epsilon_T=0$ (el punto a la derecha) y $\epsilon_T\rightarrow\infty$ (el punto ubicado arriba).}}
	\label{eSSPhi}
\end{figure}

D'ejenos considerar el mismo resultado desde una perspectiva diferente. Debido a que la primera ley en el ensamble gran can'onico puede ser escrita como
\begin{equation}
dG=-SdT-Qd\Phi
\label{GCensm}
\end{equation}
podemos fijar $T$, para obtener
$dG=-Qd\Phi$. Ahora, integrando, uno obtiene
\begin{equation}
\Delta G 
=-\int Q d\Phi
=-\int_{\Phi=\frac{1}{\sqrt{2}}}^{\Phi=\Phi_{m}} Q d\Phi
-\int_{\Phi=\Phi_m}^{\Phi=0} Q d\Phi
\end{equation}
donde $\Phi_m$ es el m'aximo valor que $\Phi$ asume para una temperatura $T\neq 0$ fija. Por lo tanto, la Fig. \ref{StabilityRegions} provee informaci'on, salvo un factor constante, del potencial termodin'amico como funci'on de $\Phi$ y la comparaci'on es hecha en la Fig. \ref{eSSPhi}.
Como se mencion'o antes, la estabilidad el'ectrica est'a dada por una concavidad negativa en el potencial termodin'amico (como una funci'on de $\Phi$) que puede ser visualizada en la Fig. \ref{eSSPhi}. Los puntos rojos, en aquella figura, indican $\epsilon_T=0$ y $\epsilon_T\rightarrow\infty$. Entre $\Phi=0$ y el primer punto rojo (arriba, en la figura), la concavidad $\mathcal{G}$ es negativa (esto es, $\epsilon_T>0$). Entre ambos puntos rojos, $\epsilon_T<0$ y la concavidad es positiva. Finalmente, entre el segundo punto rojo (abajo a la derecha, en la imagen) y el l'imite $\Phi=1/\sqrt{2}$ ($Q\rightarrow 0$), la concvidad se vuelve positiva tambi'en y $\epsilon_T>0$. Por lo tanto, el gr'afico de $\mathcal{G}$ vs $\Phi$ es consistente con el comportamiento de la soluci'on obtenida de la ecuaci'on de estado.

La existencia de agujeros negros con pelo asint'oticamente planos y termodin'amicamente estables abre la posibilidad de investigar no solamente los diagramas de fase y posibles transiciones de fase, sino tambi'en chequear la estabilidad cl'asica (vea, por ejemplo, \cite{Gross:1982cv, Prestidge:1999uq, Gregory:2001bd, Hertog:2004bb, Gubser:2000ec, Gubser:2000mm, Reall:2001ag}) en este nuevo contexto.




\appendix

\chapter{El t'ermino de Gibbons-Hawking}
\label{apeA}

En la secci'on (\ref{principioaccion}), vimos que, con respecto a la m'etrica, el principio de acci'on $\delta I=0$ est'a garantizado siempre que
\be
\delta\(\frac{1}{2\kappa}\int_{\mathcal{M}}{d^4x\sqrt{-g}g^{\mu\nu}\delta R_{\mu\nu}}+I_{GH}\)=0
\label{a10}
\ee
donde
\be
\label{gh0}
I_{GH}=\frac{1}{\kappa}\int_{\pa\mathcal{M}}{d^3x\sqrt{|h|}K}
\ee
Mostraremos en detalle que (\ref{gh0}) es correcto para condiciones de borde espec'ificas para la m'etrica.

Tomando la variaci\'on del tensor de Ricci $R_{\mu\nu}=\pa_\alpha\Gamma^{\alpha}_{\mu\nu}
-\pa_\nu\Gamma^{\alpha}_{\alpha\mu}
+\Gamma^{\beta}_{\beta\alpha}\Gamma^{\alpha}_{\mu\nu}
-\Gamma^{\beta}_{\nu\alpha}\Gamma^{\alpha}_{\beta\mu} 
$, se tiene que
\be
\delta R_{\mu\nu}=\nabla_\alpha\(\delta\Gamma^{\alpha}_{\mu\nu}\)
-\nabla_\nu\(\delta\Gamma^{\alpha}_{\alpha\mu}\)
\label{r1}
\ee
de manera que
\begin{align*}
\sqrt{-g}g^{\mu\nu}\delta R_{\mu\nu}
&=\sqrt{-g}\,\nabla_\alpha\(g^{\mu\nu}\delta\Gamma^{\alpha}_{\mu\nu}\)
-\sqrt{-g}\,\nabla_\nu\(g^{\mu\nu}\delta\Gamma^{\alpha}_{\alpha\mu}\)\\
&=\pa_\alpha\(\sqrt{-g}g^{\mu\nu}\delta\Gamma^{\alpha}_{\mu\nu}\)
-\pa_\nu\(\sqrt{-g}g^{\mu\nu}\delta\Gamma^{\alpha}_{\alpha\mu}\) \\
&=\pa_\alpha \(\sqrt{-g} v^\alpha\)
\end{align*}
donde $v^\alpha\equiv g^{\mu\nu}\delta\Gamma^{\alpha}_{\mu\nu}
-g^{\mu\alpha}\delta\Gamma^{\nu}_{\nu\mu} $ y, entonces,
\begin{equation}
\frac{1}{2\kappa}\int_{\mathcal{M}}
{d^4x\sqrt{-g}g^{\mu\nu}\delta R_{\mu\nu}}
=\frac{1}{2\kappa}\int_{\pa\mathcal{M}}
{d^3x\,\sqrt{|h|}\epsilon n_\alpha  v^\alpha}
\label{gauss}
\end{equation}
donde $h_{\mu\nu}$ es la m'etrica sobre el borde del espaciotiempo $\pa\mathcal{M}$ que es una hiper-superficie dada por una ecuaci\'on del tipo $\Phi=\Phi(x^\mu)$, con un vector unitario normalizado
\begin{equation}
n_\mu
\equiv \frac{\nabla_\mu\Phi}
{|\nabla_\alpha\Phi\nabla^\alpha\Phi|}
\quad \rightarrow \quad
n^\mu n_\mu=\frac{\nabla_\alpha\Phi\nabla^\alpha\Phi}
{|\nabla_\alpha\Phi\nabla^\alpha\Phi|}=\epsilon\equiv\pm 1
\end{equation}
La relaci\'on de completitud entre $g_{\mu\nu}$ y $h_{\mu\nu}$ es
\begin{equation}
\label{normal}
g^{\mu\nu}=\epsilon\, n^\mu n^\nu+h^{\mu\nu} 
\end{equation}
$h^{\mu\nu}=h^{ab}e^{\mu}_{a}e^{\nu}_{b}$ ($a=1,2,3$) es la extensi\'on de la m\'etrica 3-dimensional $h_{ab}$ sobre $\pa\mathcal{M}$, por medio de las bases $e^\mu_a\equiv \frac{\pa x^\mu}{\pa y^a}$, donde $y^a$ son las coordenadas sobre $\pa\mathcal{M}$,
$$ds^2_{\pa\mathcal{M}}
=g_{\mu\nu}dx^\mu dx^\nu
=g_{\mu\nu}\(e^\mu_ady^a \)\(e^\nu_bdy^b\)=h_{ab}dy^ady^b$$

Ahora, revisaremos brevemente qu\'e condiciones de borde para la m\'etrica son compatibles con el t\'ermino de Gibbons-Hawking (\ref{gh0}). 

Una de aquellas condiciones que debe demandarse es que la m\'etrica est\'e fija en el borde, esto es, $\left.\delta g_{\mu\nu}\right|_{\pa\mathcal{M}}=\left.\delta g^{\mu\nu}\right|_{\pa\mathcal{M}}=0$. Esto nos permite escribir
\begin{align}
\label{varc}
\left.\delta \Gamma^{\alpha}_{\mu\nu}\right|_{\pa\mathcal{M}}=
\frac{1}{2}g^{\alpha\beta}
\(\delta g_{\beta\mu,\nu}
+\delta g_{\beta\nu,\mu}-\delta g_{\mu\nu,\beta}\)
\end{align}
y entonces $\left. v_\alpha\right|_{\pa\mathcal{M}}=g^{\mu\nu}\(\delta g_{\alpha\mu}{,_\nu}
-\delta g_{\mu\nu,\alpha}\)$. Luego, tenemos
\begin{align}
\left.n_\alpha {v^\alpha}\right|_{\pa\mathcal{M}}
&=n^\alpha g^{\mu\nu}
\(\delta g_{\alpha\mu,\nu}-\delta g_{\mu\nu,\alpha}\)\notag\\
&=n^\alpha \(\epsilon n^{\mu}n^{\nu}+h^{\mu\nu}\)
\(\delta g_{\alpha\mu,\nu}-\delta g_{\mu\nu,\alpha}\)\notag\\
&=n^\alpha h^{ab}e^\mu_ae^\nu_b
\(\delta g_{\alpha\mu,\nu}-\delta g_{\mu\nu,\alpha}\)\notag\\
&=n^\alpha h^{ab}e^\mu_a\delta g_{\alpha\mu,b}
-n^\alpha h^{\mu\nu}\delta g_{\mu\nu,\alpha}
\end{align}
En este punto, una segunda condici\'on de borde necesita imponerse para la m\'etrica: que sus derivadas tangenciales sobre el borde sean id'enticamente cero. Esto garantiza que la m'etrica, fija al borde, tome el mismo valor en cada punto de la hipersuperficie $\pa\mathcal{M}$, y se expresa proyectando la derivada parcial de $\delta g_{\mu\nu}$ en las coordenadas $y^a$,
\begin{equation}
\delta g_{\mu\nu,\gamma}e^{\gamma}_{c}=0
\end{equation}
de manera que
\begin{align}
\left.n_\alpha {v^\alpha}\right|_{\pa\mathcal{M}}
=-n^\alpha h^{\mu\nu}\delta g_{\mu\nu,\alpha} \notag
\end{align}
Con este resultado, hemos mostrado que
\be
\frac{1}{2\kappa}\int_{\mathcal{M}}{d^4x\sqrt{-g}g^{\mu\nu}\delta R_{\mu\nu}}
=-\frac{1}{2\kappa}\int_{\pa\mathcal{M}}
{d^3x\sqrt{|h|}\epsilon h^{\mu\nu} n^\alpha \delta g_{\mu\nu,\alpha}}
\label{btmetri}
\ee
\bigskip

En el t'ermino de borde de Gibbons-Hawking, $K$ es la traza del tensor de curvatura extr'inseca, 
$K_{\alpha\beta}\equiv n_{\alpha;\beta}$. 
Entonces, $K=n^\alpha{}_{;\alpha}=\(\epsilon n^\alpha n^\beta+h^{\alpha\beta}\)n_{\alpha;\beta} 
=h^{\alpha\beta}n_{\alpha;\beta}$,
donde usamos $\(n^\alpha n_\alpha\)_{;\beta}
=2n^{\alpha}n_{\alpha;\beta}
=0$. Por lo tanto,
\begin{align}
K=h^{\alpha\beta}
\(n_{\alpha,\beta}-\Gamma^\gamma_{\alpha\beta}n_{\gamma} \)
\end{align}
Ahora, puesto que la variaci'on $\delta K$ afecta s'olamente a los campos din'amicos, en este caso, $g_{\mu\nu}$\footnote{Puesto que $\delta g_{\mu\nu}=0$ en el borde, esto inmediatamente fija $h_{ab}$.} es sencillo ver que, aplicando las condiciones de borde consideradas ya para la m'etrica,
\begin{equation}
\delta K=
-h^{\alpha\beta}\delta\Gamma^\gamma_{\alpha\beta}n_{\gamma} 
=\frac{1}{2}\,h^{\alpha\beta}\delta g_{\alpha\beta,\mu}n^{\mu}\notag
\end{equation}
con lo cual tenemos que la variaci'on del t'ermino de Gibbons-Hawking, suplementada con las condiciones de borde acordadas para la m'etrica es
\begin{equation}
\delta I_B
=\frac{1}{2\kappa}\int_{\pa\mathcal{M}}
{d^3x\sqrt{|h|} \epsilon h^{\mu\nu}
	n^{\alpha} \delta g_{\mu\nu,\alpha}}
\label{btmetri2}
\end{equation}
Sumando las variaciones (\ref{btmetri}) y (\ref{btmetri2}), vemos que (\ref{a10}) se satisface y el principio de acci'on permanece v'alido.


\chapter{El m\'etodo de contrat'erminos y el formalismo de Brown y York}
\label{aB}

Presentamos un breve repaso del formalismo cuasilocal y el m'etodo de los contrat'erminos para espaciotiempos asint'oticamente planos. Luego, calcularemos las cantidades termodin'amicas para el agujero negro de Reissner-Nordstr\"{o}m. Probaremos que la primera ley de la termodin'amica y la relaci'on estad'istica-cu'antica son satisfechas.

Consideramos soluciones est'aticas de agujeros negros en 4 dimensiones en espaciotiempos asint'oticamente planos, y el infinito espacial, que es parte del infinito alcanzado a lo largo de geodesicas espaciales, es lo 'unico relevante para nuestro an'alisis. Brown y York propusieron un tensor de superficie de energ'ia-momento, denominado el `tensor de estr'es cuasilocal', para el campo gravitacional \cite{Brown:1992br}, que es obtenido variando la acci'on con respecto a la m'etrica inducida sobre el borde de la regi'on cuasilocal. 
Una expresi'on concreta para el tensor de estr'es cuasilocal cuando el borde espacial es empujado al infinito fue dado en \cite{Astefanesei:2005ad}:
\begin{equation}
\label{stress}
\tau_{ab}=\frac{2}{\sqrt{-h}}
\frac{\delta I}{\delta h^{ab}} = \frac{1}{\kappa}\left[
K_{ab}-h_{ab}K-\Psi
\left(\mathcal{R}^{(3)}_{ab}-\mathcal{R}^{(3)}h_{ab}
\right)-h_{ab}\Box \Psi+\Psi_{;ab}\right]
\end{equation}
donde $\Psi=\sqrt{{2}/{\mathcal{R}^{(3)}}}$. Este fue obtenido variando la acci'on suplementada con el contrat'ermino gravitacional  \cite{Lau:1999dp,Mann:1999pc,Kraus:1999di} en 4 dimensiones:
\begin{equation}
\label{actcounterterm}
I=I_{\text{bulk}} + I_{\text{GH}} + I_{\text{ct}}
, \qquad
I_{ct}=
-\frac{1}{\kappa}\int_{\pa\mathcal{M}}
{d^3x\sqrt{-h}\sqrt{2\mathcal{R}^{(3)}}}
\end{equation}
donde $I_{GH}$ es el t'ermino de borde de Gibbons-Hawking.

Con este m'etodo, las dificultades asociadas con la elecci'on del background de referencia es evitada.

Una vez que el tensor de estr'es cuasilocal es conocido, las cantidades conservadas pueden ser obtenidas provisto que la superficie cuasilocal tenga una isometr'ia generada por un vector de Killing $\xi^{\mu}$.
Si el vector de Killing es $\xi=\partial/\partial t$, la energ'ia total del sistema gravitacional es \cite{Brown:1992br}
\begin{equation}
\label{energy1}
E=\oint_{s_\infty^2}
{d^2\sigma\sqrt{\sigma}n^a\xi^b\tau_{ab}}
\end{equation}
donde $n^a$ es el vector unitario normal a la superficie $s_\infty^2$ en el borde y a $t=constant$. $\xi^a=\delta^a_t$ es, entonces, el vector de Killing debido a la simetr'ia de traslaci'on temporal del tensor m'etrico y $\sigma$ es el determinante de la m'etrica sobre $s_\infty^2$. Desde un punto de vista f'isico, la existencia de la isometr'ia de la hipersuperficie con la m'etrica inducida $h_{ab}$ significa que una colleci'on de observadores sobre aquella hipersuperficie, todos ellos miden el mismo valor para la energ'ia cuasilocal.

Antes de presentar ejemplos concretos, quisi'eramos enfatizar que el tensor de estr'es cuasilocal puede ser calculado en la secci'on Lorentziana. Por otra parte, ya que necesitamos un rango finito para la coordenada temporal  para obtener una acci'on regularizada, el c'alculo de la acci'on es siempre hecho en la secci'on Euclidiana.

La acci'on Euclidiana, $I^E$, est'a relacionada con el potencial termodin'amica del ensemble gran can'onico, que corresponde a $\Phi=constant$, donde $\Phi$ es el potencial conjugado a la carga el'ectrica $Q$, por medio de la relaci'on estad'istico-cu\'antica
\begin{equation}
\label{freeenrgy1}
\mathcal{G}(T, \Phi)=\beta^{-1}{I^E}=M-TS-\Phi Q
\end{equation}
donde $\beta$, como se discuti'o en cap'itulos previos, es la periodicidad del tiempo Euclidiano. El potencial termodin'amico en el ensemble can'onico ($Q=constant$) puede ser obtenido mediante una transformaci'on de Legendre de (\ref{freeenrgy1}), 
\begin{equation}
\mathcal{F}(T,Q)=\mathcal{G}(T,\Phi)+\Phi Q
\end{equation}
Geom'etricalmente, el potencial termodin'amico en el ensemble can'onico se obtiene al agregar un t'ermino de borde extra en la acci'on, compatible con la condici'on de carga el'ectrica fija,
\begin{equation}
\label{canterm}
I_A=\frac{2}{\kappa}\int_{\pa\mathcal{M}}
{d^3x\sqrt{-h}\,n_\nu F^{\mu\nu}A_\nu}
\end{equation}
En tal caso, obtenemos la nueva acci'on $\bar I=I+I_A$ tal que
\begin{equation}
\label{freeenrgy3}
\mathcal{F}(T, Q)=\beta^{-1}{\bar I^E} = M-TS \,,\qquad 
\end{equation}

%
%

En seguida, obtendremos la acci'on regularizada en la secci'on Euclidiana, el tensor de estr'es cuasilocal y las cargas conservadas para el agujero negro de Reissner-Nordstr\"om. La m'etrica y el potencial de gauge son
\begin{align}
\label{rnmetric}
ds^2&=-f(r)dt^2+f(r)^{-1}dr^2+r^2\(d\theta^2+\sin^2\theta d\varphi^2\) \\
\label{rnpotential}
A&=\left(\frac{q}{r}-\frac{q}{r_+} \right)dt
\end{align}
donde $f(r)=1-2m/r+q^2/r^2$. La coordenada del horizonte de eventos $r_+$ satisface la ecuaci'on $f(r_+)=0$. 

Es sencillo probar que la carga el'ectrica, obtenida mediante la ley de Gauss, es
\begin{equation}
Q
=\frac{1}{4\pi}\oint_{s_\infty^2}{\star F}
=\frac{1}{4\pi}
\oint_{s_\infty^2}{\frac{1}{4}\sqrt{-g}\epsilon_{\mu\nu\alpha\beta}
	F^{\mu\nu}dx^\alpha\wedge dx^\beta}
={q}
\end{equation}
donde {$\epsilon$} es el s'imbolo de Levi-Civita (totalmente antisim'etrico), con $\epsilon_{tr\theta\varphi}=1$. 

Las cantidades termodin'amicas asociadas con este agujero negro, temperatura, entrop'ia y potencial conjugado, son obtenidas como es tradicional
\begin{align}
\label{thermod}
T&=\frac{1}{4\pi}\left.\frac{df(r)}{dr}\right|_{r=r_+}
=\frac{1}{4\pi r_+}\(1-\frac{q^2}{r_+^2}\) , \quad
S=\frac{\mathcal{A}}{4}
=\pi r_+^2  \\
\Phi&=A_t(r_+)-A_t(\infty)=\frac{Q}{r_+}
\end{align}
donde $\mathcal{A}$ es el 'area del horizonte de eventos.

Consideramos una foliaci'on del espaciotiempo compuesta por hypersuperficies esf'ericas $r=constant$. La normal unitaria a estas hypersuperficies, la curvatura extr'inseca y su traza son
\begin{equation}
n_{\mu}=\frac{\delta_{\mu}^{r}}{\sqrt{g^{rr}}}, \qquad K_{\mu\nu}=\nabla_\mu n_{\nu}, \qquad 
K=g^{\mu\nu}K_{\mu\nu}
\end{equation}
o, concretamente,
\begin{equation}
K_{tt}=-\frac{1}{2}f^{1/2}f', \qquad
K_{\theta\theta}=\frac{K_{\phi\phi}}{\sin^2\theta}=rf^{1/2}, \qquad
K=\frac{1}{2f^{1/2}}\(\frac{4f}{r}+f'\)
\end{equation}
mientras que las componentes del tensor de Ricci y su trata, para estas foliaciones, son
\begin{equation}
\mathcal{R}^{(3)}_{tt}=0, \qquad
\mathcal{R}^{(3)}_{\theta\theta}=\frac{\mathcal{R}^{(3)}_{\phi\phi}}{\sin^2\theta}=1, \qquad
\mathcal{R}^{(3)}=\frac{2}{r^2}
\end{equation}
Es ahora sencillo encontrar las componentes del tensor de estr'es cuasilocal (\ref{stress}),
\begin{align}
\tau_{tt}&=
\frac{2f}{\kappa{r}}\(f^{1/2}-1\)
=\frac{1}{\kappa}\(-\frac{2M}{r^2}
+\frac{3M^2+Q^2}{r^3}\)+\mathcal{O}(r^{-4})\\
\tau_{\theta\theta}
&=-\frac{1}{\kappa}\[r\(f^{1/2}-1\)+\frac{r^2f'}{2f^{1/2}} \]
=-\frac{1}{\kappa}\(\frac{1}{2 r}
+\frac{M}{ r^{2}}\)\(M^2-Q^2\)+\mathcal{O}(r^{-3}) \\
\tau_{\phi\phi}&={\sin^2\theta}\,{\tau_{\theta\theta}}
\end{align}

Ahora calculamos la energ'ia (\ref{energy1}), que es una cantidad conservada asociada al vector de Killing $\xi=\pa/\pa t$. La m'etrica de la 2-superficie relevante es $r^2 (d\theta+\sin^2\theta d\varphi^2)$ y la normal (tipo-tiempo) a la superficie $t=constant$ es $n_a=\delta_a^t/\sqrt{-g^{tt}}$. La 'unica componente del tensor de estr'es cuasilocal relevante es $\tau_{tt}$, entonces
\begin{equation}
\label{massRN}
E =\oint_{s_\infty^2}
{d^2\sigma\sqrt{\sigma}n^a\xi^b\tau_{ab}} 
=4\pi\lim_{r\rightarrow\infty}
{\frac{rf^{3/2}}{4\pi}\(1-f^{1/2}\)}
=M+ \mathcal{O}\(r^{-1}\)
\end{equation}
que, en este caso, coincide con la masa $ADM$ del agujero negro, calculada expandiendo la componente $g_{tt}$ de la m'etrica al infinito espacial. Usando las cantidades termodin'amicas (\ref{thermod}),
la masa cuasilocal (\ref{massRN}), y la ecuaci'on del horizonte, $f(r_+)=0$, uno puede verificar la primera ley de la termodin'amica para el agujero negro de Reissner-Nordstr\"om,
\begin{equation}
dM=T dS+\Phi\,dQ
\end{equation}

Para el ensemble gran can'onico el potencial conjugado es fijo, $\Phi=constant$, y la acci'on Euclidiana, calculada on-shell, satisface la relaci'on estad'istico-cu'antica (\ref{freeenrgy1}). En efecto, usando la relaci'on (\ref{actions}), tenemos los siguientes resultados,
\begin{align}
I_{bulk}^E&=-i\[\frac{4\pi}{2\kappa}\int_{0}^{\beta}d\(-it^E\)
\int_{r_+}^{\infty}{dr r^2\(-F^2\)}
\]=-\frac{8\pi\beta}{\kappa}\frac{Q^2}{2r_+}+\mathcal{O}(r^{-1}) \\
I_{GH}^E&=-i\[\frac{2\pi}{\kappa}\int_0^\beta{d\(-i\tau^E\)}
\(4rf+r^2f'\)\]=\frac{8\pi\beta}{\kappa}\(-r+\frac{3}{2}M\)
+\mathcal{O}(r^{-1}) \\
I_{ct}^E&=-i\[-\frac{8\pi}{\kappa}\int_0^\beta{d\(-i\tau^E\)}
{rf^{1/2}}\]=\frac{8\pi\beta}{\kappa}\(r-M\)+\mathcal{O}(r^{-1})
\label{ctRN}
\end{align}
Agregando estos resultados, y usando las expresiones para las cantidades termodin'amicas, se puede verificar f'acilmente que
\begin{equation}
{I^E}=\beta\(-\frac{Q^2}{2r_+}+\frac{1}{2}M\)=\beta\mathcal{G}
=\beta\(M-TS-\Phi Q\)
\end{equation}
Para calcular el potencial termodin'amica en el ensemble can'onico, necesitamos agregar el t'ermino de borde (\ref{canterm}) en la secci'on Euclidiana,
\begin{equation}
I_A^E=-i\[\frac{8\pi}{\kappa}\int_0^\beta{d\(-i\tau^E\)}
{Q\(\frac{Q}{r}-\frac{Q}{r_+}\)}\]=\frac{8\pi\beta}{\kappa}\frac{Q^2}{r_+}+\mathcal{O}(r^{-1})
\end{equation}
pero $Q^2r_+^{-1}=\Phi Q$, por lo tanto, para el ensemble can'onico
\begin{equation}
\bar I^E=\beta\mathcal{F}=\beta\(M-TS \)
\end{equation}
%


\chapter{Derivaci'on del t'ermino de borde para el campo escalar}
\label{apendaux}

Por simplicidad, digamos que la condici'on de borde para el potencial vectorial es $A=0$ al borde. En tal caso, la variaci'on de la acci'on
\begin{equation}
I=\frac{1}{2\kappa}\int_{\mathcal{M}}
{d^4x\sqrt{-g}
	\left[R-e^{\alpha\phi}F^2-2(\pa\phi)^2\right]}+I_{GH}
\end{equation}
no se anula. Concretamente, se obtiene
\begin{equation}
\delta{I}
=-\frac{2}{\kappa}\int_{\pa\mathcal{M}}
{d^3x\sqrt{-h}n_\mu g^{\mu\nu}\pa_\nu\phi
	\delta\phi}
\label{contribution}
\end{equation}
Considerando el comportamiento del campo escalar en la regi'on asint'oticamente plana,
\begin{equation}
\phi(r)=\phi_{\infty}+\frac{\Sigma}{r}
+\mathcal{O}\(\frac{1}{r^2}\)
\end{equation}
junto con el comportamiento asint'oticamente plano de la m'etrica, se puede verificar que, evaluando on-shell la acci'on en la secci'on Euclidiana (en el l'imite $r\rightarrow\infty$),
\begin{equation}
\label{var}
\delta I^E=-\beta \Sigma \delta\phi_\infty
\end{equation}

Mostraremos a continuaci'on que, para condiciones de borde generales,
\begin{equation}
\Sigma(\phi_\infty)=\frac{dW(\infty)}{d\phi_{\infty}}
\end{equation}
el t'ermino de borde
\begin{equation}
I_\phi=-\frac{2}{\kappa}\int_{\pa\mathcal{M}}
{d^3x\sqrt{-h}\left[\frac{(\phi-\phi_\infty)^2}
	{\Sigma^2}W(\phi_\infty)\right]}
\label{propuesta}
\end{equation}
cancela la variaci'on (\ref{var}). 

En efecto, tomando la variaci'on
\begin{align}
\delta I_\phi&
=-\frac{2}{\kappa}\int_{\pa\mathcal{M}}
{d^3x\sqrt{-h}\left[
	\frac{2(\phi-\phi_\infty)
		(\delta\phi-\delta\phi_\infty)}{\Sigma^2}W
	+\frac{(\phi-\phi_\infty)^2}{\Sigma^2}\delta W
	-\frac{2(\phi-\phi_\infty)^2}{\Sigma^3}W\delta \Sigma\right]} \notag \\
&=-\frac{2}{\kappa}\int_{\pa\mathcal{M}}
{d^3x\sqrt{-h}\(\frac{1}{r^2}\)\Sigma\delta\phi_\infty}\notag
+\mathcal{O}\(r^{-1} \)
\end{align}

Evaluando ahora este resultado en la secci'on Euclidiana (y en el l'imite $r\rightarrow\infty$), obtenemos
\begin{equation}
\delta I_\phi^E=\beta{\Sigma\delta\phi_\infty}
\end{equation}

Por lo tanto, el t'ermino de borde (\ref{propuesta}) es adecuado para la condici'on de borde general para el campo escalar considerada.


\chapter{Estabilidad termodin'amica de las teor'ias sin potencial, $\alpha=0$}
\label{apeC}

En esta secci'on, discutimos brevemente la estabilidad termodin'amica de soluciones est'aticas de las teor'ias de Einstein-Maxwell-dilat\'on, sin potencial para el campo escalar, tanto para $\gamma=1$ como para $\gamma=\sqrt{3}$, con el fin de comparar los resultados con las soluciones equivalentes con el potencial no cero. Estas teor'ias pueden verse como el l'imite $\alpha=0$, en los potencial correspondientes, para las cuales 'unicamente la rama positiva, en cada caso, contiene configuraciones de agujeros negros.

\newpage
\section{$\gamma=1$}

De la ecuaci'on del horizonte (\ref{1gamma}) con $\alpha=0$, $x_+$ puede ser despejado y las cantidades termodin'amicas puede ser escribas de la siguiete manera
\begin{equation}
M=\frac{1}{8\pi T}, \qquad
S=\frac{1-32\pi^2Q^2T^2}{16\pi T^2},\qquad
\Phi=4\pi QT
\end{equation}
Ellas satisfacen la primera ley $dM=TdS+\Phi dQ$ y la tercera ecuaci'on es, de hecho, la ecuaci'on de estado de donde se sigue f'acilmente que $\epsilon_T\geq 0$. Para el agujero negro de RN, la permitividad el'ectrica a temperatura contante contiene dos sectores, uno donde los agujeros negros son el'ectricamente inestables. El campo escalar, sin embargo, cuando no presenta auto-interacci'on, vuelve a todas las configuraciones el'ectricamente estables. Por otra parte, tambi'en se sigue que $\epsilon_S\geq 0$.

En seguida, repasamos la estabilidad t'ermica en cada ensemble, analizando las capacidades cal'oricas correspondientes.

En el gran can'onico, el potencial termodin'amico y la capacidad cal'orica son
\begin{equation}
\mathcal{G}(T,\Phi)=\frac{1-2\Phi^2}{16\pi T^2},
\qquad
C_\Phi=-\frac{(1-2\Phi^2)}{8\pi T^2}
\end{equation}
Puesto que $C_\Phi$ es irremediablemente negativa. No existen configuraciones en equilibrio estable.
 
Pasemos brevemente al ensemble can'onico. En este caso, el potencial termodin'amico y la capacidad cal'orica toman la forma
\begin{equation}
\mathcal{F}(T,Q)=\frac{1+32\pi^2Q^2T^2}{16\pi T^2} \;,\qquad
C_Q=-\frac{1}{8\pi T^2}
\end{equation}

Por lo tanto, aunque el campo escalar trae modificaciones en el comportamiento termodin'amico de los agujeros negros, favoreciendo la estabilidad el'ectrica pero arruinando cualquier posibilidad de equilibrio t'ermico, las configuraciones no gozan de una completa estabilidad termodin'amica.

\section{$\gamma=\sqrt{3}$}

En esta teor'ia, es tambi'en sencillo escribir de una manera simple las cantidades termodin'amicas, eliminando $x_+$ de la ecuaci'on del horizonte (\ref{3gamma}),
\begin{align}
S&={\frac {2\pi \sqrt {2} \left( {M}^{2}+M\sqrt {{M}^{2}+2{Q}^{2}}-{Q}^{2} \right) ^{3/2}} 
{M+\sqrt {{M}^{2}+2\,{Q}^{2}}}} \\
T&={\frac {\sqrt {2}}{8\pi \sqrt {{M}^{2}+M\sqrt {{M}^{2}+2{Q}^{2}}-{Q}^{2}}}} \\
\Phi&={\frac {Q}{M+\sqrt {{M}^{2}+2{Q}^{2}}}}
\label{quant4}
\end{align}
las que satisfacen la primera ley. En ete caso, el potencial conjugado tambi'en est'a restringido al intervalo $0<\Phi<1/\sqrt{2}$. Esto se puede ver resolviendo $Q$ de la (\ref{quant4}),
\begin{equation}
\label{charg3}
Q=\frac {2M\Phi}{1-2\Phi^2}
\end{equation}
o, en otras palabras, cuando $M$ se aproxima a cero, entonces $\Phi\rightarrow 1/\sqrt{2}$. La ecuaci'on de estado se puede obtener anal'iticamente,
\begin{equation}
\Phi=\frac{4\pi T Q}{\sqrt{1+64\pi^2Q^2T^2}}
\end{equation}
de donde se sigue que la permitividad el'ectrica a temperatura fija, $\epsilon_T=\frac{Q}{\Phi\(1-4\Phi^2 \)}$ es positiva. Poniendo $M$ de la ecuaci'on (\ref{charg3}) en la expresi'on para la entrop'ia, podemos obtener la permitividad el'ectrica a entrop'ia fija, $\epsilon_S=\frac{\(1+2\Phi^2\)Q}{1-4\Phi^2}$, que tambi'en es positiva. 

Ahora, revisemos las capacidades cal'oricas. En el ensemble gran can'onico, el potencial termodin'amico y la capacidad cal'orica son
\begin{equation}
\mathcal{G}(T,\Phi)=\frac{\sqrt{1-4\Phi^2}}{16\pi T}\;,
\qquad C_\Phi=-\frac{\sqrt{1-4\Phi^2}}{8\pi T^2}
\end{equation}
indicando que no exiten configuraciones estables. 

En el ensemble can'onico, el potencial termodin'amico y la capacidad cal'orica es
\begin{equation}
\mathcal{F}(T,Q)=\frac{\sqrt{1+64\pi^2Q^2T^2}}
{16\pi T}\;, \qquad
C_Q=-\frac{1+96\pi^2Q^2T^2}
{8\pi T^2\(1+64\pi^2Q^2T^2\)^{3/2}}
\end{equation}
y por lo tanto, obtenemos, de nuevo, que no existen configuraciones termodin'amicamente estables.

Para los casos estudiados, podemos concluir que la existencia de configuraciones de equilibrios estables, como los presentados en el cap'itulo \ref{estabilidad}, est'a relacionada con la auto-interacci'on no trivial del campo escalar, cuando $\alpha\neq 0$. Incluso aunque el acoplamiento entre el campo escalar y el campo de Maxwell mejora la estabilidad el'ectrica de los agujeros negros, las capacidades cal'oricas asumen, en todos los casos vistos, valores negativos, haci'endolos inestables desde un punto de vista termodin'amico.


\chapter{Estabilidad termodin'amica de las teor'ias con potencial, $\gamma=\sqrt{3}$}
\label{apeC2}

En este ap'endice, por completitud, presentamos el an'alisis de la estabildiad local para la soluci'on $\gamma=\sqrt{3}$ presentada en la secci'on (\ref{sec:sols}). Ya que el procedimiento es completamente equivalente al presentado en la secci'on (\ref{sec:therm2}), aqu'i deberemos escribir las expresiones relevantes y presentar los resultados importantes.

\newpage
\section{Ensemble gran can'onico en la rama negativa}
\label{gcnb2}

\begin{figure}[h]
	\centering
	\includegraphics[width=7.4 cm]{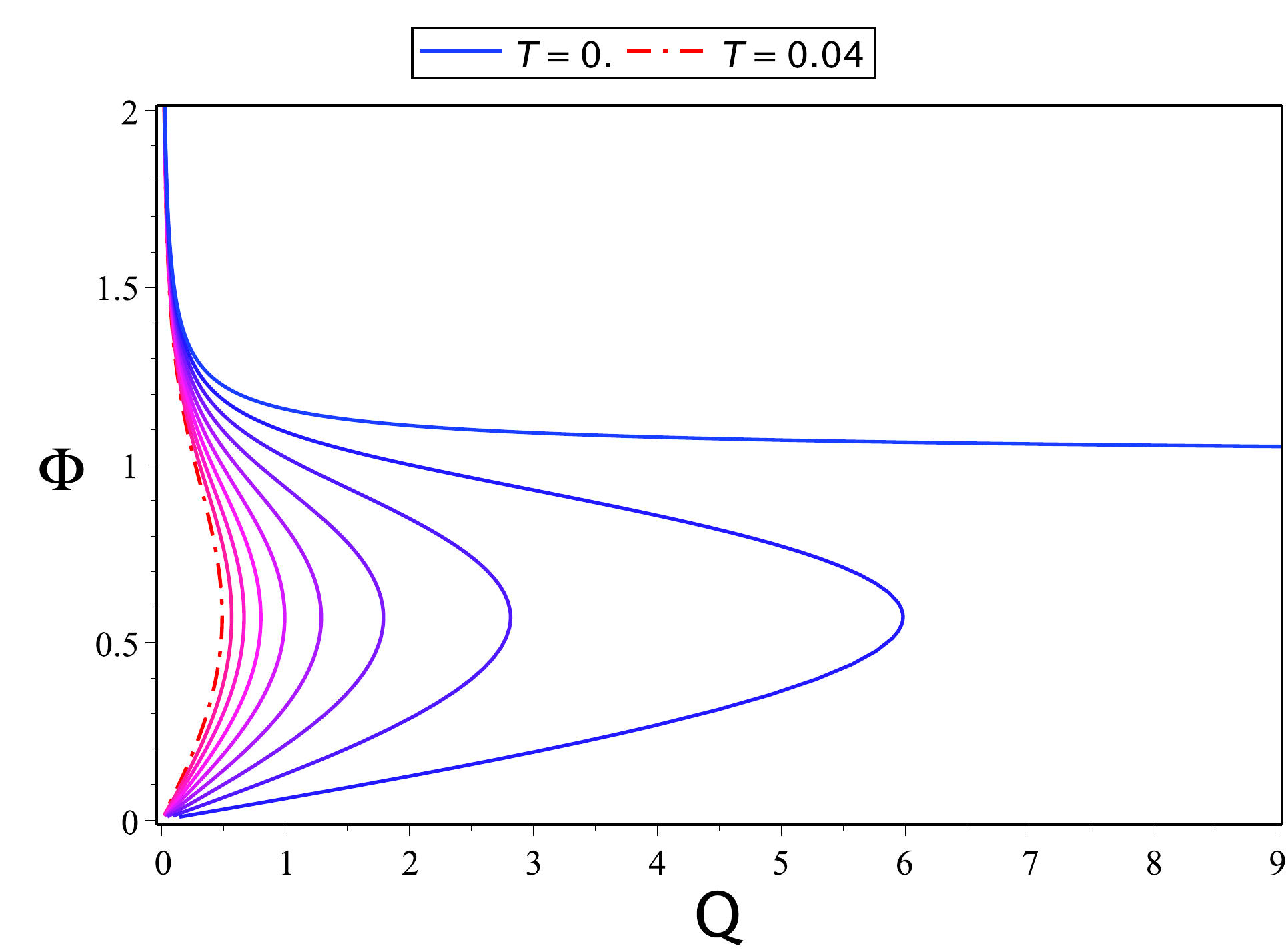}
	\includegraphics[width=7.4 cm]{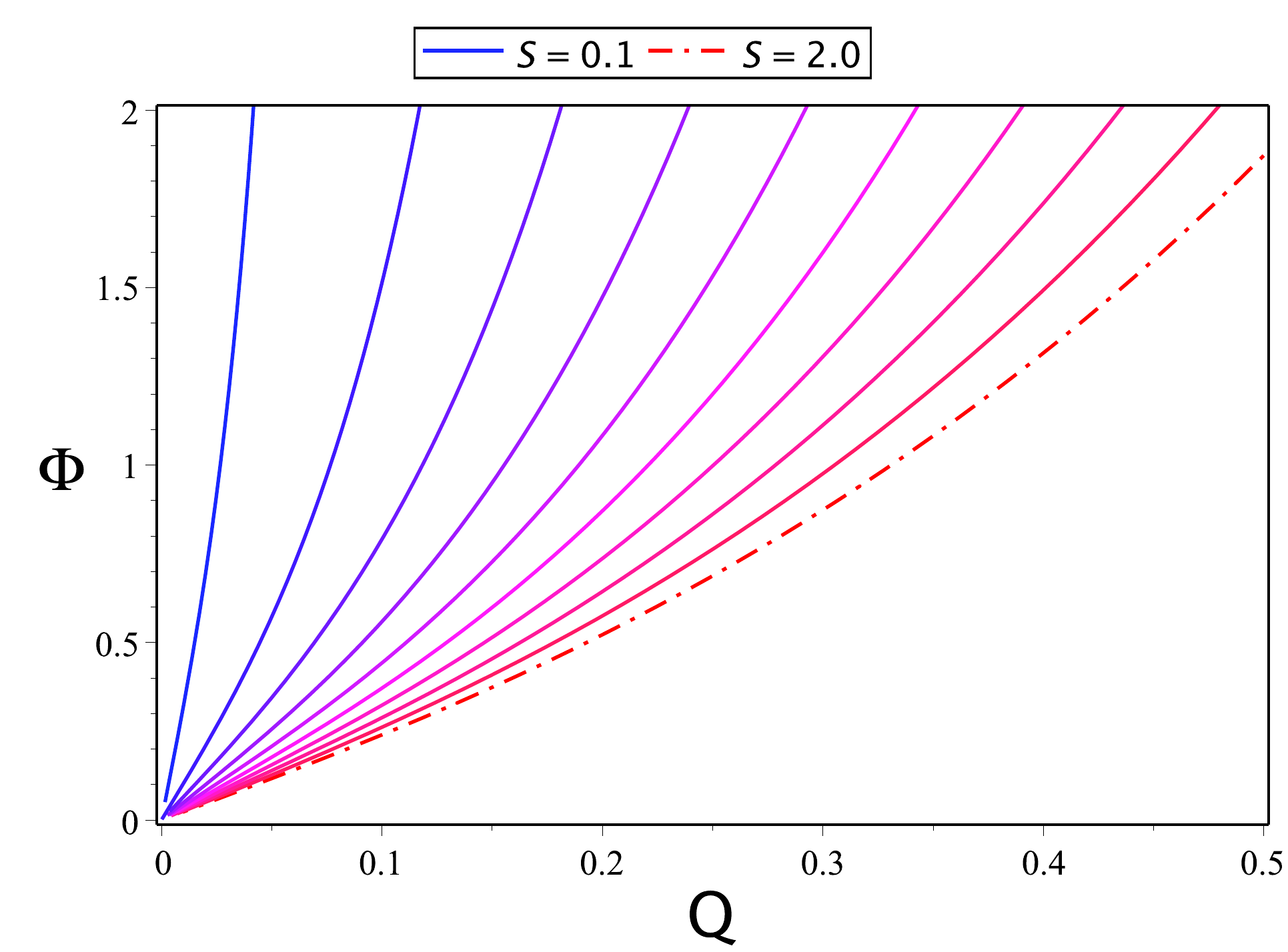}
	\caption{Izquierda: Ecuaci'on de estado, $\gamma=\sqrt{3}$ y $\alpha=10$, en la rama negativa. Derecha: Curvas a entrop'ia fija.}
	\label{state3}
\end{figure}

El potencial dilat'onico para el cual se ha escrito la soluci'on para $\gamma=\sqrt 3$ es dado en (\ref{dilaton2}). La ecuaci'on de estado puede ser estudiada param'etricamente usando la dependencia de la coordenada del horizonte y la temperatura $Q=Q(x_+,T)$ y $\Phi=\Phi(x_+,T)$. Es 'util tambi'e tener las expresiones $Q=Q(x_+,S)$ y $\Phi=\Phi(x_+,S)$. En la Fig. \ref{state3}, se han representado gr'aficamente.

\begin{figure}[h]
	\centering
	\includegraphics[width=4.8 cm]{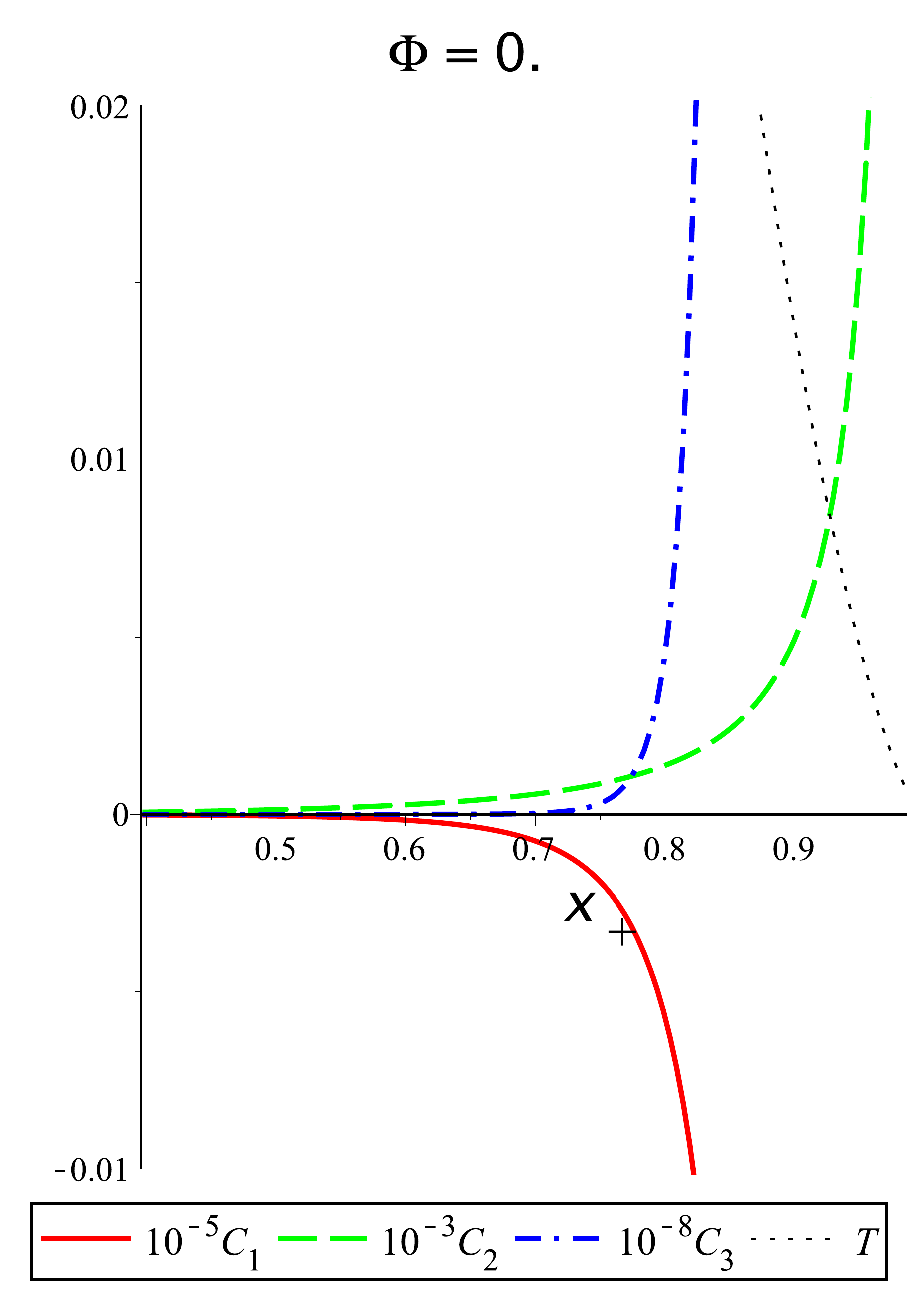}
	\includegraphics[width=4.8 cm]{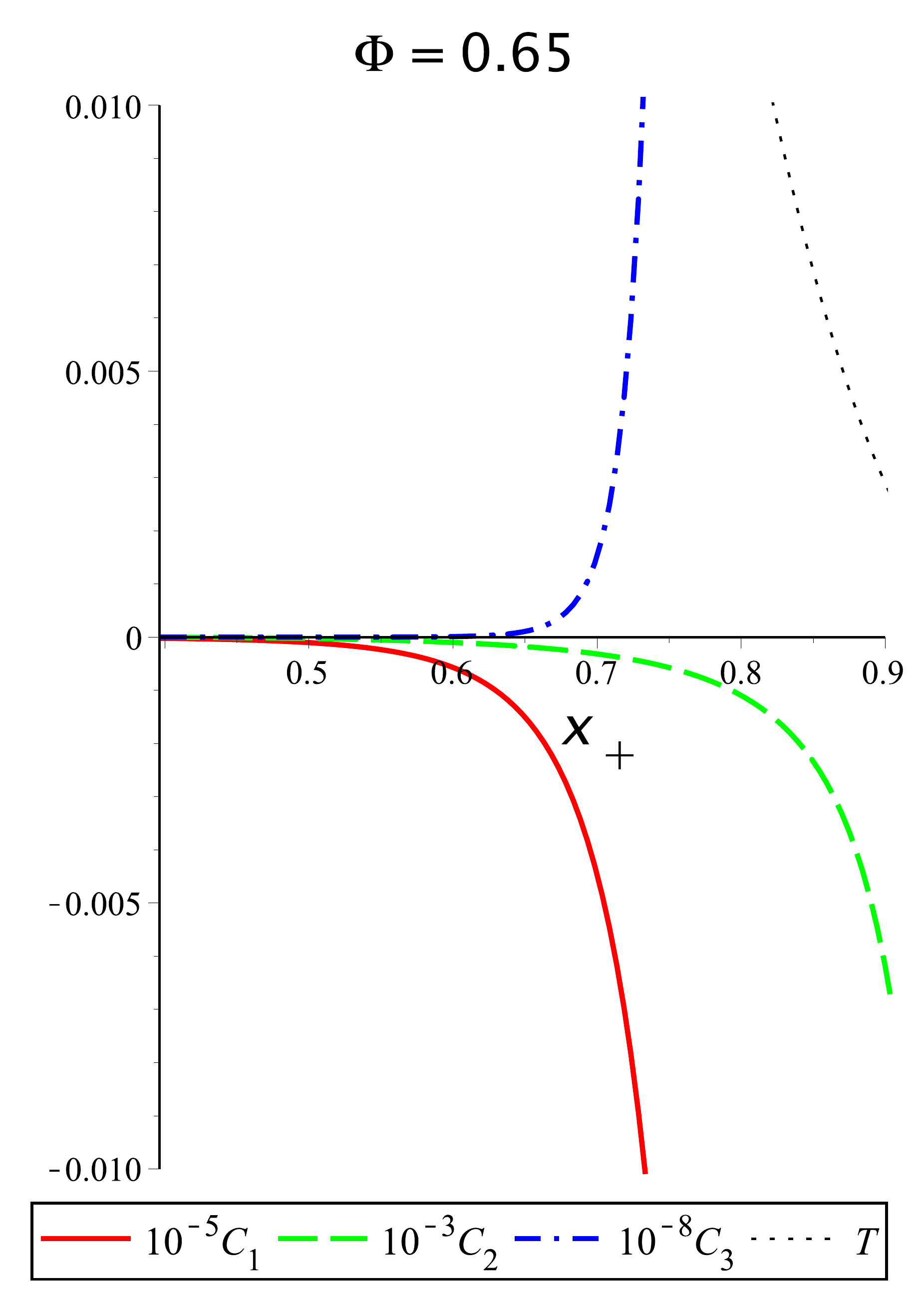}
	\includegraphics[width=4.8 cm]{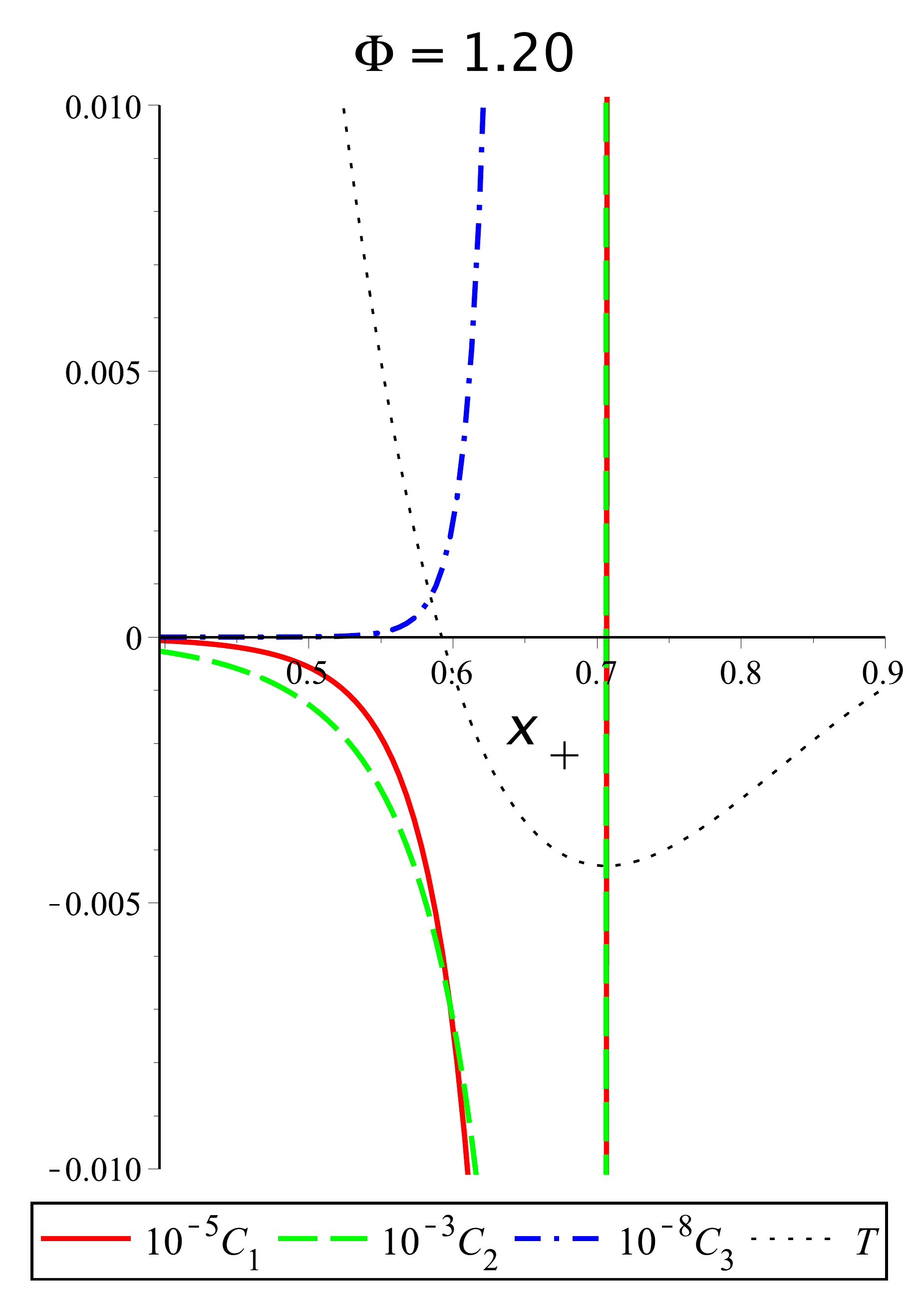}
	\caption{Funciones respuesta en t'erminos de las segundas derivadas de $\mathcal{G}$, para la rama negativa en el gran can'onico, $\gamma=\sqrt{3}$ y $\alpha=10$. La l'inea negra punteada representa $T$, la curva en rojo representa $10^{-5}C_1$; la curva verde $10^{-2}C_2$ y la azul $10^{-6}C_3$.}
	\label{resp5}
\end{figure}

Ahora, usamos $q=2x_+^2\Phi/(1-x_+^2)$ de la ecuaci'on (\ref{quant2}) en la ecuaci'on del horizonte $f(x_+)=0$ para obtener la ra'iz positiva $\eta=\eta(x_+,\Phi)$. Una vez hecho, podemos escribir todas las cantidades termodin'amicas como funciones de $x_+$ y $\Phi$. El potencial termodin'amico es
\begin{equation}
\mathcal{G}(x_+,\Phi)=-\frac{2\alpha}{3\eta^3}
+\frac{2\Phi^2x_+^4}{\eta\(x_+^2-1\)^2}
+\frac{1+x_+^2}{2\eta(1-x_+^2)}
\end{equation}
y la capacidad cal'orica $C_\Phi$, junto con las otras funciones respuestas, ha sido graficada en la Fig. \ref{resp5}, donde se ha usado la misma notaci'on que las dadas en las ecuaciones (\ref{cc2}) y (\ref{cc3}).

\newpage
Puesto que no existe una regi'on en el espacio de par'ametros donde ambas, $\epsilon_S$ y $C_\Phi$, sean positivas a la vez, no existen agujeros negros termodin'amicamente estable en la rama negativa. Las funciones respuesta tienen el mismo comportamiento esquem'atico que para el agujero negro de RN, es decir, tienen signos opuestos para cada configuraci'on.

\newpage
\section{Ensemble gran can'onico en la rama positiva}

\begin{figure}[H]
	\centering
	\includegraphics[width=7.2 cm]{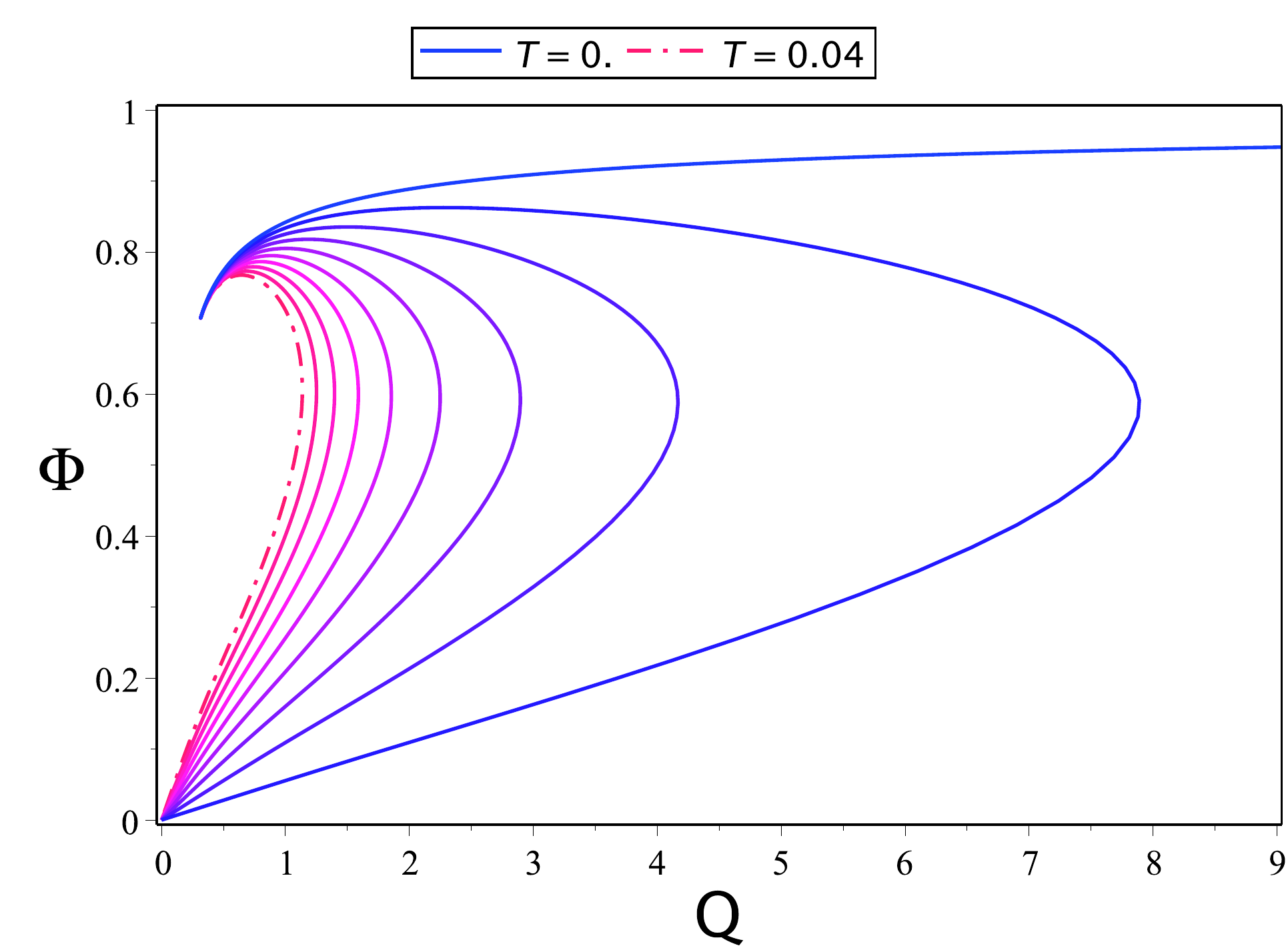}
	\includegraphics[width=7.2 cm]{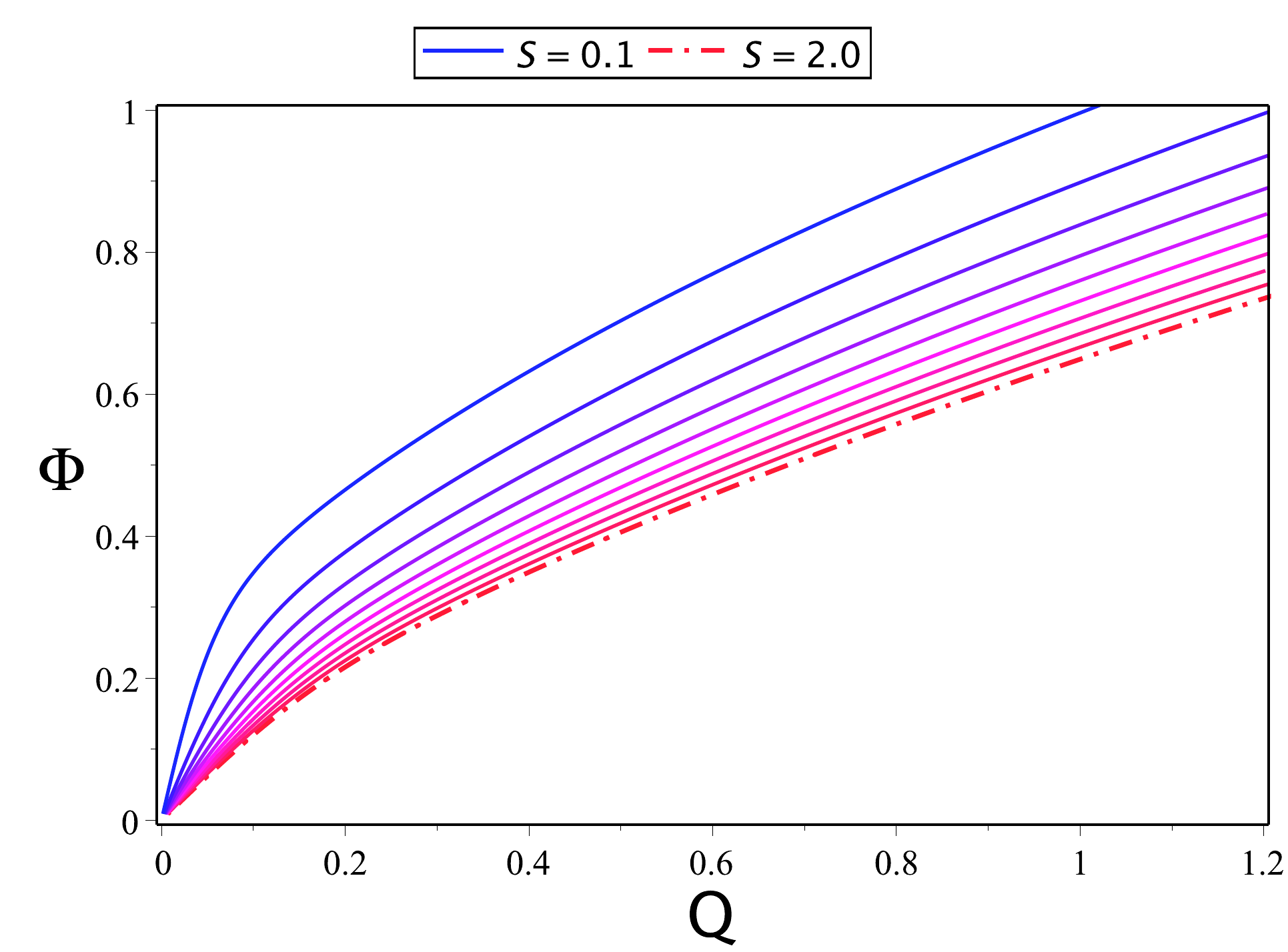}
	\caption{Izquierda: Ecuaci'on de estado en la rama positiva $\gamma=\sqrt{3}$ y $\alpha=10$. Derecha:: $\Phi$ vs $Q$ a entrop'ia fija.}
	\label{state4}
\end{figure}

\begin{figure}[H]
	\centering
	\includegraphics[width=4.86 cm]{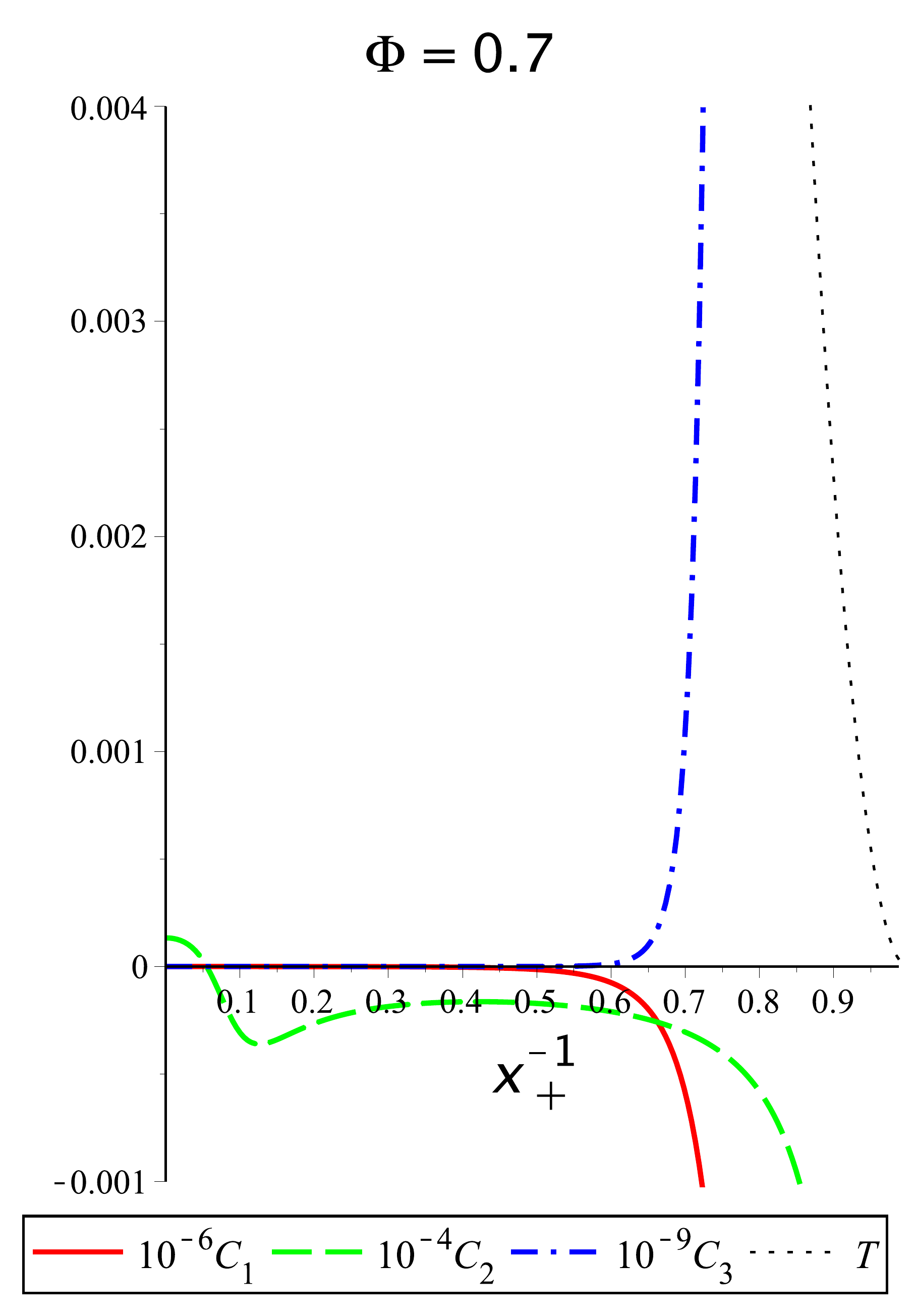}
	\includegraphics[width=4.86 cm]{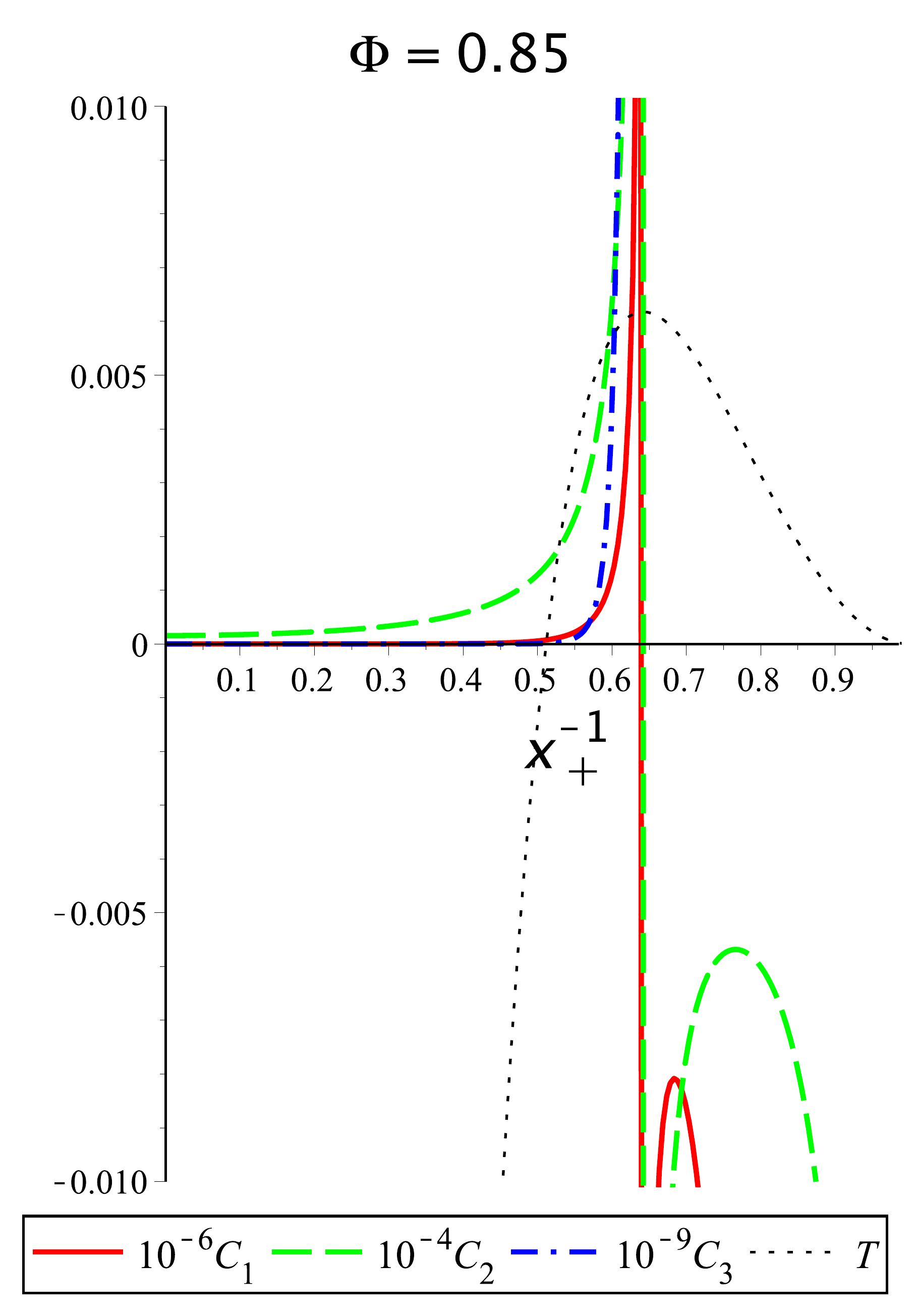}
	\includegraphics[width=4.86 cm]{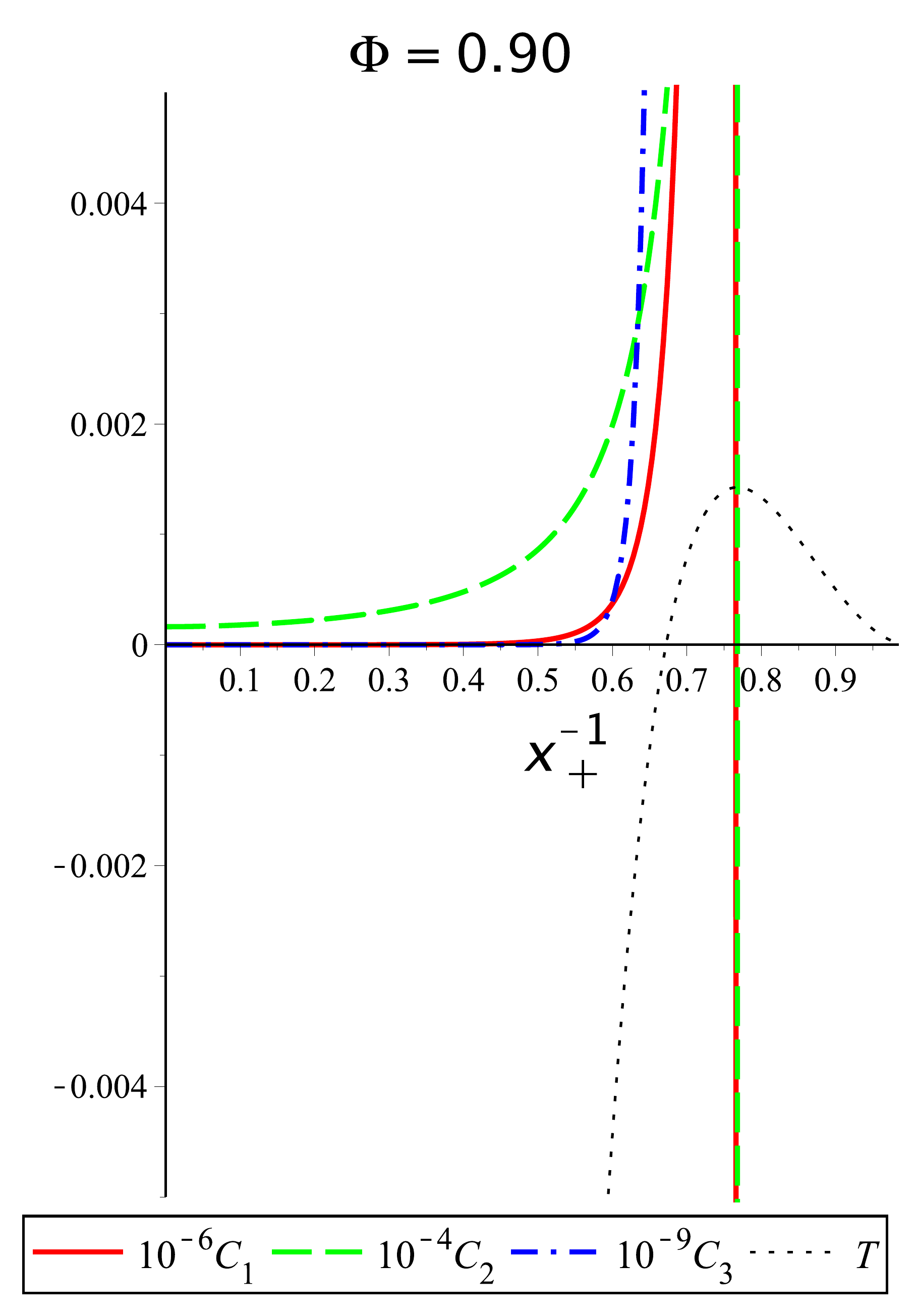}
	\caption{Funciones respuesta en t'erminos de las derivadas de $\mathcal{G}$, para la rama positiva en el gran can'onico. $\gamma=\sqrt{3}$ y $\alpha=10$.}
	\label{resp6}
\end{figure}

La ecuaci'on de estado y tambi'en $\Phi$ vs $Q$ a entrop'ia constante est'an representadas gr'aficamente en la Fig. \ref{state4}. Las funciones respuesta relevantes est'an graficadas en la Fig. \ref{resp6}.

Un aspecto interesante de esta teor'ia, en la rama positiva, es que las isotermas comienzan y terminan en $Q=0$, $\Phi=0$ y $Q= 1/\sqrt{\alpha}$ y $\Phi=1/\sqrt{2}$, respectivamente (esto puede ser probado en una manera similar como en el caso $\gamma=1$).

\begin{figure}[H]
	\centering
	\includegraphics[width=13 cm]{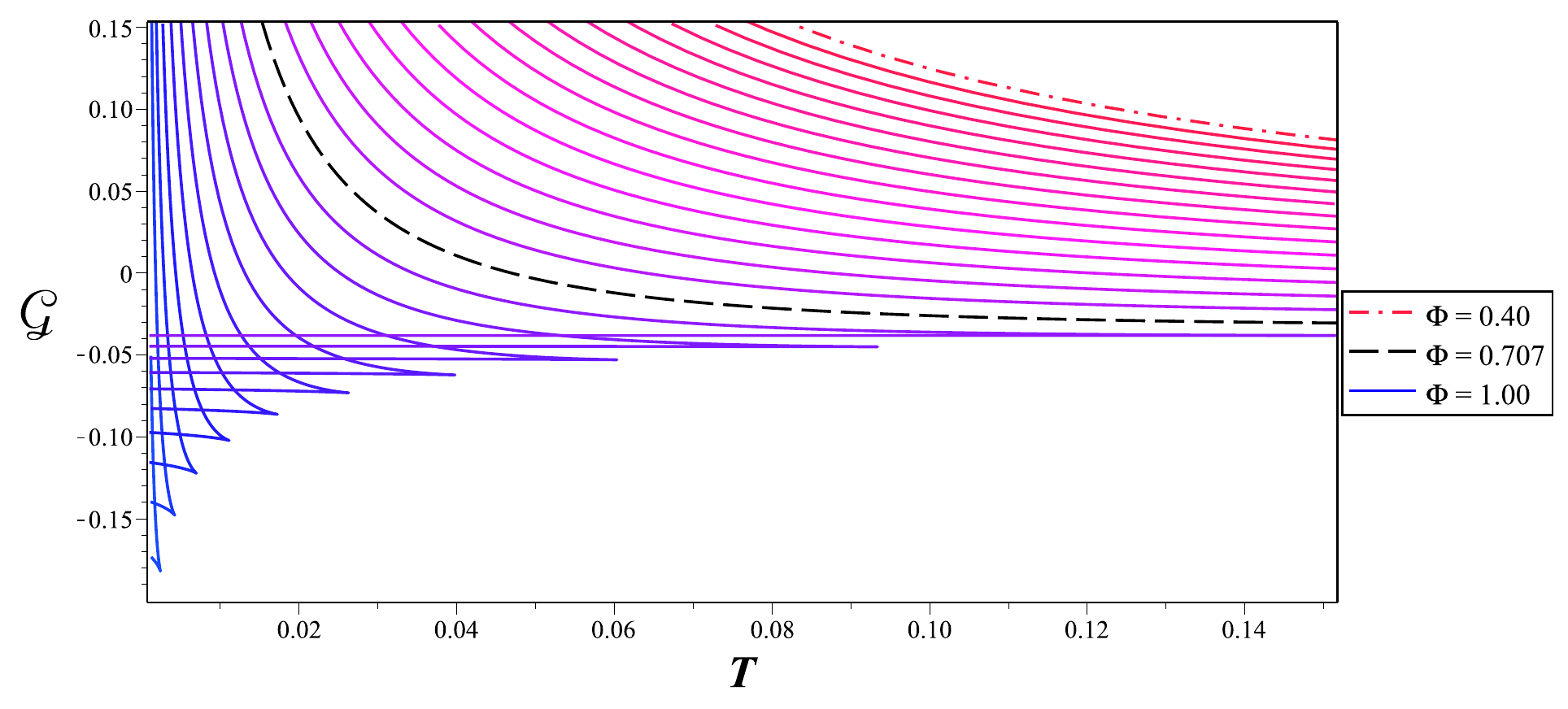}
	\caption{Energ'ia libre $\mathcal{G}$ vs $T$, para $\gamma=\sqrt{3}$ y $\alpha=10$, en la rama positiva.}
	\label{GT2}
\end{figure}

En esta rama, entonces, y tal como en la teor'ia $\gamma=1$ estudiada antes, existen configuraciones cuyas funciones respuesta son ambas positivas, dentro del intervalo $\Phi>\frac{1}{\sqrt{2}}$.
Estos agujeros negros estables tienen un potencial termodin'amico negativo, como puede ser visto en la Fig. \ref{GT2}, siguiendo un mismo comportamiento esquem'atico como en la teor'ia $\gamma=1$.

\newpage
\section{Ensemble can'onico en la rama negativa}
\label{cnb2}

\begin{figure}[H]
	\centering
	\includegraphics[width=6.4 cm]{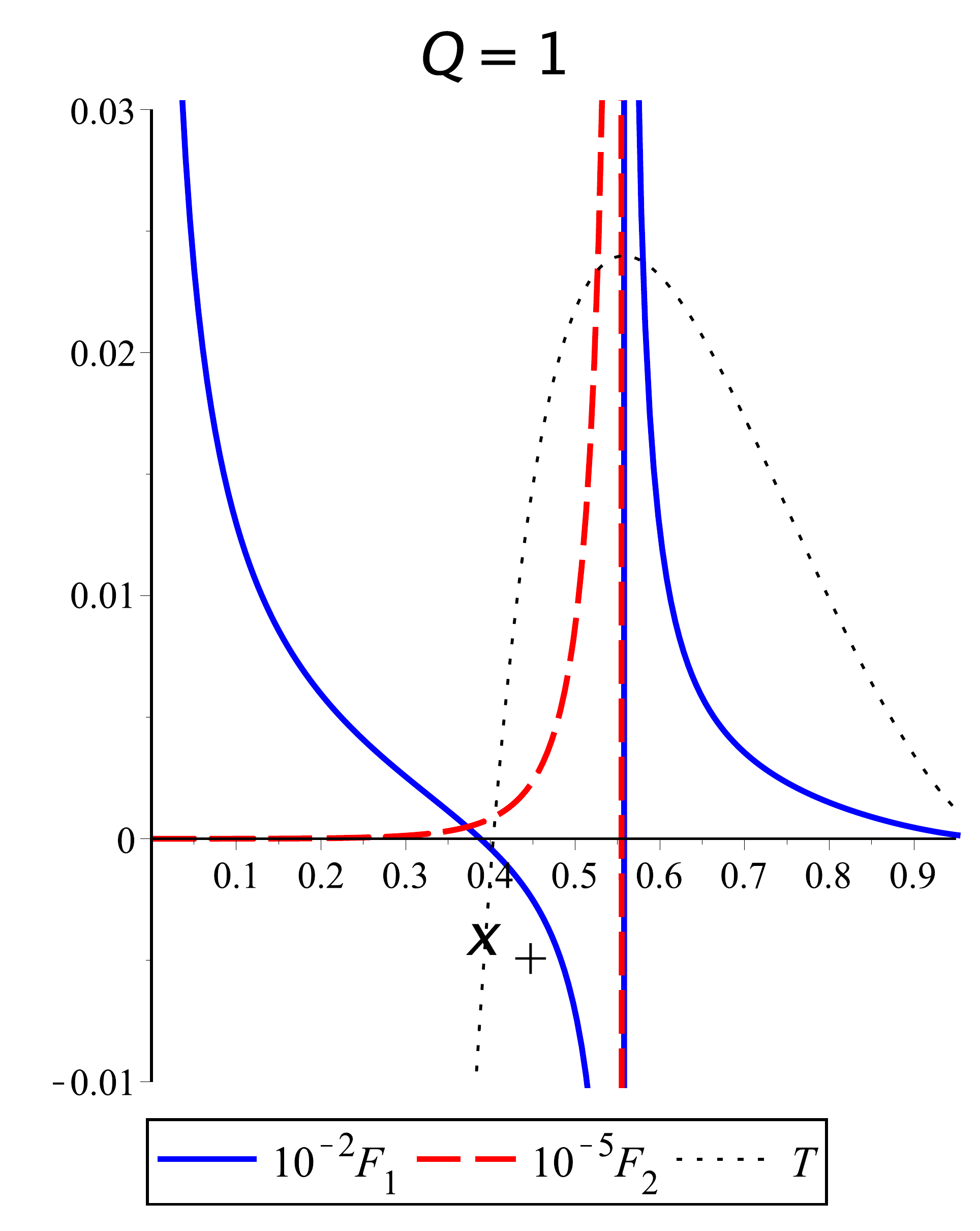}\qquad
	\includegraphics[width=6.4 cm]{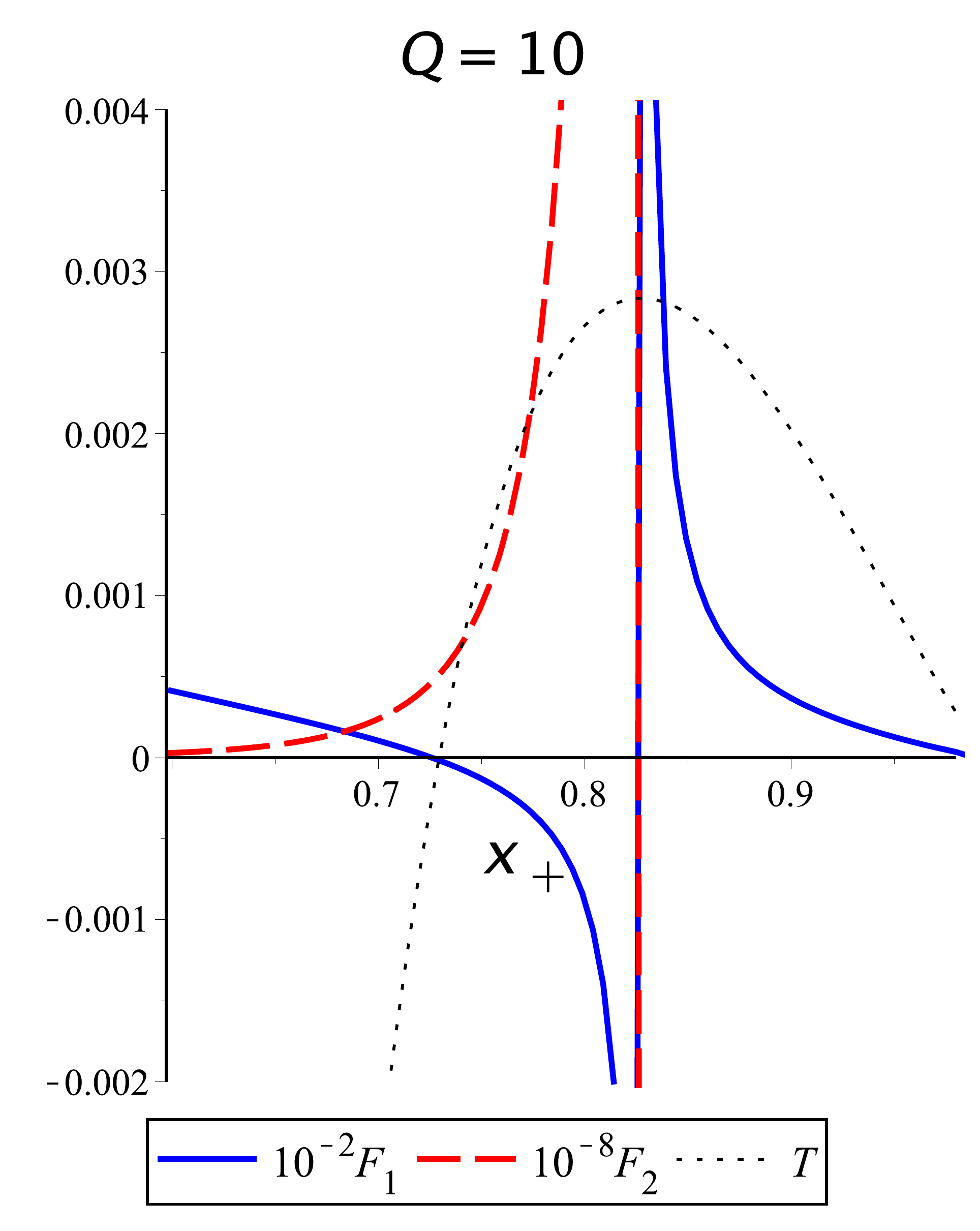}
	\caption{Segundas derivadas para la rama negativa en el ensemble can'onico, $\gamma=\sqrt{3}$ y $\alpha=10$.  $F_1:=\left(\pa^2\mathcal{F}/\pa{Q}^2\right)_T$ y $F_2:=\left(\pa^2\mathcal{F}/\pa{T}^2\right)_Q$.}
	\label{resp7}
\end{figure}

De la Fig. \ref{resp7}, podemos observar que $C_Q$ y $\epsilon_T$ no son simult'aneamente positivos en ninguna regi'on f'isica. Incluso, el agujero negro extremo es el'ectricamente inestable en esta rama, como vimos antes (vea Fig. \ref{state3}) y, por lo tanto, entre $T=0$ y $T_{max}$, para una carga el'ectrica dada, las funciones respuestas tienen signos opuestos. Concluimos que en la rama negativa no existen agujeros negros estables.


\newpage
\section{Ensemble can'onico en la rama positiva}
\label{cpb2}

\begin{figure}[h]
	\centering
	\includegraphics[width=6 cm]{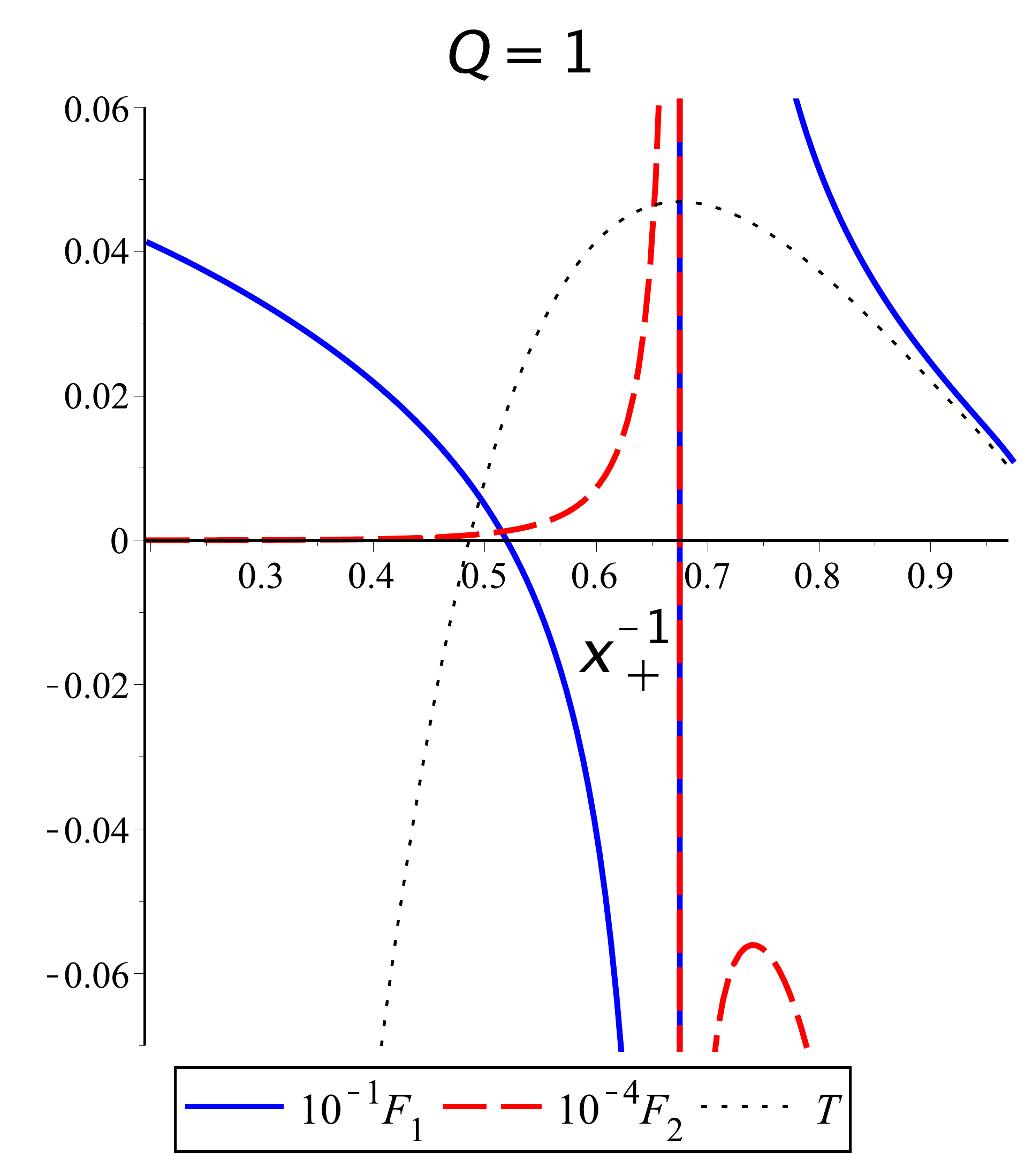}\qquad
	\includegraphics[width=6 cm]{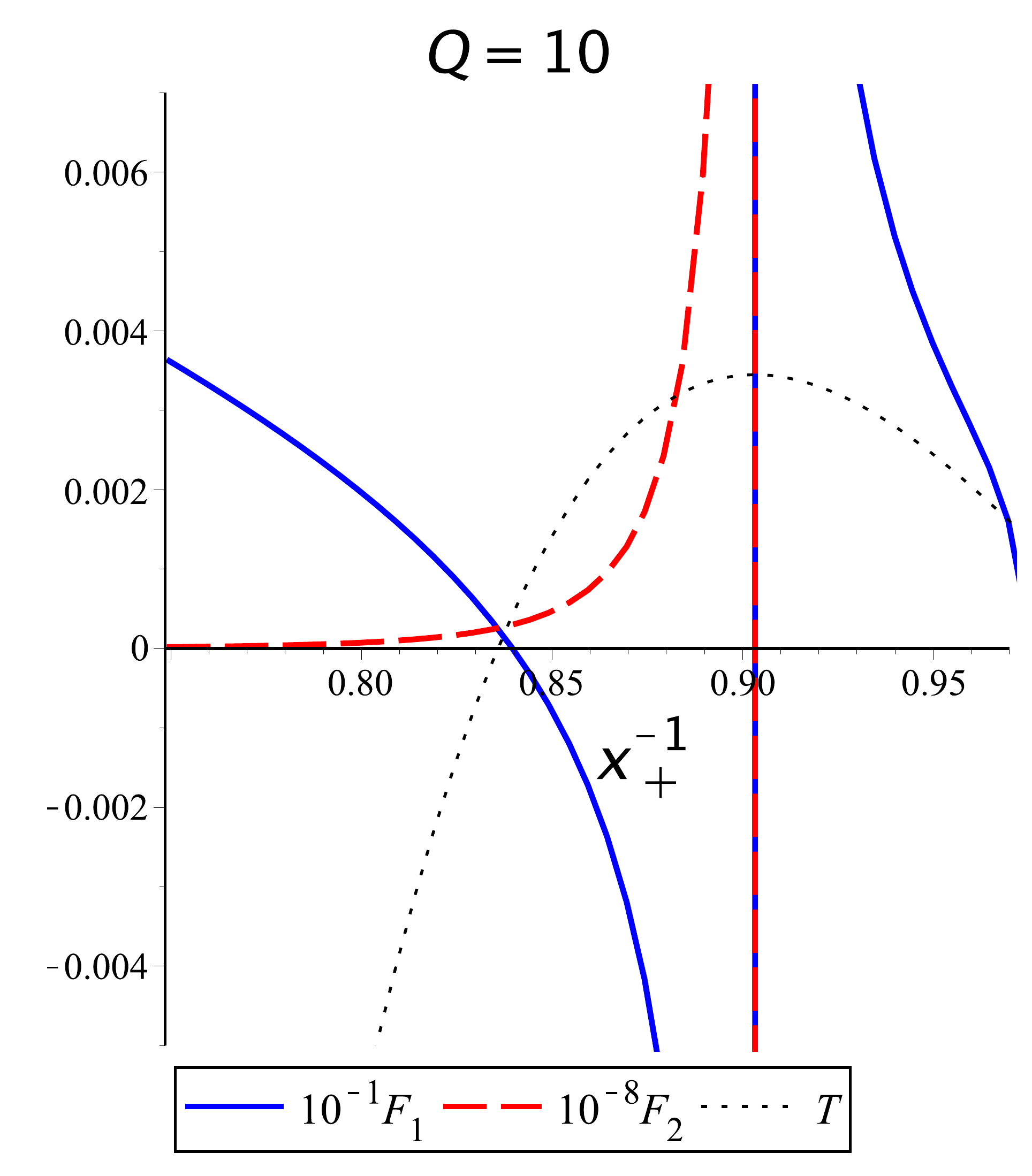}
	\caption{Segundas derivadas para la rama positiva en el ensemble can'onico, para $\gamma=\sqrt{3}$ y $\alpha=10$. La l\'inea negra punteada representa la temperatura, y las convenciones son $F_1:=\left(\pa^2\mathcal{F}/\pa{Q}^2\right)_{T}$ y $F_2:=\left(\pa^2\mathcal{F}/\pa{T}^2\right)_{Q}$, como antes.}
	\label{resp8}
\end{figure}

\begin{figure}[H]
	\centering
	\includegraphics[width=13 cm]{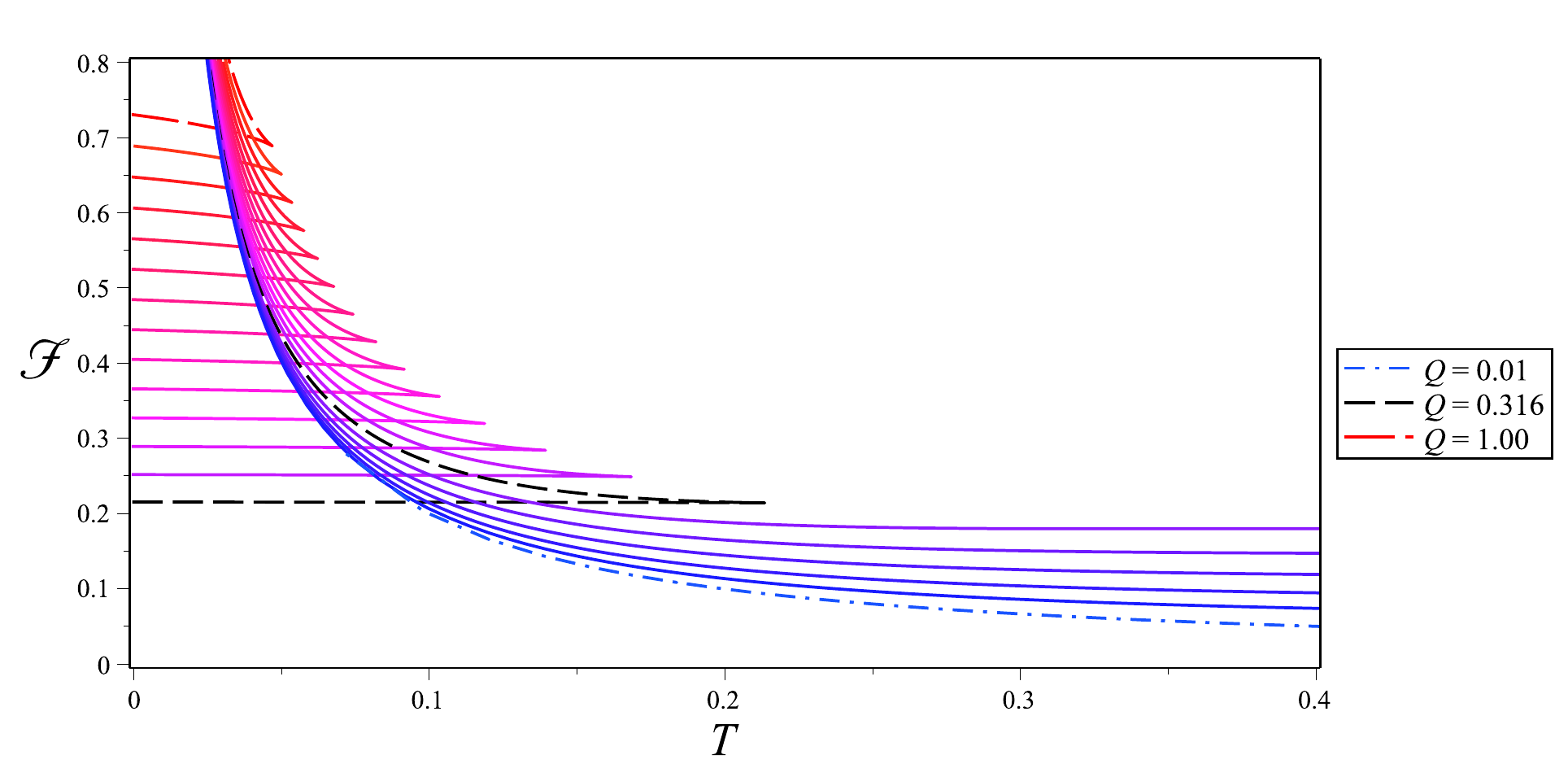}
	\caption{Potencial termodin\'amico $\mathcal{F}$ vs $T$, para $\gamma=\sqrt{3}$. El sector con concavidad negativa ($C_Q>0$) existe para $Q>1/\sqrt{\alpha}$. Como muestra la figura, fijando $\alpha=10$, los agujeros negros estables aparecen para $Q\gtrsim  0.316$.}
	\label{FT2}
\end{figure}

Finalmente, mostramos que existen tambi'en agujeros negros estables en la rama positiva para el ensemble can'onico. El potencial termodin'amico es
\begin{equation}
\mathcal{F}(x_+,Q)=\frac{2\alpha}{3\eta^3}
-\frac{\eta Q^2}{2x_+^2}
+\frac{x_+^2+1}{2\eta\(x_+^2-1\)}
\end{equation}
donde $\eta=\eta(x_+,Q)$ es obtenido de la ecuaci'on del horizonte. Las funciones respuestas est'an graficadas en la Fig. \ref{resp8}, y en la Fig. \ref{FT2} se puede verificar que solamente los agujeros negros con $Q>1/\sqrt{\alpha}$ pueden ser termodin'amicamente estables.

La principal conclusi'on es que, incluso para la teor'ia $\gamma=\sqrt 3$, la auto-interacci'on del campo escalar estabiliza, desde un punto de vista termodin'amico, los agujeros negros con campo escalar.


\end{document}